\documentclass[reprint,superscriptaddress,nofootinbib,onecolumn,amsmath,amssymb,aps,pra,longbibliography]{revtex4-2}

\usepackage{graphicx}
\usepackage{dcolumn}
\usepackage{bm}
\usepackage[colorlinks]{hyperref}
\usepackage{amsmath}
\usepackage{makecell}
\usepackage[mathscr]{eucal}
\usepackage{multirow}

\graphicspath{ {pictures} }

\begin{document}
\title{Quantum interferometers: principles and applications}
\author{Rui-Bo Jin}
\author{Zi-Qi Zeng}
\affiliation{Hubei Key Laboratory of Optical Information and Pattern Recognition, Wuhan Institute of Technology, Wuhan 430205, China}
\author{Chenglong You}
\affiliation{Quantum Photonics Laboratory, Department of Physics and Astronomy, Louisiana State University, Baton Rouge, Louisiana, USA}
\author{Chenzhi Yuan}
\email{chenzhi.yuan@wit.edu.cn}
\affiliation{Hubei Key Laboratory of Optical Information and Pattern Recognition, Wuhan Institute of Technology, Wuhan 430205, China}

\date{\today }

\begin{abstract}
Interference, which refers to the phenomenon associated with the superposition of waves, has played a crucial role in the advancement of physics and finds a wide range of applications in physical and engineering measurements. Interferometers are experimental setups designed to observe and manipulate interference. With the development of technology, many quantum interferometers have been discovered and have become cornerstone tools in the field of quantum physics. Quantum interferometers not only explore the nature of the quantum world but also have extensive applications in quantum information technology, such as quantum communication, quantum computing, and quantum measurement. In this review, we analyze and summarize three typical quantum interferometers: the Hong-Ou-Mandel (HOM) interferometer, the N00N state interferometer, and the Franson interferometer. We focus on the principles and applications of these three interferometers. In the principles section, we present the theoretical models for these interferometers, including single-mode theory and multi-mode theory. In the applications section, we review the applications of these interferometers in quantum communication, computation, and measurement. We hope that this review article will promote the development of quantum interference in both fundamental science and practical engineering applications.
\end{abstract}



\maketitle

\newpage
\tableofcontents
\newpage

\section{Introduction}
Interference plays a crucial role in advancing our understanding of the wave nature of light. Interferometers are designed to exploit the interference of light, allowing us to observe and analyze interference patterns and gain valuable information about the properties of light. Classical interferometers, such as the Michelson interferometer, the Mach-Zehnder interferometer, and the double-slit interferometer, have contributed to experiments that led to significant discoveries in the history of optics \cite{Born1980}. Interferometers are vital not only for fundamental research, but also for various industrial applications. They can accurately measure optical properties such as wavelength, phase, amplitude, coherence, and polarization. Therefore, interferometers serve as indispensable tools for precision measurements, quality control, and the development of advanced optical systems \cite{Hariharan2003}.

With the development of quantum technology, many quantum interferometers, such as the Hong-Ou-Mandel (HOM) interferometer, the N00N state interferometer, and the Franson interferometer have been discovered \cite{Hong1987PRL,Edamatsu2002PRL,Franson1989PRL}. Unlike their classical counterparts, these interferometers rely on the fundamental properties of photons and the principles of second quantization, making their behavior inexplicable by classical theories alone. Quantum interferometers are pivotal in unraveling the mysteries of the quantum realm, revealing key principles and phenomena intrinsic to quantum mechanics.  Moreover, they find extensive applications in quantum information technology, where their ability to precisely manipulate and measure quantum states facilitates breakthroughs in various domains \cite{Zhang2023arXiv,Wei2022LPR,Bongs2019NRP,Dowling2003PTRSA,Pan2012RMP,Luo2021PRA}. For example, in quantum communication, interferometers play a critical role in the encoding and decoding of quantum information, ensuring the secure transmission of cryptographic keys over long distances \cite{Xu2020RMP,Wengerowsky2018Nature,Vajner2022AQT,Kuzmich2003Nature,Luo2023LSA}. 
In the realm of quantum computing, interferometers are instrumental in the implementation of logic gates and the facilitation of quantum computations by taking advantage of the interference patterns of quantum states \cite{2022NRP,OBrien2003Nature,OBrien2007Science,Gyongyosi2019CSR,Lu2023JIII}.
Furthermore, quantum metrology benefits from the unparalleled precision of quantum interferometers, which significantly surpass the limits of classical measurement devices \cite{Eddins2022PRX,Ou2020APL}.

In this review, we provide a comprehensive overview of three prominent examples of quantum interferometers: the HOM interferometer, the N00N state interferometer, and the Franson interferometer. Our discussion begins with the theoretical foundations of each interferometer, covering both single-mode theory and multi-mode theory. We then proceed to categorize the interferometers based on their distinct characteristics and operational principles, aiming to elucidate their unique properties. Finally, we will review the wide range of applications of interferometers across various fields, such as quantum communication, quantum computing, quantum metrology. To provide readers with an overall picture of these interferometers, we include a summary table which outlines the setups, theoretical models (single-mode and multi-mode), principles, advantages, and applications of each interferometer (see Tab. \ref{tab:0}). This summary is designed to help the reader quickly grasp the key aspects of each interferometer.

\begin{table}[htbp]
\caption{The summary of HOM interferometer, the N00N state interferometer, and the Franson interferometer.}
\centering
\begin{tabular}{cccc} 
\hline
  &\makecell{HOM interferometer} & N00N state interferometer&Franson interferometer\\
\hline

Setup&
\makecell{\includegraphics[width= 0.15\textwidth]{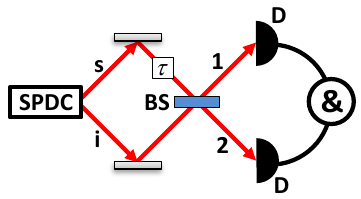}}
& \makecell{\includegraphics[width= 0.20\textwidth]{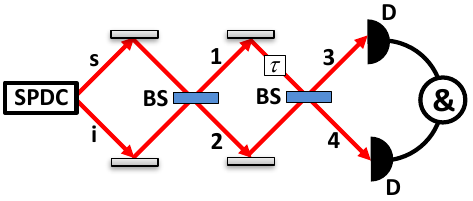}}&\makecell{\includegraphics[width= 0.25\textwidth]{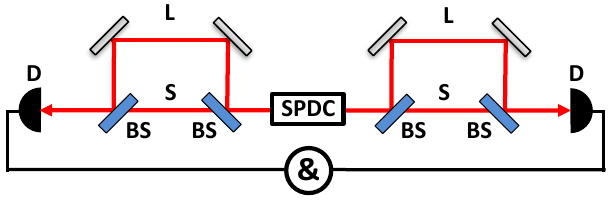}}\\ \\

\makecell{Single-mode \\ model}  &$P = \frac{1}{2}[1 - \delta (t)]$& $P = \frac{1}{2}[1 + \cos (2\omega t)]$ & $P = \frac{1}{4}[1 + \cos (\omega t)]^2$\\ \\

\makecell{Multi-mode \\ model}
 & $\begin{array}{l}
P = \frac{1}{2} - \frac{1}{2}\int  {\int_0^\infty  {d{\omega _1}d{\omega _2}} }  \times \\ \\
{\left| {f({\omega _1},{\omega _2})} \right|^2}\cos [({\omega _1} - {\omega _2})t]
\end{array}$ & $\begin{array}{l}
P = \frac{1}{2} + \frac{1}{2}\int  {\int_0^\infty  {d{\omega _1}d{\omega _2}} }  \times \\\\
\begin{array}{*{20}{c}}
{}
\end{array}{\left| {f({\omega _1},{\omega _2})} \right|^2}\cos [({\omega _1} + {\omega _2})t]
\end{array}$ & $\begin{array}{l}
P = \frac{1}{4}\int  {\int_0^\infty  {d{\omega _1}d{\omega _2}} } {\left| {f({\omega _1},{\omega _2})} \right|^2} \times  \\\\
\begin{array}{*{20}{c}}
{}
\end{array}[1 + \cos ({\omega _1}{t_1})][1 + \cos ({\omega _2}{t_2})]
\end{array}$\\ \\

Principle& \makecell{Anti-bunching effect \\ of bosons}
 & \makecell{Multi-photon  wave-packet \\interference  }& \makecell{Interference of time-frequency \\entanglement} \\ \\
Advantages &\makecell{ Phase  fluctuation \\ insensitive}& \makecell{ Super-resolution  \\ Super-sensitivity }& Nonlocal interference

\\ \\
Applications& \makecell{Quantum communication\\Quantum computation\\Quantum imaging} & \makecell{Quantum measurement\\Quantum  sensing}&\makecell{Quantum communication \\Quantum imaging}

\\
\hline
\end{tabular}
\label{tab:0}
\end{table}

\section{HOM interferometer}
The HOM interferometer is a fundamental tool in quantum optics that allows investigation of quantum interference phenomena. It is named after the physicists C. K. Hong, Z. Y. Ou, and L. Mandel, who first proposed and demonstrated the interferometer in 1987 \cite{Hong1987PRL}. In a typical HOM interferometer, two photons are injected into the separate input ports of a beam splitter. Under certain conditions, the photons can exhibit constructive or destructive interference at the output ports of the beam splitter. The interference pattern depends on the overlap of the two photons as they propagate through the beam splitter. A key characteristic of the HOM interferometer is that when the two photons are perfectly indistinguishable and their arrival times at the beam splitter coincide, they always emerge together from the same output port of the beam splitter. This effect leads to complete suppression of coincidence counts at the two separate output ports, resulting in a dip in the coincidence measurement. This effect is known as the HOM dip.

In this section, we will first introduce the theory of HOM interferometers, covering both single-mode theory and multi-mode theory. Subsequently, we will explore HOM interferometers implemented with various physical systems, such as atoms, photons, phonons, molecules, plasmons. Next, we will highlight several typical HOM interferometers, including HOM interferometer of independent sources, multi-photon HOM interferometer, detection dependent multi-photon HOM interferometer, and spectrally resolved HOM interferometer. Finally, we will discuss the applications of HOM interferometers in quantum communication, quantum computation, quantum metrology, and quantum imaging.

\subsection{Principles of HOM interferometer}
In order to make it easier for the readers to understand the concept of the HOM interferometer, we will present the theoretical models of the HOM interferometer step by step, from the simplest to the most complex. These models include the single-mode model, uncorrelated multi-frequency model, correlated multi-frequency model, multi-temporal model, and four-fold coincidence counting model. Since the beam splitter (BS) is a crucial element in all quantum interferometers, we will begin by discussing the theory of a BS.

\subsubsection{Quantum description of a beam splitter}
%
\begin{figure}[htbp]
\centering
\includegraphics[width= 0.85\textwidth]{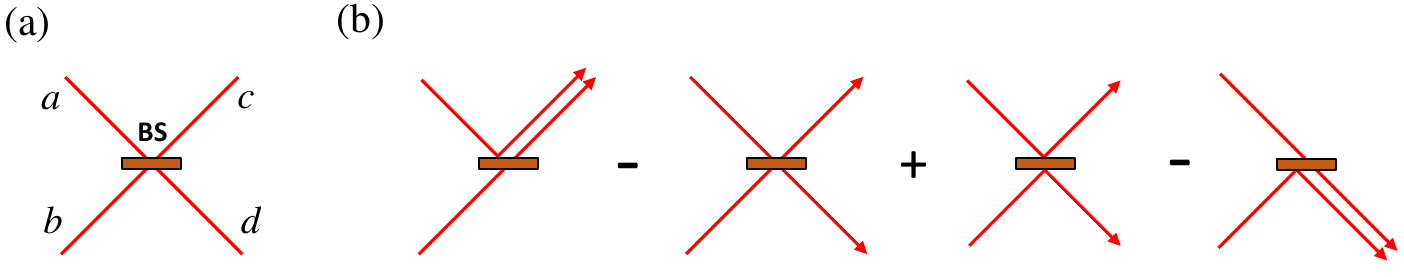}
\caption{(a) The quantum model of a beam splitter. (b) Four combinations of reflections and transmissions.}
\label{11BS}
\end{figure}
%

A depiction of the BS is shown in Fig.\,\ref{11BS}(a). The quantum description of a
50/50 BS can be written as \cite{Gerry2004, Zhang2015}:
\begin{equation}\label{Eq:1-1-1}
\hat{a}^{\dagger}=\frac{1}{\sqrt{2}}\left(\hat{c}^{\dagger}+\hat{d}^{\dagger}\right), \hat{b}^{\dagger}=\frac{1}{\sqrt{2}}\left(\hat{c}^{\dagger}-\hat{d}^{\dagger}\right),
\end{equation}
or in a matrix form:
\begin{equation}\label{Eq:1-1-2}
\left( {\begin{array}{*{20}c}
   {\hat a^\dag  }  \\
   {\hat b^\dag  }  \\
\end{array}} \right) = \frac{1}{{\sqrt 2 }}\left( {\begin{array}{*{20}c}
   1 & 1  \\
   1 & { - 1}  \\
\end{array}} \right)\left( {\begin{array}{*{20}c}
   {\hat c^\dag  }  \\
   {\hat d^\dag  }  \\
\end{array}} \right).
\end{equation}
Here, $ \hat a^\dag$, $ \hat b^\dag$, $ \hat c^\dag$ and $ \hat d^\dag$ are the creation operators for each port. We note that the reflection off the bottom side of the beam splitter introduces a relative phase shift by a factor of $\exp{(i \pi)}=-1$. When two photons are sent to the input ports of the beam splitter, there are four combinations of transmission and reflection as shown in Fig.\,\ref{11BS}(b).

\subsubsection{The single-mode theory of HOM interferometer}
%
\begin{figure}[htbp]
\centering
\includegraphics[width=0.95\textwidth]{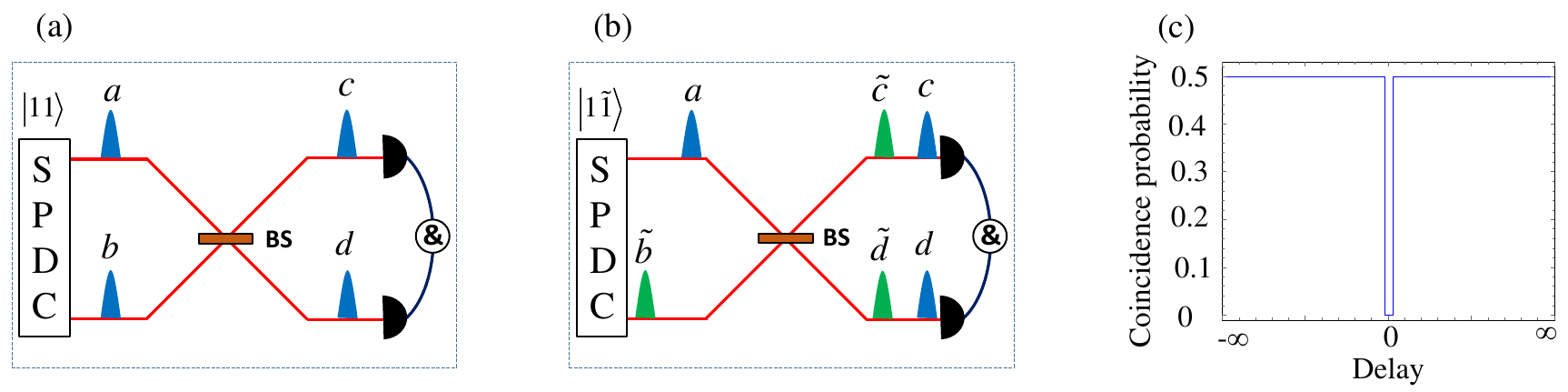}
\caption{The setup of single-mode HOM interferometer when (a) two photons are completely overlapped and (b) two photons are not overlapped. (c) The corresponding HOM interference pattern.}
\label{HOMIsimple}
\end{figure}
%
In the single-mode theory of the HOM interferometer, we assume the photons only have one mode in all degree of freedoms.  Here, the input single photons can be the signal and idler photons from a spontaneous parametric down-conversion (SPDC) source. 
For simplicity, we assume the single photons are ideal, e.g., the temporal width of the single photons is a Dirac delta function.
First, we consider the case where the two input photons completely overlap, as shown in Fig. \ref{HOMIsimple}(a). The input state is given by ${\left| \psi  \right\rangle _{\text{in}}} = \hat a^\dag \hat b^\dag {\left| {00} \right\rangle }={\left| {11} \right\rangle}.$ 
The output state $|\psi\rangle_{\text{out}}$ after the BS can be calculated as

\begin{equation}\label{Eq:1-2-2}
\begin{aligned}
|\psi\rangle_{\text {out }} & =\frac{1}{2}\left(\hat{c}^{\dagger}+\hat{d}^{\dagger}\right)\left(\hat{c}^{\dagger}-\hat{d}^{\dagger}\right)|00\rangle \\
& =\frac{1}{2}\left(\hat{c}^{\dagger} \hat{c}^{\dagger}-\hat{c}^{\dagger} \hat{d}^{\dagger}+\hat{d}^{\dagger} \hat{c}^{\dagger}-\hat{d}^{\dagger} \hat{d}^{\dagger}\right)|00\rangle \\
& =\frac{1}{2}( \sqrt{2} |20\rangle-|11\rangle+|11\rangle- \sqrt{2} |02\rangle) \\
& = \frac{1}{\sqrt{2}}(|20\rangle-|02\rangle).
\end{aligned}
\end{equation}
Here, $ \hat c^\dag \hat c^\dag$ and $ \hat d^\dag \hat d^\dag$ correspond to the case of two photons exiting from the same port. The terms $ \hat c^\dag \hat d^\dag$ and $ \hat d^\dag \hat c^\dag$ correspond to the case in which photons exit from different ports. In this case, these $\left| {11} \right\rangle$ states are completely indistinguishable, therefore canceling each other. 
Thus, the probability of coincidence $P=0$.

Let us now consider the scenario where the signal and idler photons do not overlap, as depicted in Fig.\,\ref{HOMIsimple}(b). In this case, the two photons are distinguishable from each other, and the input state is given by ${\left| \psi  \right\rangle _{\text{in}}} = \hat a^\dag \hat{\tilde{b}}^\dag {\left| {00} \right\rangle }={\left| {1\tilde{1}} \right\rangle}.$ Here, 1 and $\tilde{1}$ represent distinguishable single photon states. The output state $|\psi\rangle_{\text{out}}$ after the BS is
%
\begin{equation}\label{Eq:1-2-6}
\begin{aligned}
|\psi\rangle_{\text {out }} & =\frac{1}{2}\left(\hat{c}^{\dagger}+\hat{d}^{\dagger}\right)\left(\hat{\tilde{c}}^{\dagger}-\hat{\tilde{d}}^{\dagger}\right)|00\rangle \\
& =\frac{1}{2}\left(\hat{c}^{\dagger} \hat{\tilde{c}}^{\dagger}-\hat{c}^{\dagger} \hat{\tilde{d}}^{\dagger}+\hat{d}^{\dagger} \hat{\tilde{c}}^{\dagger}-\hat{d}^{\dagger} \hat{\tilde{d}}^{\dagger}\right)|00\rangle \\
& =\frac{1}{2}(|20\rangle-|1\tilde{1}\rangle+|\tilde{1}1\rangle-|02\rangle). \\
\end{aligned}
\end{equation}
Note that $\hat{c}^{\dagger} \hat{\tilde{c}}^{\dagger}|00\rangle=|1_c, 1_{\tilde{c}}, 0\rangle=|20\rangle$, and this state do not contribute to the coincidence count.  In addition, $|1 \tilde{1}\rangle$ and $|\tilde{1} 1\rangle$ are distinguishable and cannot be eliminated; thus, they contribute to the coincidence count. The probability of coincidence $P=\left|-\frac{1}{2}\right|^2+\left|\frac{1}{2}\right|^2=\frac{1}{2}$. 
In the above calculation, we utilized the post-selection technique, which is widely used in quantum optics. At the output of the BS,  several states may be generated, such as the $|20\rangle$ state, $|11\rangle$ state, and $|02\rangle$ state. By performing post-selection, the $|11\rangle$ can be recorded, while $|02\rangle$ and $|20\rangle$ are not. We also need to consider the coincidence window in the above calculation. Coincidence window is an important parameter in coincidence counting techniques. In coincidence counting, it is crucial to determine whether two signals arrive simultaneously at two different ports. This requires a criterion. We define that as long as the two signals arrive within a certain time interval, they are considered to have arrived simultaneously. This time interval is known as the coincidence counting window. In this model, the coincidence window can be very narrow.

According to Eq. (\ref{Eq:1-2-2}) and Eq. (\ref{Eq:1-2-6}), the probability of coincidence is 0.5 when the two input photons do not overlap, and it reduces to 0 when the two input photons overlap. Based on these results, we can construct the corresponding HOM interference fringes, as depicted in Fig.\,\ref{HOMIsimple}(c).
The overall probability of coincidence can be written as:
\begin{equation}\label{Eq:1-2-7}
P(\tau) = \frac{1}{2}[1 - \delta (\tau)],
\end{equation}
where $\delta (\tau)$ is the Dirac delta function, and $\tau$ represents the delay time.

In the above calculation, it is important to note that the state $\left| {00} \right\rangle$ remains unchanged while the operator evolves, which is a characteristic feature of Heisenberg's picture. Single-mode theory, as discussed above, offers an intuitive understanding of HOM interference. However, in real experiments, photons with multiple modes are commonly encountered. As a result, it is necessary to consider a theory that accounts for multi-frequency scenarios. The calculations for such scenarios will be presented in the next section.

\subsubsection{The multi-frequency theory of HOM interferometer for biphotons with no spectral correlation }
%
%
\begin{figure}[htbp]
\centering
\includegraphics[width= 0.85\textwidth]{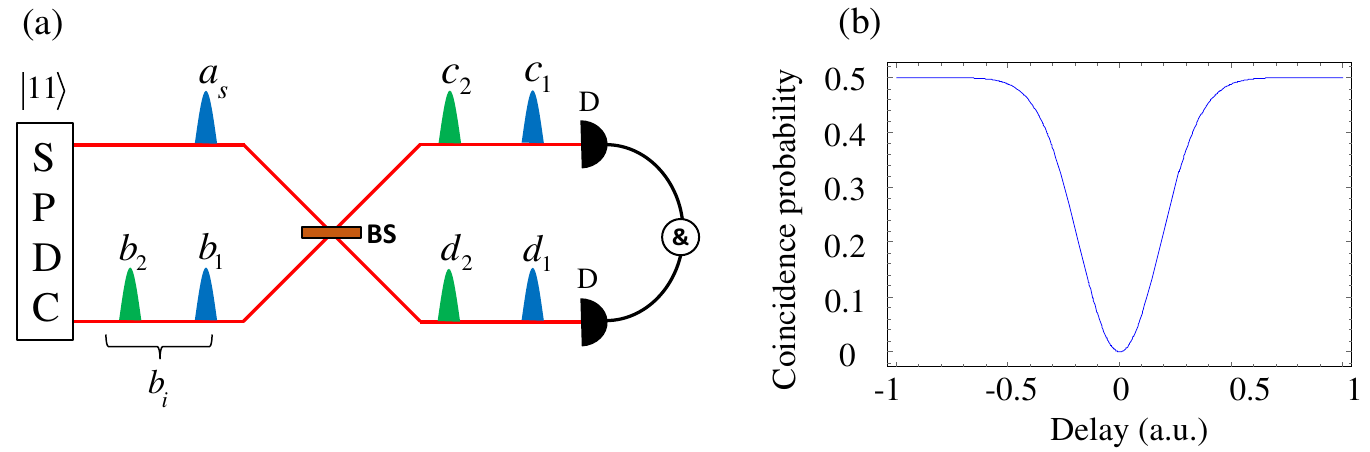}
\caption{(a) Diagram of multi-frequency HOM interference. (b) The corresponding HOM interference pattern. }
\label{11HOMItime}
\end{figure}
%
The multi-frequency HOM interference theory for biphotons with no spectral correlation was introduced in Refs.\,\cite{Tichy2011PRA, Ra2013nc, Ra2013PNAS, Tichy2014JPB, Tichy2015NJP}. We consider the two-photon interference shown in Fig.\,\ref{11HOMItime}(a). The input state is written as ${\left| \psi \right\rangle _{\text{in}}} = \hat a^\dag \hat b^\dag {\left| {00} \right\rangle }={\left| {11} \right\rangle}$. Due to the phase shift $\omega \tau $ on mode a, by scanning the optical path delay $\tau$, the operators evolve as
\begin{equation}\label{Eq:1-3-2}
\hat a^\dag \to \hat a^\dag {e^{i\omega \tau }}, \hat b^\dag \to \hat b^\dag \left( {t + \tau } \right).
\end{equation}
In the above formula, due to the optical path delay, the time information of $\hat a^\dag$ is $t$, and the time information of $\hat b^\dag$ is $(t + \tau )$. We can apply a Schmidt decomposition on the temporal modes of mode ${b}$:
\begin{equation}\label{Eq:1-3-3}
\begin{array}{lll}
\hat b^\dag \left( {t + \tau } \right) = \sqrt {I\left( \tau \right)} \hat b_1^\dag  + \sqrt {1 - I\left( \tau \right)} \hat b_2^\dag \left( {t,\tau } \right).
\end{array}
\end{equation}
Here, modes $a$ and $b$ are partially overlapped. Specifically, modes $a$ and $b_1$ are completely overlapped (indistinguishable), and modes $a$ and $b_2$ are completely separated (distinguishable).
Here $I\left( \tau  \right) = \exp \left[ { - {{\left( {\Delta \omega \tau } \right)}^2}} \right]$ represents the indistinguishability and $\Delta \omega $ is determined by the spectral width of the source. 
Thus, the input state evolves as
\begin{equation}\label{Eq:1-3-4}
\hat a^\dag {e^{i\omega \tau }}\left[ {\sqrt {I\left( \tau \right)} \hat b_1^\dag  + \sqrt {1 - I\left( \tau \right)} \hat b_2^\dag \left( {t,\tau } \right)} \right]\left| {00} \right\rangle.
\end{equation}
The output state after the BS is given by
\begin{equation}\label{Eq:1-3-7}
\begin{aligned}
{\left| \psi  \right\rangle _{\text{out}}} &= \frac{1}{2}\left( {\hat c_1^\dag  + \hat d_1^\dag } \right){e^{i\omega \tau }}[\sqrt {I\left( \tau \right)} \left( {\hat c_1^\dag  - \hat d_1^\dag } \right)
 + \sqrt {1 - I\left(\tau \right)} \left( {\hat c_2^\dag  - \hat d_2^\dag } \right)]\left| {00} \right\rangle \\
 &=\frac{1}{2}{e^{i\omega \tau }} [\sqrt {I\left( \tau \right)} \left( {\hat c_1^\dag \hat c_1^\dag  - \hat d_1^\dag \hat d_1^\dag  - \hat c_1^\dag \hat d_1^\dag  + \hat d_1^\dag \hat c_1^\dag } \right) +
\sqrt {1 - I\left( \tau \right)} \left( {\hat c_1^\dag \hat c_2^\dag  - \hat d_1^\dag \hat d_2^\dag  - \hat c_1^\dag \hat d_2^\dag  + \hat d_1^\dag \hat c_2^\dag } \right)]\left| {00} \right\rangle.
\end{aligned}
\end{equation}
Here, we omit the time label for each operator for simplicity. For coincidence detection at modes $c$ and $d$, the corresponding state is
\begin{equation}\label{Eq:1-3-8}
\frac{1}{2}{e^{i\omega \tau }}\left[ {\sqrt {1 - I\left( \tau \right)} \left( -{{{\left| {11} \right\rangle }_{{c_1}{d_2}}} + {{\left| {11} \right\rangle }_{{d_1}{c_2}}}} \right)} \right].
\end{equation}
Therefore, the probability of coincidence $P(\tau)$ is 
\begin{equation}\label{Eq:1-3-9}
P(\tau) = {\left| -{\frac{1}{2}{e^{i\omega \tau }}\sqrt {1 - I\left( \tau \right)} } \right|^2} + {\left| {\frac{1}{2}{e^{i\omega \tau }}\sqrt {1 - I\left( \tau \right)} } \right|^2}
   = \frac{1}{2}\left( {1 - I\left( \tau \right)} \right).
\end{equation}
Therefore, we can plot $P(\tau)$, as shown in Fig.\,\ref{11HOMItime}(b). 

Note that this multi-frequency approach offers the advantage of simplicity, making it suitable for handling multi-photon interference, as demonstrated in Refs. \cite{Ra2013nc, Ra2013PNAS}. However, the limitation of this theory is that the signal and idler photons are not correlated. Therefore, this model does not work with biphotons with arbitrary correlations. In the next section, we will introduce a multi-frequency model for biphotons with arbitrary frequency correlations.

\subsubsection{The multi-frequency theory of HOM interferometer for biphotons with arbitrary distribution}
%
%
\begin{figure}[htbp]
\centering
\includegraphics[width= 0.95\textwidth]{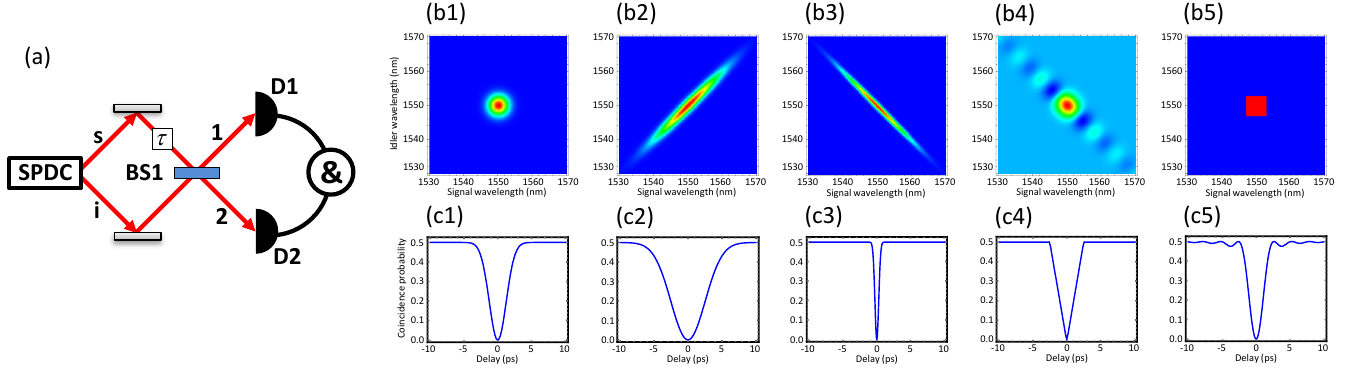}
\caption{(a) Setup of multi-frequency HOM interference. (b1-c5) The joint spectral amplitudes and the corresponding HOM interference patterns.
}
\label{11HOMarbitrary}
\end{figure}
%
In this section, we consider the multi-frequency model with arbitrary joint spectral amplitude, with the setup shown in Fig.\,\ref{11HOMarbitrary}(a). In this model, we can calculate HOM interference using joint spectral amplitudes, as shown in Fig.\,\ref{11HOMarbitrary}(b1-c5). The two-photon state from the SPDC process can be described as
\begin{equation}\label{Eq:1-4-1}
\left| \psi  \right\rangle  = \int_0^\infty  {\int_0^\infty  {d\omega _s d\omega _i } } f(\omega _s ,\omega _i )\hat a_s^\dag  (\omega _s )\hat a_i^\dag  (\omega _i )\left| {00} \right\rangle,
\end{equation}
where $\omega$ is the angular frequency; $\hat a^\dag$ is the creation operator and the subscripts $s$ and $i$ denote the signal and idler photons from SPDC, respectively; $f(\omega _s ,\omega _i )$ is the two-photon spectral amplitude (also called joint spectral amplitude) of the signal and idler photons, which can be arbitrary.

The field operators of detector 1 (D1) and detector 2 (D2) are given by $\hat E_1^{( + )} (t_1 ) = \frac{1}{{\sqrt {2\pi } }}\int_0^\infty  {d\omega _1 } \hat a_1 (\omega _1 )e^{ - i\omega _1 t_1 } $ and $\hat E_2^{( + )} (t_2 ) = \frac{1}{{\sqrt {2\pi } }}\int_0^\infty  {d\omega _2 \hat a_2 (\omega _2 )} e^{ - i\omega _2 t_2 }$. After the delay $\tau$ and the BS transformation, the mode operators are given by $\hat a_1 (\omega _1 ) = \frac{1}{{\sqrt {2} }}[\hat a_s (\omega _1 ) + \hat a_i (\omega _1 )e^{ - i\omega _1 \tau } ]$ and $\hat a_2 (\omega _2 ) = \frac{1}{{\sqrt {2} }}[\hat a_s (\omega _2 ) - \hat a_i (\omega _2 )e^{ - i\omega _2 \tau } ]$. Therefore, the detector field operators can be written as
\begin{equation}\label{Eq:1-4-2}
\begin{aligned}
  \hat E_1^{( + )} (t_1 ) &= \frac{1}{{\sqrt {4\pi } }}\int_0^\infty  {d\omega _1 } [ \hat a_s (\omega _1 )e^{ - i\omega _1 t_1 }  +  \hat a_i (\omega _1 )e^{ - i\omega _1 (t_1  + \tau )} ],\\
\hat E_2^{( + )} (t_2 ) &= \frac{1}{{\sqrt {4\pi } }}\int_0^\infty  {d\omega _2 [ \hat a_s (\omega _2 )} e^{ - i\omega _2 t_2 }  - \hat a_i (\omega _2 )e^{ - i\omega _2 (t_2  + \tau )} ].  
\end{aligned}
 \end{equation}
The probability of coincidence $P(\tau)$ can be expressed as
\begin{equation}\label{Eq:1-4-4}
P(\tau ) = \int {\int {dt_1 dt_2 } } \left\langle {\psi \left| {\hat E_1^{( - )} \hat E_2^{( - )} \hat E_2^{( + )} \hat E_1^{( + )} } \right|\psi } \right\rangle.
\end{equation}
Consider $\hat E_2^{( + )} \hat E_1^{( + )} \left| \psi  \right\rangle$, only 2 out of 4 terms exist. The first term is
\begin{equation}\label{Eq:1-4-5}
\begin{aligned}
  & - \frac{1}{{4\pi }}\int_0^\infty  {\int_0^\infty  {d\omega _1 d\omega _2 } } \hat a_s (\omega _1 )\hat a_i (\omega _2 )e^{ - i\omega _1 t_1 } e^{ - i\omega _2 (t_2  + \tau )} \int_0^\infty  {\int_0^\infty  {d\omega _s d\omega _i } } f(\omega _s ,\omega _i )\hat a_s^\dag  (\omega _s )\hat a_i^\dag  (\omega _i )\left| {00} \right\rangle \\
  &=  - \frac{1}{{4\pi }}\int_0^\infty  {\int_0^\infty  {d\omega _1 d\omega _2 } } f(\omega _1 ,\omega _2 )e^{ - i\omega _1 t_1 } e^{ - i\omega _2 (t_2  + \tau )} \left| {00} \right\rangle. 
 \end{aligned}
\end{equation}
And the second term is
\begin{equation}\label{Eq:1-4-6}
\begin{aligned}
& \frac{1}{{4\pi }}\int_0^\infty  {\int_0^\infty  {d\omega _1 } } d\omega _2 \hat a_i (\omega _1 )\hat a_s (\omega _2 )e^{ - i\omega _1 (t_1  + \tau )} e^{ - i\omega _2 t_2 } \int_0^\infty  {\int_0^\infty  {d\omega _s d\omega _i } } f(\omega _s ,\omega _i )\hat a_s^\dag  (\omega _s )\hat a_i^\dag  (\omega _i )\left| {00} \right\rangle  \\ 
  &= \frac{1}{{4\pi }}\int_0^\infty  {\int_0^\infty  {d\omega _1 } } d\omega _2 f(\omega _2 ,\omega _1 )e^{ - i\omega _1 (t_1  + \tau )} e^{ - i\omega _2 t_2 } \left| {00} \right\rangle.  \\
 \end{aligned}
\end{equation}
In the above calculations, we have used $\hat a_s (\omega _1 )\hat a_s^\dag  (\omega _s )  \left| {0} \right\rangle = \delta (\omega _1  - \omega _s ) \left| {0} \right\rangle$  and $\hat a_i (\omega _2 )\hat a_i^\dag  (\omega _i )\left| {0} \right\rangle = \delta (\omega _2  - \omega _i ) \left| {0} \right\rangle$.
Combining these two terms yields
\begin{equation}\label{Eq:1-4-7}
 \hat E_2^{( + )} \hat E_1^{( + )} \left| \psi  \right\rangle = \frac{1}{{4\pi }}\int_0^\infty  {\int_0^\infty  {d\omega _1 } } d\omega _2 e^{ - i\omega _1 t_1 } e^{ - i\omega _2 t_2 } [f(\omega _2 ,\omega _1 )e^{ - i\omega _1 \tau }  - f(\omega _1 ,\omega _2 )e^{ - i\omega _2 \tau } ] \left| {00} \right\rangle.
\end{equation}
Therefore,
\begin{equation}\label{Eq:1-4-8}
\begin{aligned}
     \left\langle {\psi \left| {\hat E_1^{( - )} \hat E_2^{( - )} \hat E_2^{( + )} \hat E_1^{( + )} } \right|\psi } \right\rangle
 = &   (\frac{1}{{4\pi }})^2 \int_0^\infty \int_0^\infty \int_0^\infty \int_0^\infty  {d\omega _1 } d\omega _2 d\omega _1^\prime d\omega _2^\prime e^{ - i(\omega _1  - \omega _1^\prime )t_1 } e^{ - i(\omega _2  - \omega _2^\prime )t_2 }\\
 &  \times [ {f^* (\omega _2^\prime ,\omega _1^\prime )e^{i\omega _1^\prime \tau }  - f^* (\omega _1^\prime ,\omega _2^\prime )e^{i\omega _2^\prime \tau } } ] [ {f(\omega _2 ,\omega _1 )e^{ - i\omega _1 \tau }  - f(\omega _1 ,\omega _2 )e^{ - i\omega _2 \tau } }].  \\
\end{aligned}
\end{equation}
Finally,
\begin{equation}\label{Eq:1-4-9}
\begin{aligned}
 P(\tau ) =&\int {\int {dt_1 dt_2 } } \left\langle {\psi \left| {\hat E_1^{( - )} \hat E_2^{( - )} \hat E_2^{( + )} \hat E_1^{( + )} } \right|\psi } \right\rangle  \\
   =&\frac{1}{4}\int_0^\infty \int_0^\infty \int_0^\infty \int_0^\infty {d\omega _1 } d\omega _2 d\omega _1^\prime d\omega _2^\prime \delta (\omega _1  - \omega _1^\prime )\delta (\omega _2  - \omega _2^\prime ) \\
  &  \times [ {f^* (\omega _2^\prime ,\omega _1^\prime )e^{i\omega _1^\prime \tau }  - f^* (\omega _1^\prime ,\omega _2^\prime )e^{i\omega _2^\prime \tau } }][
   {f(\omega _2 ,\omega _1 )e^{ - i\omega _1 \tau }  - f(\omega _1 ,\omega _2 )e^{ - i\omega _2 \tau } } ] \\
 =& \frac{1}{4}\int_0^\infty   \int_0^\infty   {d\omega _1 } d\omega _2 {\rm{|}}f(\omega _1 ,\omega _2 ) - f(\omega _2 ,\omega _1 )e^{ - i(\omega _1  - \omega _2 )\tau } {\rm{|}}^{\rm{2}}  \\
 =& \frac{1}{4}\int_0^\infty  { \int_0^\infty  {d{\omega _1}} } d{\omega _2}\{ {\left| {f({\omega _1},{\omega _2})} \right|^2} + {\left| {f({\omega _2},{\omega _1})} \right|^2} - 2{\mathop{\rm Re}\nolimits} [f({\omega _1},{\omega _2}){f^*}({\omega _2},{\omega _1}){e^{i({\omega _1} - {\omega _2})\tau }}]\}.
 \end{aligned}
\end{equation}
Equation (\ref{Eq:1-4-9}) is the general form of HOM interference with arbitrary joint spectral amplitude.  

For simplicity, we assume that $f(\omega _1 ,\omega _2 )$ is normalized ($\int_0^\infty \int_0^\infty  {d\omega _1 } d\omega _2 {\rm{|}}f(\omega _1 ,\omega _2 ){\rm{|}}^{\rm{2}}=1$) and real ($f^* = f$).
In addition, we can also assume that $f(\omega _1 ,\omega _2 )$ satisfies the exchange symmetry of $f(\omega _1 ,\omega _2 )= f(\omega _2 ,\omega _1 )$. Then, we can further simplify the two-photon detection probability $P(\tau)$ as
\begin{equation}\label{Eq:1-4-10}
 P(\tau)= \frac{1}{2} -  \frac{1}{2}  \int_0^\infty \int_0^\infty  {d\omega _s } d\omega _i {\rm{|}} f(\omega _s ,\omega _i ) {\rm{|}}^{\rm{2}} \cos(\omega _s  - \omega _i )\tau.
\end{equation}

This multi-frequency model has the ability to deal with biphotons with arbitrary frequency correlations. However, this treatment is only valid in the frequency domain. It is also necessary to model the HOM interference in the time domain. Therefore, it is necessary to consider the multi-temporal model theory.

\subsubsection{The multi-temporal theory of HOM interferometer}
Here, we derive the equations for HOM interference based on the function of joint temporal amplitude. Assuming that SPDC biphotons have a joint spectral amplitude of $f(\omega _1,\omega _2 )$, they have a joint temporal distribution of $F(t_1 ,t_2 )=\mathscr{F}\{ f(\omega _1,\omega _2 )\}$ in the time domain, where $\mathscr{F}$ is the Fourier transform.

The probability of coincidence $P(\tau)$ of an HOM interference in the frequency domain is given by \cite{Jin2018Optica}
\begin{equation}\label{Eq:1-5-1}
P(\tau ) = \frac{1}{4}\int_{ - \infty }^{ + \infty } {} \int_{ - \infty }^{ + \infty } {d\omega _1 } d\omega _2 {\rm{|}}f(\omega _1 ,\omega _2 ) - f(\omega _2 ,\omega _1 )e^{ - i(\omega _1  - \omega _2 )\tau } {\rm{|}}^{\rm{2}}.
\end{equation}
If we have the Fourier transform of a function as $\mathscr {F} \left\{ {g(x,y)} \right\} = G(\xi ,\eta )$, and both integrals $\int_{ - \infty }^{ + \infty } {\int_{ - \infty }^{ + \infty } {dx} dy} {\rm{|}}g(x,y){\rm{|}}^{\rm{2}}$ and $\int_{ - \infty }^{ + \infty } {\int_{ - \infty }^{ + \infty } {d\xi } d\eta } {\rm{|}}G(\xi ,\eta ){\rm{|}}^{\rm{2}}$ exist, then
\begin{equation}\label{Eq:1-5-2}
\int_{ - \infty }^{ + \infty } {\int_{ - \infty }^{ + \infty } {dx} dy} {\rm{|}}g(x,y){\rm{|}}^{\rm{2}}  = \int_{ - \infty }^{ + \infty } {\int_{ - \infty }^{ + \infty } {d\xi } d\eta } {\rm{|}}G(\xi ,\eta ){\rm{|}}^{\rm{2}}, 
\end{equation}
which is the Parseval's Theorem \cite{Grice1997Phdtheis}. Equation (\ref{Eq:1-5-2}) can be intuitively interpreted as the conservation of energy. Here, the total energy by summing the power-per-sample across time equals the energy by summing the spectral power across frequency.
Therefore,
\begin{equation}\label{Eq:1-5-3}
\begin{aligned}
P(\tau ) &= \frac{1}{4}\int_{ - \infty }^{ + \infty } {} \int_{ - \infty }^{ + \infty } {d\omega _1 } d\omega _2 {\rm{|}}f(\omega _1 ,\omega _2 ) - f(\omega _2 ,\omega _1 )e^{ - i(\omega _1  - \omega _2 )\tau } {\rm{|}}^{\rm{2}}  \\
&= \frac{1}{4}\int_{ - \infty }^{ + \infty } {} \int_{ - \infty }^{ + \infty } {dt_1 } dt_2 {\rm{| \mathscr {F}\{ }}f(\omega _1 ,\omega _2 ) - f(\omega _2 ,\omega _1 )e^{ - i(\omega _1  - \omega _2 )\tau } {\rm{\} |}}^{\rm{2}}  \\
&= \frac{1}{4}\int_{ - \infty }^{ + \infty } {} \int_{ - \infty }^{ + \infty } {dt_1 } dt_2 {\rm{|}}F(t_1 ,t_2 ) - F(t_2  - \tau ,t_1  + \tau ){\rm{|}}^{\rm{2}},
\end{aligned}
\end{equation}
where the translation (time shifting) property of Fourier transformation $\mathscr {F}\{ f(\omega _2 ,\omega _1 )e^{ - i(\omega _1  - \omega _2 )\tau } \}  = F(t_2  - \tau ,t_1  + \tau )$  is utilized. Equation (\ref{Eq:1-5-3}) describes the HOM interference in the time domain.

\subsubsection{The multi-frequency theory of four-fold HOM interferometer between two independent sources}
\begin{figure}[htbp]
\centering
\includegraphics[width= 0.65\textwidth]{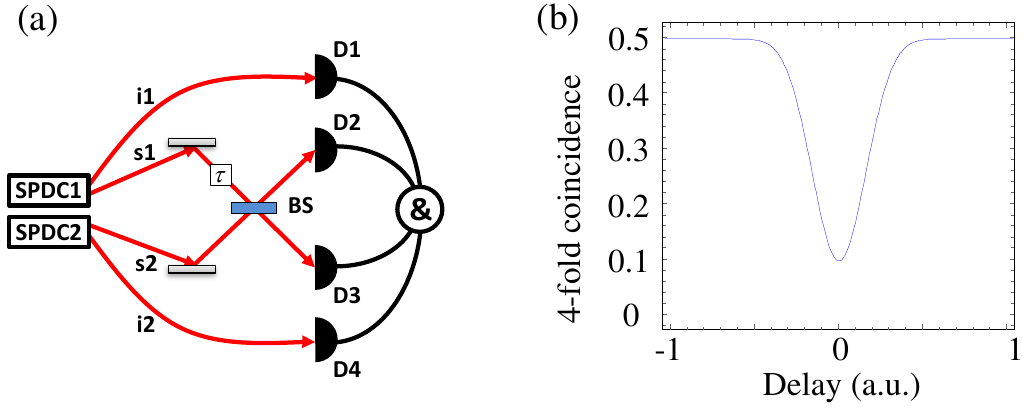}
\caption{(a) Setup of the four-fold HOM interference between two independent sources. (b) The four-fold HOM interference dip.
}
\label{11HOM4fold}
\end{figure}
%
We can also consider the HOM interference between two independent sources. Here, we take two independent SPDC sources as an example, with the setup shown in Fig.\,\ref{11HOM4fold}(a). 
We assume that the signal and idler photons from the first and second SPDC source have the joint spectral amplitude of $f_1 (\omega _{s1} ,\omega _{i1})$ and $f_2 (\omega _{s2} ,\omega _{i2})$. The signal photons ($s1$ and $s2$) are sent to a 50/50 BS, while the idler photons ($i1$ and $i2$) act as herald, i.e., detected by single-photon detectors to claim the existence of $s1$ and $s2$.
The four-fold coincidence probability $P_4 (\tau )$ as a function of the time delay $\tau $ can be written as \cite{Jin2015OE, MosleyPhD}:
\begin{equation}\label{eq:P4}
   P_4 (\tau ) = \frac{1}{4}\int\limits_0^\infty  {\int\limits_0^\infty  {\int\limits_0^\infty  {\int\limits_0^\infty  {d\omega _{s1} d\omega _{s2} d\omega _{i1} d\omega _{i2} } } } }      {\rm{|}}f_1 (\omega _{s1} ,\omega _{i1} )f_2 (\omega _{s2} ,\omega _{i2} ) - f_1 (\omega _{s2} ,\omega _{i1} )f_2 (\omega _{s1} ,\omega _{i2} )e^{ - i(\omega _{s2}  - \omega _{s1} )\tau } {\rm{|}}^{\rm{2}}.    
\end{equation}
Using Eq. (\ref{eq:P4}), we can plot the HOM interference pattern and Fig.\,\ref{11HOM4fold}(b) shows a typical four-fold HOM interference dip \cite{Jin2015OE}.

We can compare the four-fold HOM interference described by Eq. (\ref{eq:P4}) and the two-fold HOM interference described by Eq. (\ref{Eq:1-4-9}). To achieve perfect HOM interference according to Eq. (\ref{Eq:1-4-9}), the joint spectral distribution needs to satisfy exchange symmetry, meaning that $f(\omega_s, \omega_i) = f(\omega_i, \omega_s)$. However, a high spectral purity is not required. On the other hand, to achieve high visibility four-fold HOM interference as stated in Eq. (\ref{eq:P4}), exchange symmetry is not required. Instead, a high spectral purity is necessary. In Eq. (\ref{eq:P4}), to obtain perfect visibility, the condition $f_1 (\omega _{s1} ,\omega _{i1} )f_2 (\omega _{s2} ,\omega _{i2} ) = f_1 (\omega _{s2} ,\omega _{i1} )f_2 (\omega _{s1} ,\omega _{i2} )$ must be satisfied. To achieve this condition for all $\omega$, two requirements must be met. First, the two SPDC sources must be indistinguishable: $f_1(\omega_s, \omega_i) = f_2(\omega_s, \omega_i) = f(\omega_s, \omega_i)$. Second, the sources must be spectrally pure: $f(\omega_s, \omega_i) = g_s(\omega_s)g_i(\omega_i)$ \cite{MosleyPhD, Jin2015OE}.

So far, we have discussed five distinct situations regarding the HOM interferometer, beginning with the simplest and progressing to the most intricate. However, our discussion is not exhaustive, and the readers may refer to Refs. \cite{Ou2007Springer,GerryandKnight2004BOOK,Ou2017Springer,Branczyk2017arXiv,Wang2019NP,Bouchard2020RPP} for more discussion.

\subsection{HOM interferometers based on different physical systems}

\begin{table}[h]
\caption{HOM Interferometer based on different systems}
\label{Tab:HOMelectron}
\begin{tabular}{l|l|l}
\hline \hline
 \multicolumn{2}{l|}{Physical system/Process }  &  References                                        \\ \hline   
& Nonlinear  crystals and waveguides     &  \cite{Hong1987PRL}  \\  \cline{2-3} 
&  Quantum dots     &   \cite{Sanaka2009PRL,Flagg2010PRL,Patel2010NP,Wei2014NL,Senellart2017NN,Grobe2020APL,Koong2021PRL,Ollivier2021PRL,Steindl2021PRL,Appel2022PRL,Dusanowski2023Photonics}  \\  \cline{2-3} 
&   Nitrogen/silicon-vacancy (NV/SiV) centers in diamond   &   \cite{Bernien2012PRL,Sipahigil2012PRL,Sipahigil2014PRL,Waltrich2023}  \\  \cline{2-3}
Photon &  Molecule    &   \cite{Kiraz2005PRL,Lettow2010PRL,Lombardi2021APL}  \\  \cline{2-3}
&  Four-wave mixing in atoms    &   \cite{Felinto2006NP,Yuan2007PRL,Yuan2008Nature,Li2016PRL}  \\  \cline{2-3}
 &  Microwave    &   \cite{Lang2013NP}  \\  \cline{2-3}
 &  2D material    &   \cite{Kaplan2023NP,Fournier2023PRAppl}  \\  \cline{2-3}
 &  Trapped iron    &   \cite{Maunz2007NP,Krutyanskiy2023PRL,Krutyanskiy2023PRL-2}  \\  \hline
\multicolumn{2}{l|}{Atom} & \cite{Beugnon2006Nature,Specht2011Nature,Kaufman2014Science,Lopes2015Nature,Kaufman2018AAMOP}                            
\\  \hline 
\multicolumn{2}{l|}{Electron} & \cite{Jonckheere2012PRB,Bocquillon2013Science}                            
\\  \hline 
\multicolumn{2}{l|}{Plasmon} & \cite{You2020Nano,Fujii2014PRB,Martino2014PRAppl,Dheur2016SA,Heeres2013Nnano,Fakonas2014NP,Cai2014PRAppl}                            
\\  \hline 
\multicolumn{2}{l|}{Phonon} & \cite{Toyoda2015Nature}                            
\\  \hline 
\multicolumn{2}{l|}{Magnon $\&$ Photon}  & \cite{Su2022PRL}                            
\\  \hline 
\multicolumn{2}{l|}{Soliton}  & \cite{Sun2014PRA}                            
\\ 
\hline \hline      
\end{tabular}
\end{table}

\begin{figure}[htbp]
\centering
\includegraphics[width= 0.95\textwidth]{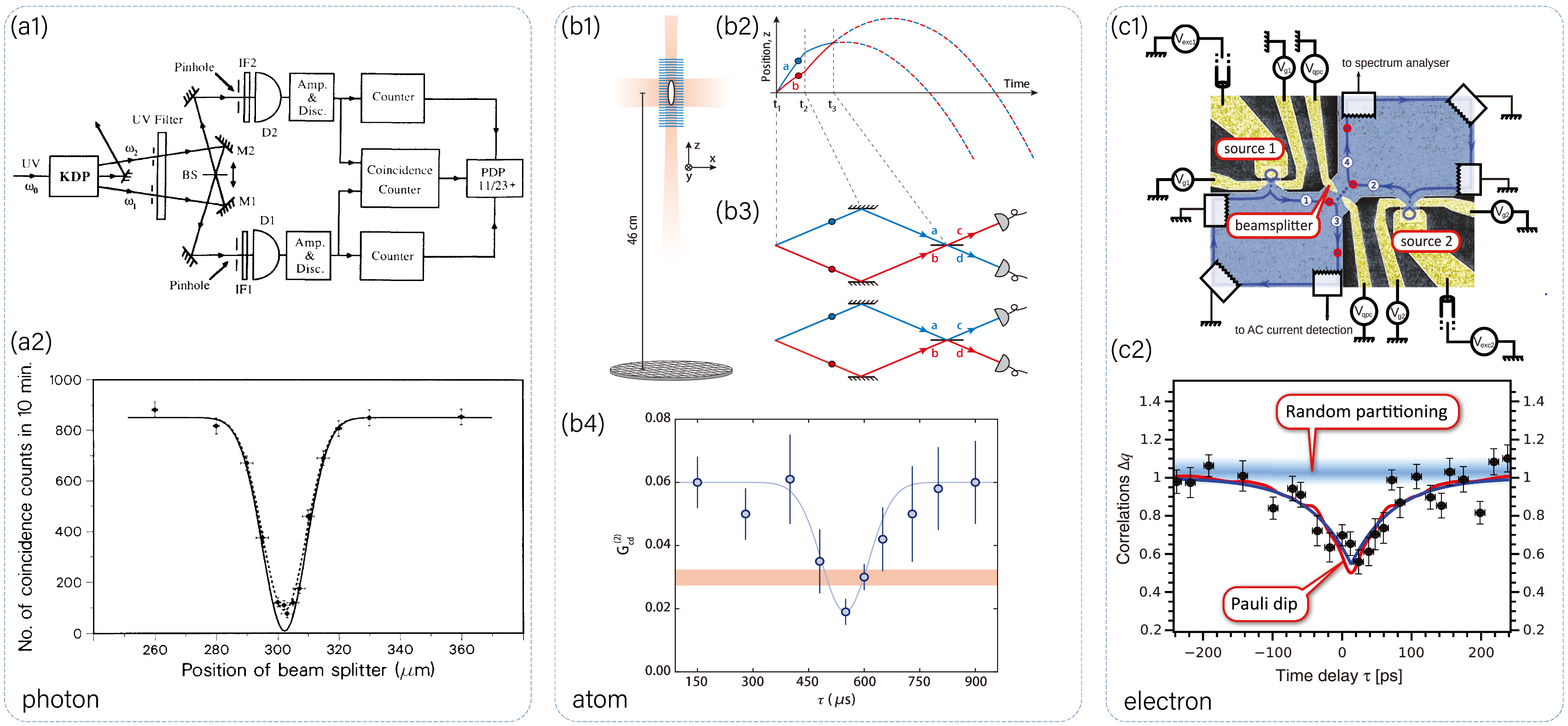}
\caption{HOM interferometers based on different physical systems. (a1-a2) Setup and interference patterns of the first HOM interferometer using photons. Reprinted from Ref.\,\cite{Hong1987PRL}. (b1-b4) Principles and experimentally measured interference patterns of the HOM interferometer using atoms. Reprinted from Ref.\,\cite{Lopes2015Nature}. (c1-c2) Setup and interference patterns of the HOM interferometer using electrons. Reprinted from Ref.\,\cite{Bocquillon2013Science}. }
\label{HOMelectron}
\end{figure}

The first HOM interferometer was demonstrated using photons generated from a SPDC source in a nonlinear crystal \cite{Hong1987PRL}. The setup and interference patterns are shown in Fig.\,\ref{HOMelectron}(a). However, it is important to note that HOM interference is not limited to photons generated through SPDC. Photons with indistinguishable characteristics can also be generated from various physical systems as well, as listed in Tab.\,\ref{Tab:HOMelectron}.
 
For instance, quantum dots, which are nanoscale semiconductor structures, have been utilized as a source of indistinguishable photons for HOM interferometers. These quantum dots can be engineered to emit single photons with high purity, allowing precise interference measurements \cite{Sanaka2009PRL,Flagg2010PRL,Patel2010NP,Wei2014NL,Senellart2017NN,Grobe2020APL,Koong2021PRL,Ollivier2021PRL,Steindl2021PRL,Appel2022PRL,Dusanowski2023Photonics}. Another promising system is the nitrogen-vacancy (NV) center in diamond. NV centers are atomic-scale defects in diamond that exhibit remarkable quantum properties \cite{Bernien2012PRL,Sipahigil2012PRL}. They have been employed as a source of indistinguishable photons for HOM interferometers, enabling studies in quantum information processing and sensing. Additionally, silicon-vacancy can also be used to realize two-photon interference \cite{Sipahigil2014PRL,Waltrich2023}. Furthermore, molecules \cite{Kiraz2005PRL,Lettow2010PRL,Lombardi2021APL}, atoms \cite{Felinto2006NP,Yuan2007PRL,Yuan2008Nature,Li2016PRL}, and even microwaves \cite{Lang2013NP} have been harnessed as sources for generating indistinguishable photons suitable for HOM interference. These systems offer unique advantages and capabilities for exploring quantum interference phenomena.

Recently, the field of quantum optics has greatly benefited from the progress made in materials science. These advances have yielded plenty of novel materials that are instrumental in the development and application of quantum optics technology. In particular, the exploration of quantum emitters within two-dimensional materials, such as $\text{CsPbBr}_3$ nanocrystals and Hexagonal Boron Nitride, has unveiled exciting prospects for the integration of quantum information into photonic devices \cite{Kaplan2023NP,Fournier2023PRAppl}. In addition, trapped ions have emerged as another fascinating system for generating indistinguishable photons for HOM interferometers. The quantum interference of two remote trapped atomic ions paves the way for a remote network of entangled quantum processors. Typically, these ions are confined within two identical Paul traps situated in separate vacuum chambers. By utilizing pulsed lasers to stimulate the ions, it becomes possible to generate highly coherent photons, thus facilitating precise quantum interference experiments \cite{Maunz2007NP,Krutyanskiy2023PRL,Krutyanskiy2023PRL-2}.

The HOM interference phenomenon is not limited to photons; it can also be observed with other particles and quasi-particles such as atoms, electrons, plasmons, phonons, magnons, and solitons. Atoms, for example, have been used in experiments to demonstrate HOM interference \cite{Beugnon2006Nature,Specht2011Nature,Kaufman2014Science,Lopes2015Nature,Kaufman2018AAMOP}. Figure\,\ref{HOMelectron}(b) shows the principle and interference patterns of the HOM interferometer using atoms \cite{Lopes2015Nature}. A moving optical lattice superimposed on a Bose-Einstein condensate (BEC) of metastable He atoms induces the scattering of atom pairs in the longitudinal direction through a process similar to spontaneous four-wave mixing. The generated atoms enter the BS, which is achieved by using Bragg scattering on another optical lattice, and interference patterns can be observed \cite{Lopes2015Nature}. This enables the investigation of atom-based quantum technologies. Similarly, electrons, which exhibit wave-particle duality, have been used to demonstrate HOM interference \cite{Jonckheere2012PRB,Bocquillon2013Science}. The setup and interference patterns are shown in Fig.\,\ref{HOMelectron}(c) \cite{Bocquillon2013Science}. The single electron emitter is a mesoscopic capacitor, which is composed of a small quantum dot that is capacitively connected to a metallic top gate and tunnel-coupled to a single edge channel through a quantum point contact with adjustable transmission. To generate a quantized current, a square-wave periodic radio-frequency excitation is applied to the top gate.  This excitation has a peak-to-peak amplitude that matches the dot addition energy,  resulting in the emission of a single electron followed by a single hole. When two such emitters are utilized as inputs in an electronic beam splitter, interference effects can be observed. These studies contribute to our understanding of the quantum behavior of electrons and have implications for electron-based quantum technologies.

Furthermore, surface plasmon polaritons, which are a combination of light and a collective oscillation of the free-electron plasma at metal/dielectric interfaces, have been utilized to study HOM interference \cite{You2020Nano,Fujii2014PRB,Martino2014PRAppl,Dheur2016SA,Cai2014PRAppl}. Using nanofabrication techniques, plasmonic BS and integrated on-chip detectors can be made, allowing efficient detection of a single plasmon \cite{Vest2017Science}. Plasmon-based HOM interferometers could be a versatile platform for studying the quantum behavior of light, enabling advanced functionalities for quantum technologies \cite{Heeres2013Nnano,Fakonas2014NP,Cai2014PRAppl, Hong2024NP}. Last but not least, HOM interference has been investigated with other particle-like entities, such as phonons (quantized vibrations in solids) \cite{Toyoda2015Nature}, magnons (quanta of magnetic excitations) \cite{Su2022PRL}, and solitons (stable, localized wave packets) \cite{Sun2014PRA}. These studies provide insights into the fundamental behavior of these quasi-particles and have implications for applications in quantum acoustics, spintronics, and nonlinear optics.

\subsection{Typical HOM interferometers}
\subsubsection{HOM interferometer using independent photon sources}
%
%
\begin{figure}[!htbp]
\centering
\includegraphics[width= 0.92\textwidth]{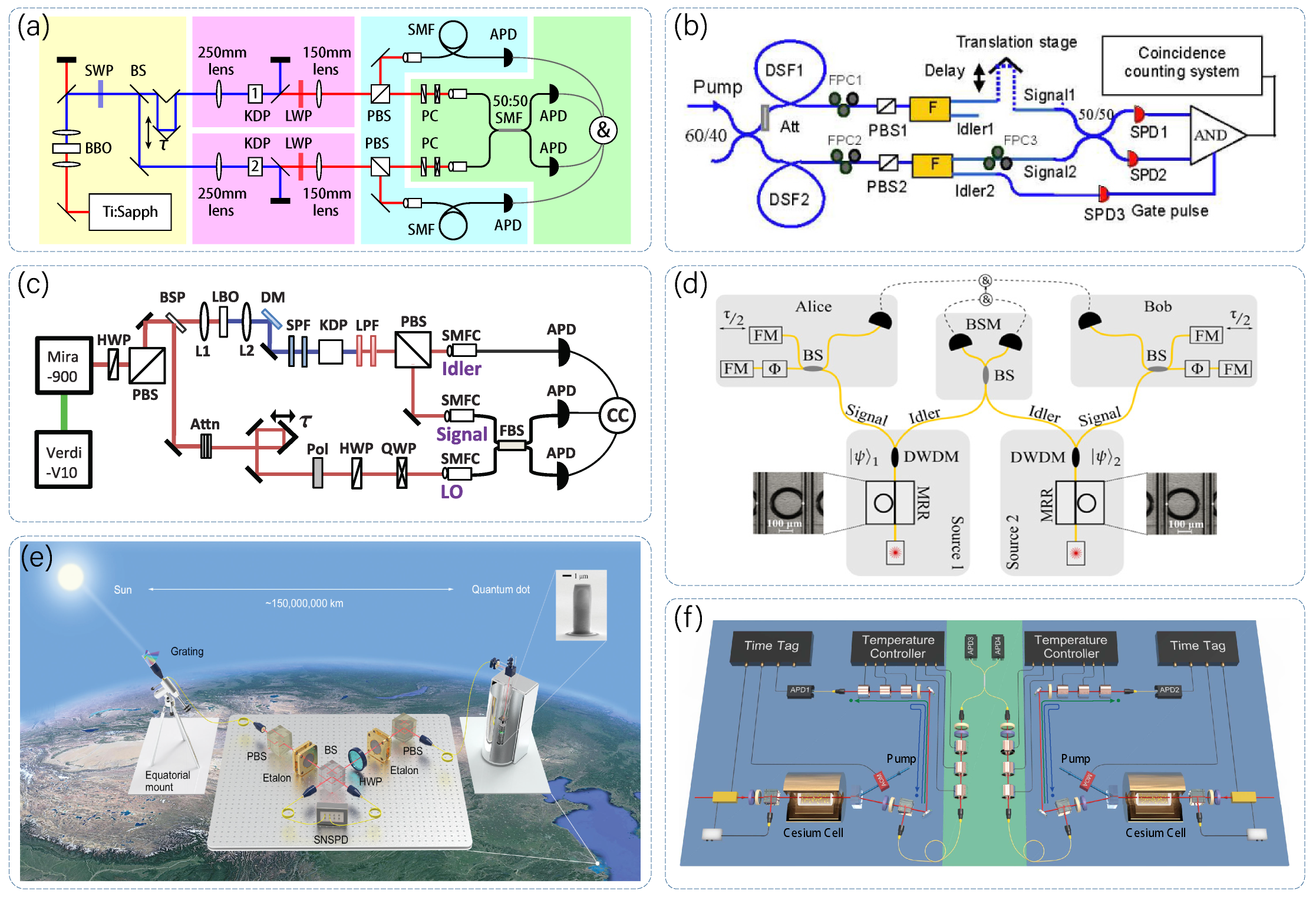}
\caption{Examples of HOM interferometer using independent photon sources (HOMI-IPS). (a) HOMI-IPS between two heralded single-photon states. Reprinted from Ref.\,\cite{Mosley2008PRL}. (b) HOMI-IPS between the heralded single-photon state and the thermal state of the dispersion-shifted fibers. Reprinted from Ref.\,\cite{Li2008OE}. (c) HOMI-IPS between a heralded single-photon state and a weak coherent state. Reprinted from Ref.\,\cite{Jin2011}. (d) HOMI-IPS between two heralded single-photon states from integrated sources. Reprinted from Ref.\,\cite{Samara2021QST}. (e) HOMI-IPS between a thermal photon from the Sun and a single photon from the semiconductor quantum dot on Earth. Reprinted from Ref.\,\cite{Deng2019PRL}. (f) HOMI-IPS between between two heralded single photons from two independent room-temperature quantum memories (two vapor cells containing cesium atoms). Reprinted from Ref.\,\cite{Zhang2022PR}.}
\label{HOMindependent}
\end{figure}

In the traditional HOM interference setup, a pair of identical photons is generated from the same source. However, it is also possible to implement an HOM interferometer using independent photon sources (HOMI-IPS). In fact, HOMI-IPS plays an important role in many quantum information processing applications, such as quantum teleportation \cite{Valivarthi2016np}, quantum computation \cite{Walmsley2005}, boson sampling \cite{Broome2013} and measurement device independent quantum key distribution (MDI-QKD) \cite{Valivarthi2017QST,Lo2012PRL,Xue2022PRAppl}. A high visibility in HOMI-IPS corresponds to high operation fidelity of the quantum processor. Therefore, achieving high visibility in HOMI-IPS is crucial to ensure the reliable and efficient performance of quantum information processing tasks. 

Many HOMI-IPS experiments have been demonstrated at near-infrared wavelengths \cite{Kaltenbaek2006, Mosley2008PRL, Mosley2008NJP, Rarity2005, Jin2011, Soller2011, Tanida2012, Zhao2014} or telecom wavelengths \cite{Takesue2007, Xue2010, Aboussouan2010, Harada2011, Jin2013PRA, Harder2013, Bruno2014}, and with heralded single-photon sources \cite{Kaltenbaek2006, Mosley2008PRL, Mosley2008NJP, Jin2013PRA2, Harder2013, Bruno2014,LiuSL2018PRA,Zeng2024COL}, weak coherent sources \cite{Rarity2005, Jin2011}, or even thermal sources \cite{Li2008OE, Jin2013PRA}. Figure\,\ref{HOMindependent}(a) presents the HOMI-IPS using heralded single photons prepared in pure quantum states from an SPDC source at 830 nm. An HOM interference visibility of 94.4\% was observed without using spectral filters \cite{Mosley2008PRL, Mosley2008NJP}. Figure\,\ref{HOMindependent}(b) demonstrates the HOMI-IPS with a heralded single-photon state and a thermal state. The photons in the 1550 nm telecom band were generated from two independent dispersion-shifted fibers via a four-wave mixing process. The observed visibility is (82 $\pm$ 11)\% \cite{Li2008OE}. Figure\,\ref{HOMindependent}(c) shows the HOMI-IPS with a heralded single-photon state and a weak coherent state at 830 nm. Without the use of bandpass filters, spectrally pure single photons can have a high-visibility of 89.4 $\pm$ 0.5\% \cite{Jin2011}. Figure\,\ref{HOMindependent}(d) illustrates the HOMI-IPS setup utilizing two independent, asynchronously pumped, integrated Si$_3$N$_4$ micro-ring resonator photon-pair sources at 1559 nm, operating in the continuous wave regime. The visibility achieved in this configuration was measured to be 93.2 $\pm$ 1.6\%. The time-resolved detection enabled high spectral purities without the need for spectral filtering \cite{Samara2021QST}. Figure \ref{HOMindependent}(e) shows the HOMI-IPS setup that utilizes a thermal photon from the Sun and a single photon from a semiconductor quantum dot on Earth, with a separation of approximately 150 million kilometers. In this configuration, a raw visibility of 79.6 $\pm$ 1.7\% was observed \cite{Deng2019PRL}. Furthermore, by using photons with no common history, the researchers demonstrated post-selected two-photon entanglement with a state fidelity of 82.6 $\pm$ 2.4\% and a violation of the Bell inequality by 2.20 $\pm$ 0.06. Figure\,\ref{HOMindependent}(f) demonstrated HOM interference between heralded single photons from two independent room-temperature quantum memories, specifically, two vapor cells containing cesium atoms \cite{Zhang2022PR}. The measured interference visibility of 75.0\% highlights the potential of these independent sources in constructing a quantum memory-enabled network.

\subsubsection{Multi-photon HOM interferometer}
%
%
\begin{figure}[htbp]
\centering
\includegraphics[width= 0.92\textwidth]{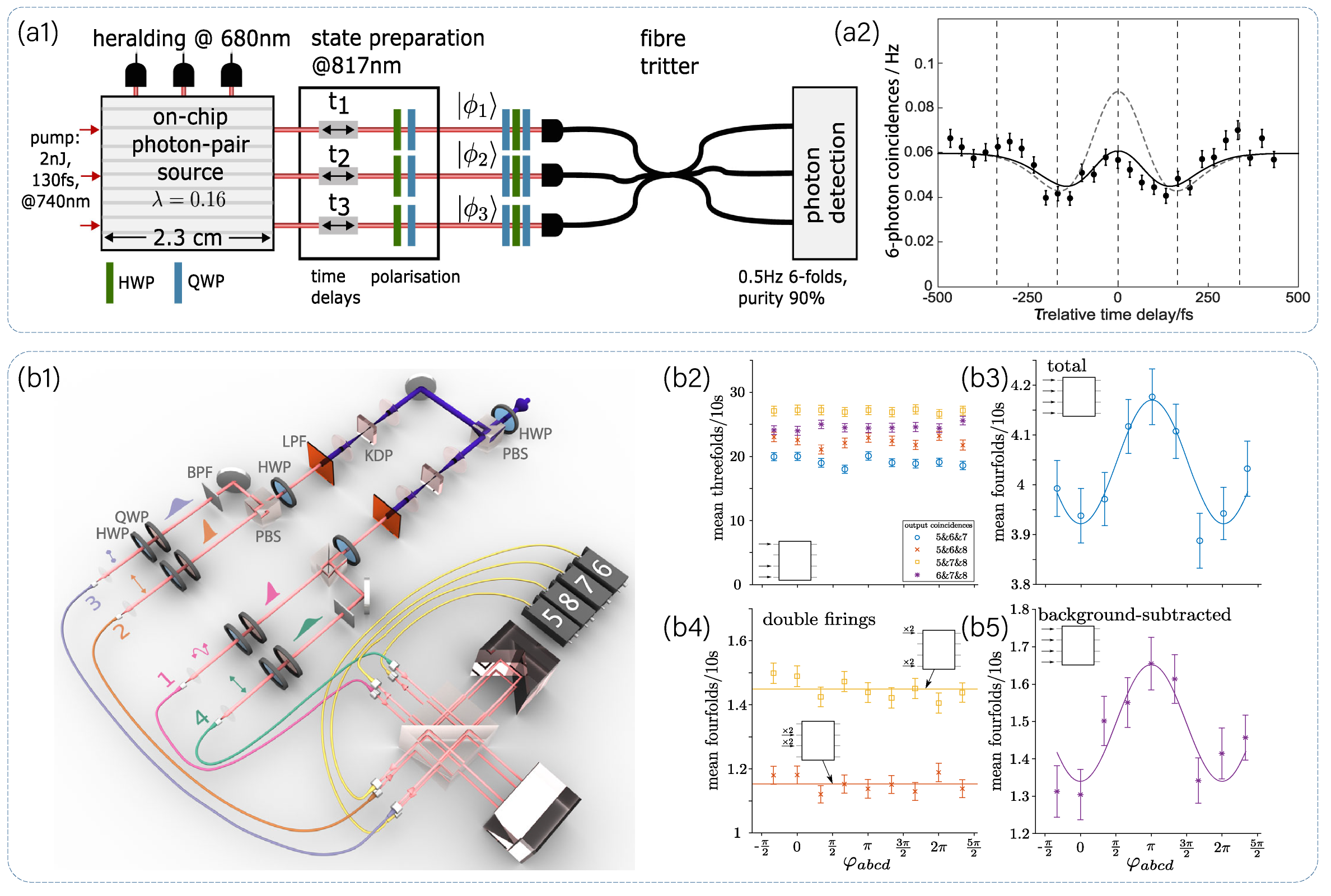}
\caption{Examples of multi-photon HOM interferometer. (a1-a2) Three-photon HOM interferometer with a tritter. Reprinted from Ref.\,\cite{Menssen2017PRL}. (b1-b5) Four-photon HOM interference with a quitter. Reprinted from Ref.\,\cite{Jones2020PRL}.}
\label{HOMmultiPhoton}
\end{figure}
%
%
In the traditional HOM interference setup, two photons interfere on a BS with two input ports and two output ports. However, when multiple photons and output ports are involved, the HOM interference pattern becomes more complex and can exhibit interesting features. The study of multiphoton interference can provide insight into the behavior of quantum systems with multiple particles and enables the development of advanced techniques for information processing, precise measurements, and the manipulation of quantum states \cite{You2023npj}. The first example is three-photon interference using a ``tritter'', which is a three-port symmetric beam splitter \cite{Spagnolo2013NC,Menssen2017PRL,Campos2000PRA}. Figure\,\ref{HOMmultiPhoton}(a1) illustrates the experimental setup for three-photon interference \cite{Menssen2017PRL}. In this setup, three heralded photons are generated through a spontaneous four-wave mixing (SFWM) process in silica-on-silicon waveguides. These photons are then directed into a fiber tritter where interference occurs. Figure \ref{HOMmultiPhoton}(a2) displays the heralded threefold coincidences ($P_{111}$) observed between the different output ports of the tritter when varying the temporal delays of the photons. The dashed gray line represents the theoretical prediction, while the black line is calculated using a model which incorporates experimental imperfections. It is evident that the three-photon interference exhibits much richer behavior compared to the two-photon HOM interference. 

The second example is four-photon interference using a ``quitter'', which is a four-port symmetric beam splitter \cite{Jones2020PRL, Tichy2011PRA}. Figure\,\ref{HOMmultiPhoton}(b1) presents the experimental setup for this four-photon interference. To generate the four photons, a pair of identical SPDC sources based on bulk potassium dihydrogen phosphate (KDP) crystals is used. The photons generated from these sources are then directed into a balanced four-port interferometer where they interfere. Figure\,\ref{HOMmultiPhoton}(b2-b5) depicts the interference patterns observed. In Fig.\,\ref{HOMmultiPhoton}(b2), the threefold coincidences are shown when different input states are sent to the quitter. Figure\,\ref{HOMmultiPhoton}(b3) represents the total fourfold coincidences obtained when the emissions from both sources are directed to the quitter. Figure\,\ref{HOMmultiPhoton}(b4) shows the four-fold coincidences of the output resulting from double emissions. Finally, Fig. \ref{HOMmultiPhoton}(b5) illustrates the background-subtracted fourfold coincidence $P_{1111}$. This experiment provides the first evidence of the interference of four pairwise distinguishable photons, which depends solely on a multiparticle phase.

\subsubsection{Detection dependent multi-photon HOM interferometer}
%
%
\begin{figure}[htbp]
\centering
\includegraphics[width= 0.92\textwidth]{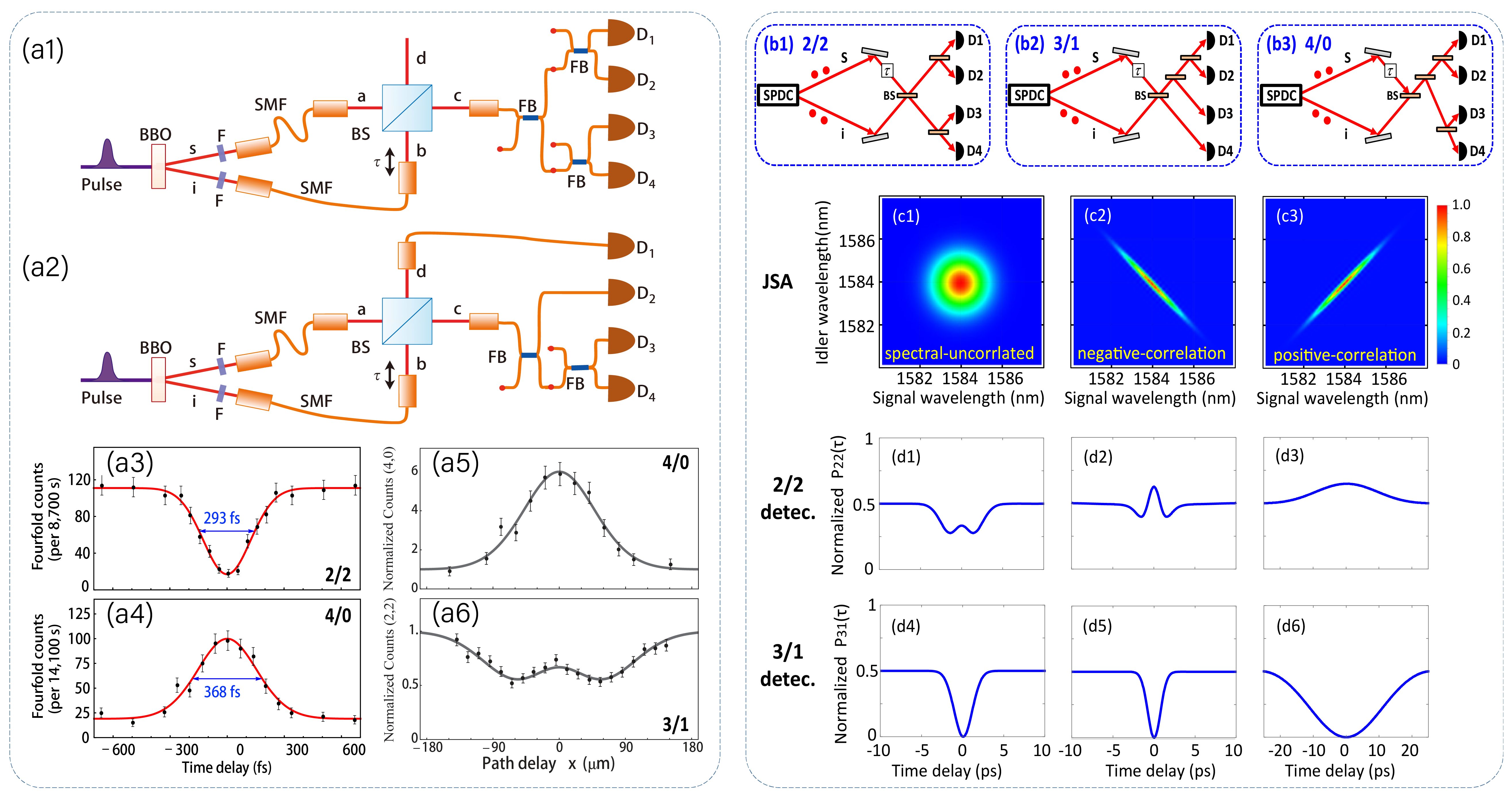}
\caption{The setups and interference patterns of detection dependent multiphoton HOM interferometer. (a1-a2) The experimental setup for the 2/2 and 3/1 detection schemes. (a3-a4) Interference patterns of 3/1 and 4/0 detections. Reprinted from Ref.\,\cite{Ra2013nc}. (a5-a6) Interference patterns of 4/0 and 2/2 detections. Reprinted from Ref.\,\cite{Ra2013PNAS}. (b1-b3) The 2/2, 3/1, and 4/0 detection schemes. (c1-c3): The joint spectral distributions of biphotons, including no spectral correlation, negative correlation, and positive correlation. (d1-d6) The interference patterns are simulated under different detection schemes and various spectral correlations. Reprinted from Ref.\,\cite{Jin2017PRA}.}
\label{HOMdependent}
\end{figure}
%
%
In multi-photon HOM interference, if the detection scheme differs, the interference patterns can vary significantly. This discrepancy arises from the postselection of the final state. Ra et al. reported an intriguing phenomenon regarding multi-photon coherence times \cite{Ra2013nc}. It was found that the coherence times, as defined by the width of the observed multi-photon signal, are not uniform. Instead, they depend on the number of interfering particles and the specific photon detection scheme employed. As illustrated in Fig.\,\ref{HOMdependent}(a1-a4), the coherence time for the 3/1 detection scheme is measured to be 293 fs, which is noticeably shorter than the 368 fs coherence time observed in the 4/0 detection scheme. This observation highlights that the strength of quantum multi-photon interference is greatly influenced by the chosen observable. The fundamental reason behind this phenomenon can be attributed to higher-order effects in the mutual indistinguishability of the photons \cite{Ra2013nc}.

Another interesting phenomenon is the nonmonotonic quantum-to-classical transition in multiphoton HOM interference. It was previously believed that as the indistinguishability increases, the multiphoton interference pattern would change monotonically. For example, in the case of the first HOM interference \cite{Hong1987PRL}, the two-fold coincidence counts show a monotonic increase when the time delay is scanned from zero to infinite. Similarly, this monotonic transition was also observed in the cases of four-photon \cite{Ou1999PRL} and six-photon \cite{Niu2009} HOM-type interference, where all the photons were detected in one output port of the BS. However, recent research challenges this notion. Ra et al. revealed that such a monotonic quantum-to-classical transition was an exception and only valid for two-photon cases and for bunching detection in multi-photon cases \cite{Ra2013PNAS}. In Fig.\,\ref{HOMdependent}(a5-a6), it can be seen that in the 4/0 detection scheme, the four-fold coincidence counts exhibited a monotonic dependence. In contrast, in the 2/2 detection scheme, a nonmonotonic dependence was observed.

The interesting phenomenon described in Ref.\,\cite{Ra2013PNAS} was observed with spectrally uncorrelated biphotons. However, when the biphotons exhibit spectral correlation, the interference patterns can be altered. In Ref.\,\cite{Jin2017PRA}, the role of spectral correlation (frequency entanglement) in the quantum-to-classical transition of a four-photon HOM interference was theoretically investigated, as shown in Fig.\,\ref{HOMdependent}(b1-d6). The study revealed that the transition can be monotonic for positively-correlated biphotons, while it can be nonmonotonic for negatively-correlated or uncorrelated biphotons in the 2/2 detection scheme. However, in the 3/1 and 4/0 detection schemes, the monotonicity of the transition remained unchanged. These findings highlight the influence of spectral correlation on the quantum-to-classical transition in multiphoton interference.
\subsubsection{Spectrally resolved HOM interferometer}
\begin{figure}[!ht]
\centering
\includegraphics[width= 0.92\textwidth]{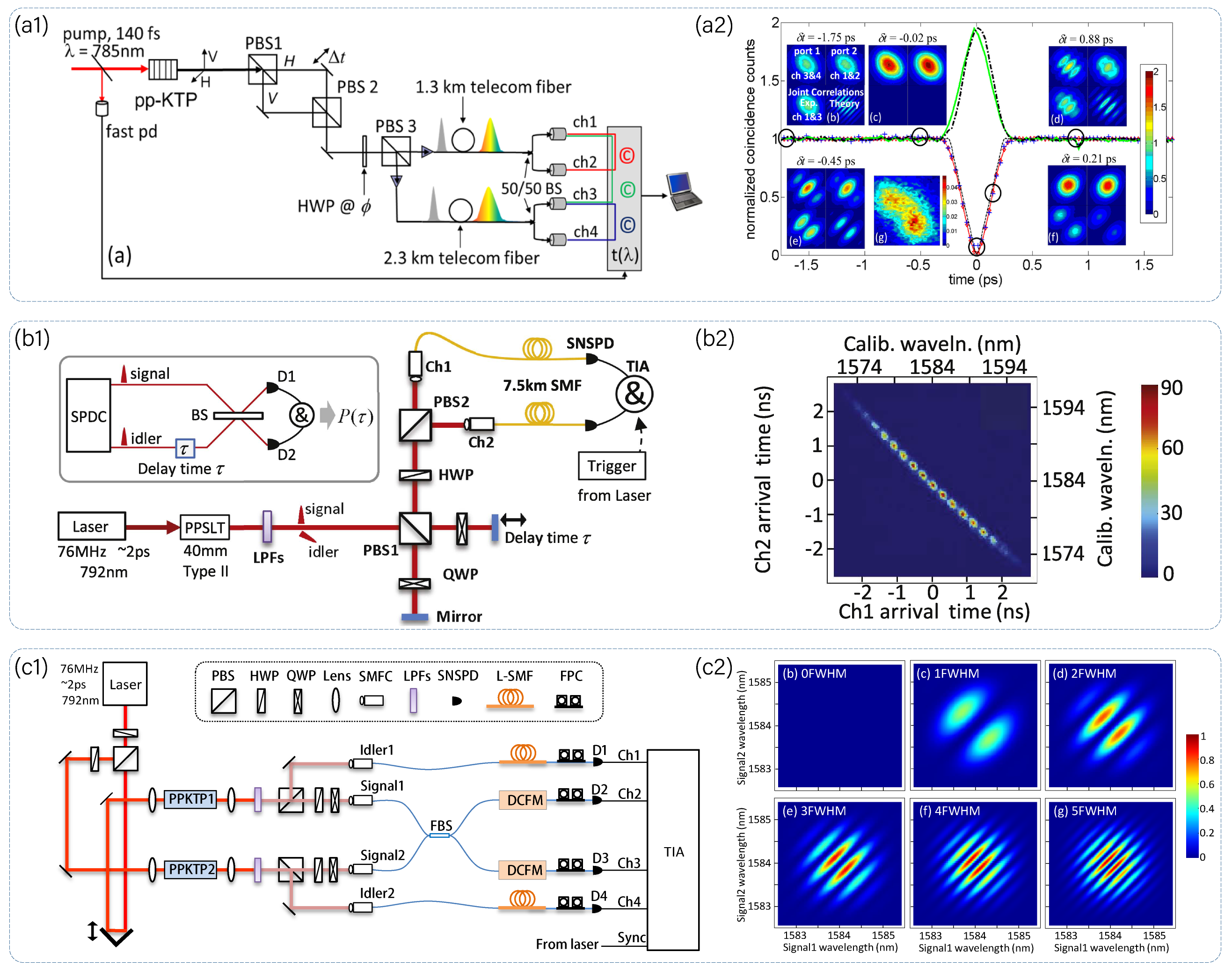}
\caption{Examples of spectrally resolved HOM interferometer. (a1-a2) The setup and interference patterns of the spectrally resolved HOM interferometer. Reprinted from Ref.\,\cite{Gerrits2015PRA}. (b1-b2) The setup and results for generating and distributing frequency-entangled qudits. Reprinted from Ref.\,\cite{Jin2016QST}. (c1-c2) The setup and interference patterns of the spectrally resolved HOM interference between two independent sources. Reprinted from Ref.\,\cite{Jin2015OE}. }
\label{HOMSpecRes}
\end{figure}
%
The traditional HOM interference is typically measured in the time domain, where the interference patterns are obtained by recording coincidence counts as a function of temporal delay. However, in time-domain measurements, the spectral information of the photons involved is integrated and lost. The knowledge of spectral correlations between interfering photons can provide valuable information on interference visibility \cite{MosleyPhD, Li2023OLT,Zhai2017JPB,Quan2015APB}. Therefore, exploring and understanding the spectral information in HOM interference is of great value.

Figure \ref{HOMSpecRes}(a1-a2) illustrates the first spectrally resolved HOM interference as presented in the study by Gerrits et al. \cite{Gerrits2015PRA}. In Fig. \ref{HOMSpecRes}(a1), the experimental setup for HOM interference is depicted, with the notable feature being the measurement of photons using a fiber spectrometer. Figure \ref{HOMSpecRes}(a2) displays the HOM interference patterns alongside the joint spectral distribution at various delay positions. It is evident that the joint spectral distribution exhibits splitting at different delay positions.

The technique of spectrally resolved HOM interference was subsequently employed for the generation and distribution of frequency-entangled qudits, as demonstrated in Fig.\,\ref{HOMSpecRes}(b1-b2) \cite{Jin2016QST}. In Fig.\,\ref{HOMSpecRes}(b1), the experimental setup for HOM interference is depicted, with a significant feature being the utilization of a type-II phase-matched periodically poled stoichiometric lithium tantalate (PPSLT) crystal. This crystal exhibits a unique distribution where the signal and idler photons possess a broad and anti-correlated spectral distribution \cite{Shimizu2009}. Fig.\,\ref{HOMSpecRes}(b2) showcases the measured joint spectral distributions of biphotons, which indeed represent spectrally entangled qudits. This method enables the generation of frequency-entangled qudits with $d>10$ without the need for spectral filters or cavities. Furthermore, the generated state can be distributed over a total length of 15 km, showcasing the potential for long-distance applications of this technique.

Finally, Fig. \ref{HOMSpecRes}(c1-c2) present the experimental setup and interference patterns of spectrally resolved HOM interference between two independent sources \cite{Jin2015OE}. The interference pattern in the joint spectral distribution is distinctly observed and closely aligns with the theoretical predictions. This experiment provides valuable insights into the spectral domain perspective of HOMI-IPS and contributes to a deeper understanding of this phenomenon. 
In addition, the spectrally resolved phenomenon is also explored in the modified HOM interferometer recently \cite{Li2023OLT}.

\subsection{Applications of HOM interferometers}
\subsubsection{Quantum communication}
\begin{figure}[!ht]
\centering
\includegraphics[width= 0.92\textwidth]{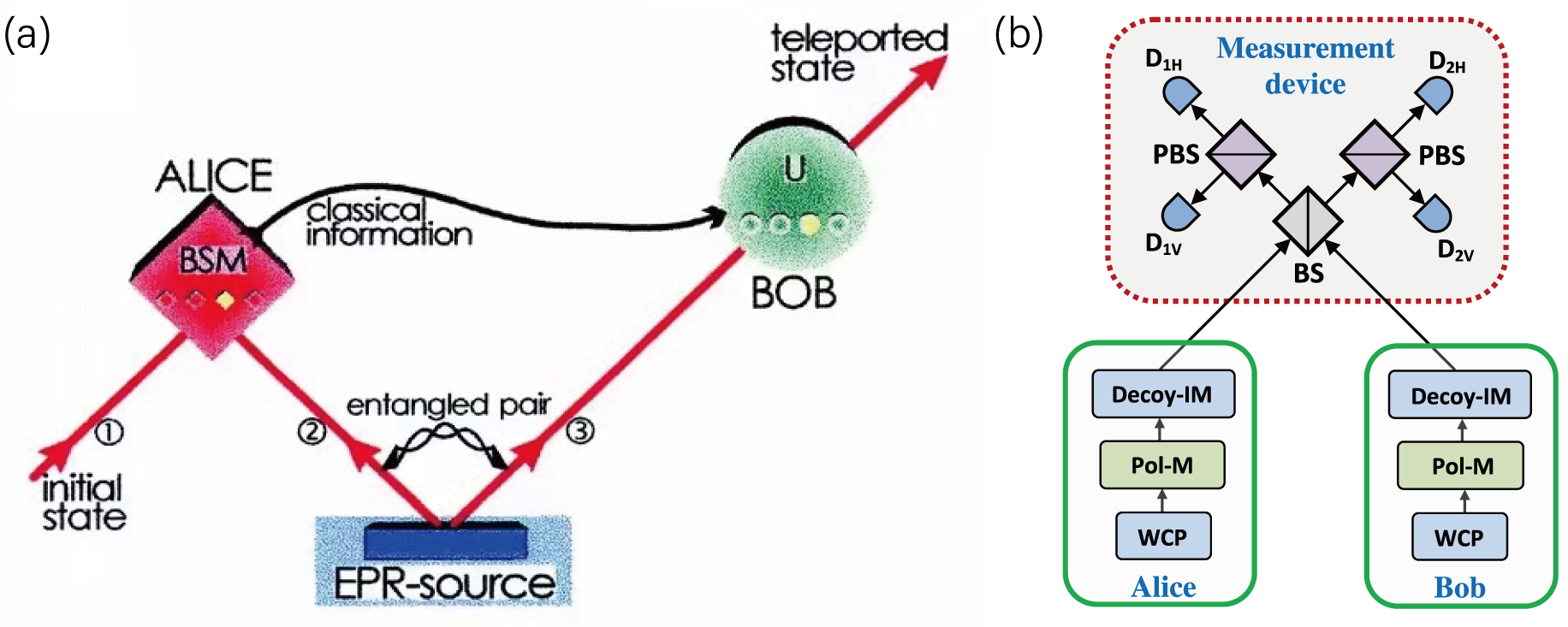}
\caption{Applications of HOM interferometer in quantum communication. (a) The quantum teleportation scheme. Reprinted from Ref.\,\cite{Bouwmeester1997Nature}. (b) Schematic of the measurement-device-independent quantum key distribution (MDI-QKD). Reprinted from Ref.\,\cite{Lo2012PRL}.  }
\label{HOMcommunication}
\end{figure}

The HOM interferometer has significant applications in quantum communication. HOM interferometer can be used as a powerful tool for achieving secure and efficient transmission of quantum information. The first example is quantum teleportation, a remarkable process that enables the transmission and reconstruction of the state of a quantum system over arbitrary distances \cite{Bouwmeester1997Nature,Hu2023NRP,Ren2017Nature,Thoma2016PRL,Jin2015SR,Valivarthi2016np}. Using the HOM interferometer to generate entanglement between remote particles, quantum teleportation protocols can be implemented, allowing for the secure and efficient transmission of quantum information \cite{Bouwmeester1997Nature,Ge2024SA}. Figure\,\ref{HOMcommunication}(a) shows the scheme of quantum teleportation. In the process of quantum teleportation, Alice possesses a quantum system called particle 1, which she intends to teleport to Bob. Additionally, Alice and Bob share an entangled pair of particles, particle 2 and particle 3, generated by an Einstein-Podolsky-Rosen (EPR) source. To initiate the teleportation, Alice performs a joint Bell-state measurement (BSM) on particle 1 and one of the entangled ancillary particles (either particle 2 or particle 3). This BSM results in the projection of both particles onto an entangled state, establishing a correlation between them. After performing the measurement, Alice transmits the outcome of her measurement as classical information to Bob. This classical information carries the necessary instructions for Bob to perform a corresponding unitary transformation on the remaining ancillary particle (the one not used for the BSM). By applying the unitary transformation to the ancillary particle, Bob can reconstruct the state of the original particle 1. This means that the state of the initial particle has been teleported from Alice to Bob, achieving the transmission of quantum information without physically transmitting the quantum state itself.

Another key application of the HOM interferometer in quantum communication is in measurement-device-independent quantum key distribution (MDI-QKD) protocols \cite{Lo2012PRL,Tang2014PRL,TangYL2014PRL,Yin2016PRL,Gisin2010PRL,Cao2020PRL,Duplinskiy2021PRA,Woodward2021npj,Prabhakar2020SA}. The HOM interferometer enables secure key generation without relying on the trustworthiness of the measurement devices. Figure\,\ref{HOMcommunication}(b) illustrates the basic setup of an MDI-QKD protocol. In this setup, Alice and Bob generate phase-randomized weak coherent pulses (WCPs) in different BB84 polarization states. The polarization state for each signal is independently and randomly selected using a polarization modulator (Pol-M). Decoy states are also generated using an intensity modulator (Decoy-IM). Inside the measurement device, the signals from Alice and Bob interfere at a 50:50 BS. At each end of the BS, there is a polarizing beam splitter (PBS) that projects the input photons into either horizontal (H) or vertical (V) polarization states. Four single-photon detectors are used to detect the photons, and the detection results are publicly announced. A successful Bell state measurement is indicated by precisely two detectors being triggered, signifying the observation of a specific outcome. The MDI-QKD protocol offers several advantages over conventional QKD methods. Firstly, it eliminates all detector side channel attacks, enhancing the security of the system. Additionally, it enables doubling the secure distance compared to conventional laser-based implementations. Another benefit of MDI-QKD is its compatibility with standard optical components, even in scenarios with low detection efficiency and highly lossy channels. This makes it a practical and feasible solution for real-world implementations. Furthermore, MDI-QKD achieves a significantly higher key generation rate compared to full device-independent QKD approaches. The key generation rate of MDI-QKD is several orders of magnitude higher, making it more efficient for practical quantum communication applications.

\subsubsection{Quantum computation}
%
%
\begin{figure}[htbp]
\centering
\includegraphics[width= 0.92\textwidth]{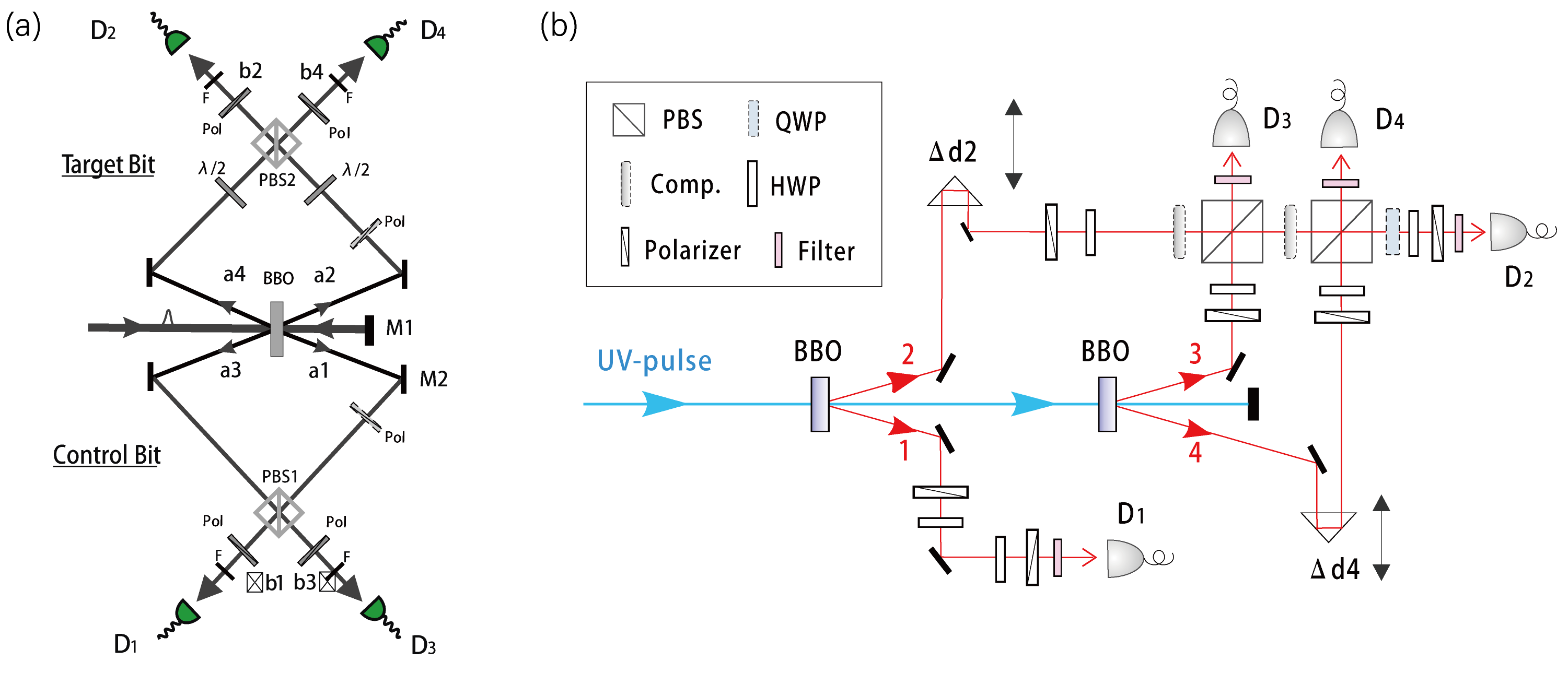}
\caption{Applications of HOM interferometer in quantum computation. (a) The setup for the realization of a photonic controlled-NOT gate. Reprinted from Ref.\,\cite{Gasparoni2004PRL}. (b) The experimental setup for verifying a compiled version of Shor’s quantum factoring algorithm. Reprinted from Ref.\,\cite{Lu2007PRL}.  }
\label{HOMcomputation}
\end{figure}
%
%
The HOM interferometer also has important applications in quantum computing. HOM interferometer can be utilized as a fundamental tool for implementing quantum gates and performing quantum operations \cite{OBrien2003Nature,Carolan2015Science, Gasparoni2004PRL,Luo2019SA, Zhong2020Science, Lu2007PRL, Lanyon2007PRL}. One of the key applications of the HOM interferometer in quantum computing is the realization of controlled-NOT (CNOT) gates, which are essential building blocks for various quantum algorithms \cite{OBrien2003Nature, Gasparoni2004PRL}. By encoding quantum information into the states of two particles, such as qubits, and directing them towards a BS, the interference between the particles can be controlled to implement CNOT gates. This allows for entangling qubits and performing quantum computations. Figure \ref{HOMcomputation}(a) shows the experimental setup that combines linear optical components, polarization entanglement, and postselection techniques to implement the CNOT gate on the input qubits carried by independent photons \cite{Gasparoni2004PRL}. Here, the HOM interference measurement is important for verifying the success of the gate operation and confirming the coherence of the output states. Another notable example is the experimental demonstration of a compiled version of Shor's algorithm by Lu et al. in 2007 \cite{Lu2007PRL}. Figure \ref{HOMcomputation}(b) illustrates the experimental setup. They utilized a simplified linear optical network to coherently implement the quantum circuits required for modular exponential execution and semi-classical quantum Fourier transformation, which are key components of Shor's algorithm. While performing the computation, genuine multiparticle entanglement was observed, providing strong evidence for the quantum nature of the system. This experiment represents a significant milestone towards the full realization of Shor's algorithm and scalable linear optics quantum computation. It demonstrates the potential to use CNOT gates and other quantum operations in complex computations \cite{Lu2007PRL, Lanyon2007PRL}.

\subsubsection{Quantum metrology}
%
%
\begin{figure}[htbp]
\centering
\includegraphics[width= 0.92\textwidth]{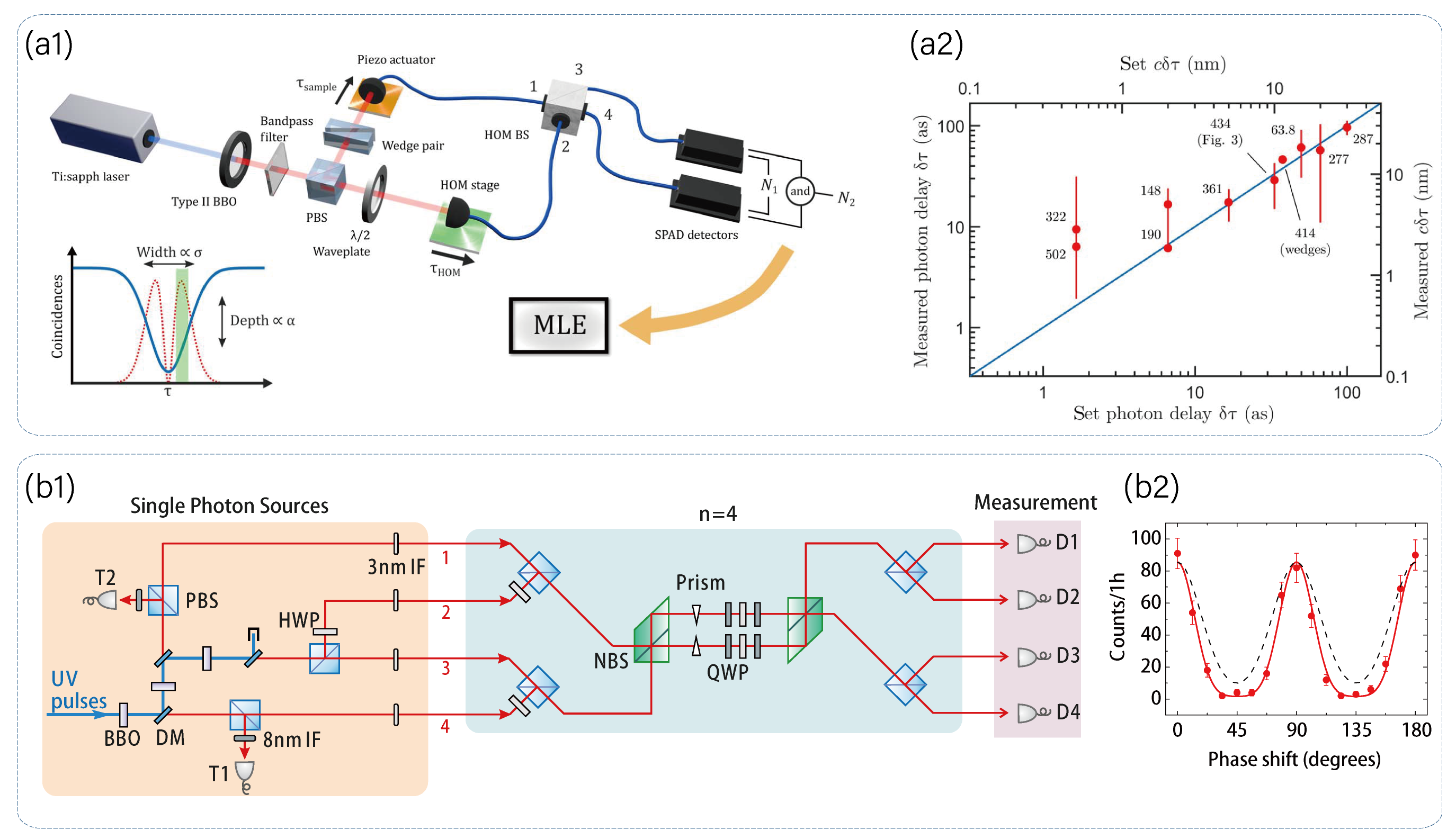}
\caption{Applications of HOM interferometer in quantum metrology. (a1) The setup for attosecond-resolution HOM interferometry. (a2) Experimentally measured photon delays are shown against the set delays. Reprinted from Ref.\,\cite{Lyons2018SA}. (b1) The setup for quantum Fourier transform (QFT) circuits. (b2) Quantum metrology results based on the QFT. Reprinted from Ref.\,\cite{Su2017PRL}. }
\label{HOMmetrology}
\end{figure}
%
%
The HOM interferometer also has significant applications in quantum metrology, which focuses on achieving high-precision measurements using quantum systems. HOM interferometer can enhance the sensitivity and precision of measurements in various scenarios \cite{Lyons2018SA, Su2017PRL, Quan2016SR, Liu2021APL, Quan2019OL}. One notable example is attosecond-resolution HOM interferometry. In 2018, Lyons et al. devised and implemented a novel measurement and estimation strategy based on Fisher information analysis \cite{Lyons2018SA}. 
The experimental setup and results are shown in Fig.\,\ref{HOMmetrology}(a).
The authors achieved a significant improvement in precision and accuracy by two orders of magnitude compared to previous HOM approaches. They accomplished this by tuning the interferometer to the delay that maximizes the information content and employing a maximum-likelihood estimation (MLE) procedure.
%
Specifically, they conducted measurements of the change in relative arrival time between two photons, represented as $\delta$t, achieving an average accuracy of 6 as (corresponding to 1.7 nm) and an average precision of 16 as (4.8 nm). The best accuracy achieved was 0.5 as (0.15 nm), and the best precision achieved was 4.7 as (0.9 nm).
This work has the potential to enable the characterization of optically transparent samples using single photons, with thicknesses and length scales relevant to fields such as cell biology.

Another important application is the use of multi-photon HOM interference-based quantum Fourier transform (QFT) circuits for quantum metrology purposes.
In Ref.\,\cite{Su2017PRL}, Su et al. presented a novel approach to simplify the construction of quantum Fourier transform (QFT) interferometers by utilizing both path and polarization modes.
This approach allows for a reduction in resources by up to 75\% compared to devices that only utilize path modes. 
The researchers reported the first experimental demonstration of the generalized HOM interference with up to four photons.
Additionally, they constructed multimode  MZIs by cascading two QFTs and observed deterministic phase supersensitivities.
Figure\,\ref{HOMmetrology}(b1) illustrates the experimental setup. 
The QFT interferometers, which can be regarded as generalized HOM interferometers, were implemented by simultaneously utilizing polarization and path modes.
Figure\,\ref{HOMmetrology}(b2) presents the measured four-fold coincidence counts as a function of the phase shift in four-photon generalized HOM interference.
The solid red line represents a fit to the case of indistinguishable single photons, while the dashed line represents the limiting distribution of distinguishable single photons.
The fringes exhibit phase superresolution and, importantly, oscillate with a higher visibility than the corresponding classically limited distributions.

\subsubsection{Quantum imaging}
\begin{figure}[htbp]
\centering
\includegraphics[width= 0.90\textwidth]{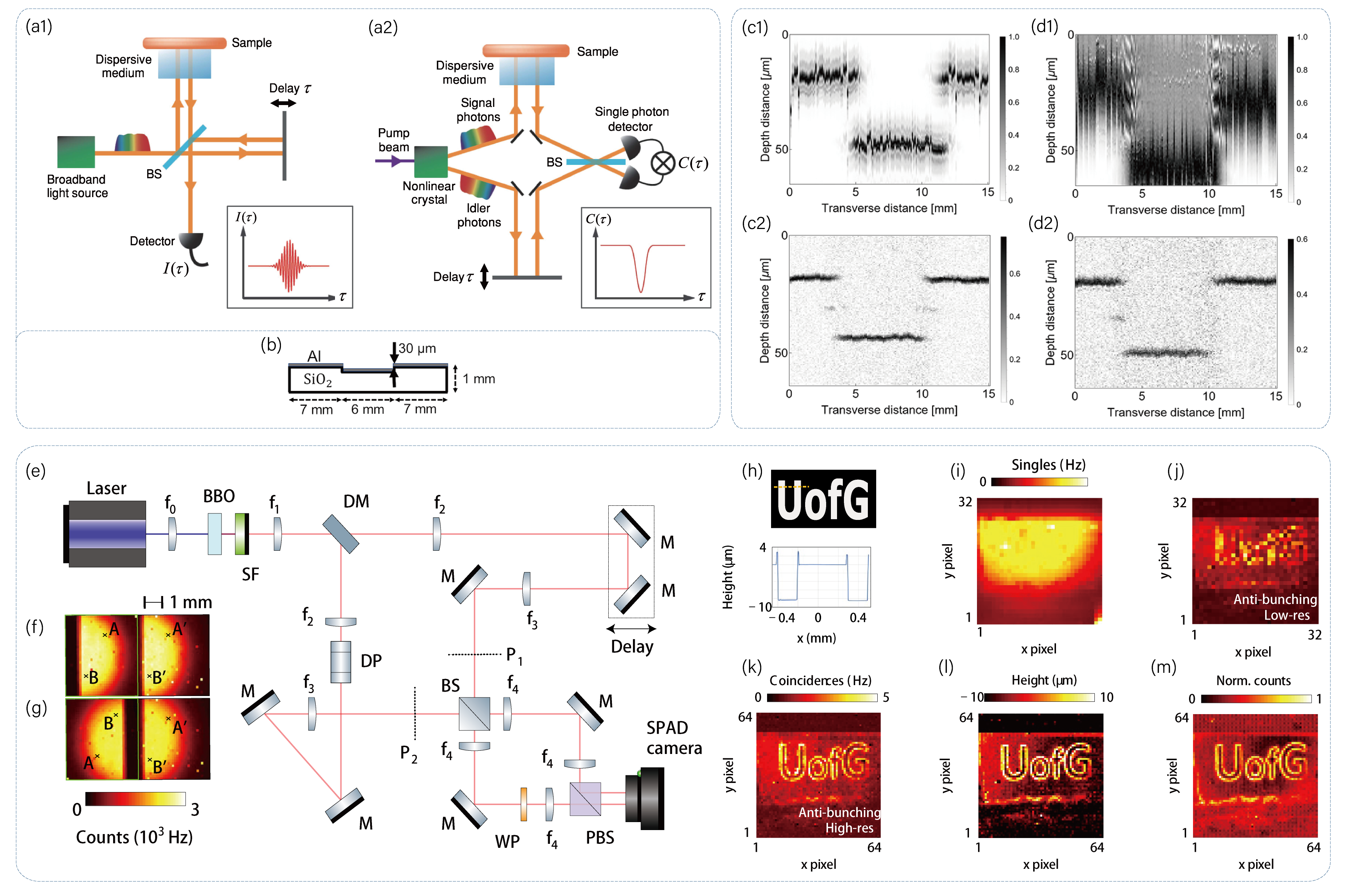}
\caption{Applications of HOM interferometer in quantum imaging. (a1-a2) Experimental setup for optical coherent tomography (OCT) and quantum optical coherent tomography (QOCT). Reprinted from Ref.\,\cite{Okano2015SR}. (b) Structure of the
aluminum-coated glass sample. (c1-c2) OCT and QOCT images without dispersion material. (d1-d2) OCT and QOCT images after the insertion of the dispersion material. Reprinted from Ref.\,\cite{Hayama2022OL}.
(e) Experimental setup of quantum microscopy based on the HOM interferometer. (f, g) Intensity image acquired by the SPAD camera.
(h) The sample. (i-m) Experimentally measured images of the sample. Reprinted from Ref.\,\cite{Ndagano2022NP}.
}
\label{HOMimaging}
\end{figure}

HOM interferometer can also be applied in quantum imaging.
One typical example is quantum optical coherence tomography (QOCT).
QOCT combines the principles of optical coherence tomography (OCT) with quantum optics to achieve high-resolution imaging and depth profiling of biological samples. OCT, with the principle shown in Fig.\,\ref{HOMimaging}(a1) \cite{Okano2015SR} is a non-invasive imaging modality widely used in medical and biological research for imaging and analyzing biological tissues  \cite{Huang1991Science,Fercher2003RPP,Brezinski2006Book}.

However, a significant challenge in OCT arises when attempting to achieve higher resolution by broadening the bandwidth of the light source. Despite the intention of improving resolution, the presence of dispersion in the medium causes the resolution to degrade instead. This phenomenon poses a severe problem in OCT systems.
On the other hand, QOCT has been proposed as a solution to the resolution degradation caused by dispersion in traditional OCT systems. QOCT utilizes HOM interference of frequency-entangled photon pairs, as depicted in Fig.\,\ref{HOMimaging}(a2) \cite{Okano2015SR}.
One key advantage of QOCT is its ability to maintain resolution, represented by the width of the HOM interference dip, even in the presence of group velocity dispersion (GVD) in the medium. This is due to the frequency correlation of entangled photon pairs, which enables a phenomenon known as ``dispersion cancellation''.

QOCT was proposed \cite{Abouraddy2002PRA} by Teich's group in 2002 and subsequently experimentally demonstrated in 2003 with a resolution of 18.5 $\mu$m   \cite{Nasr2003PRL}.
Following the initial demonstration, this research group conducted a series of experiments between 2003 and 2009 to showcase the applicability of QOCT on different samples \cite{Nasr2004OE,Booth2004PRA,Booth2011OC,Mohan2009AO,Nasr2009OC,Nasr2008OE}, for example,   dispersive medium \cite{Nasr2004OE}, polarization-sensitive materials \cite{Booth2004PRA,Booth2011OC}, and biological samples \cite{Nasr2009OC}. 
In 2008, Kaltenbaek et al.  demonstrated a completely classical technique called chirped-pulse interferometry, which is based on oppositely-chirped laser pulses, to produce an interferogram with the advantages of HOM interference \cite{Kaltenbaek2008NP}. 
The following year, they utilized their technique to solve the artifact problem in QOCT, highlighting the power of classical correlation in optical imaging \cite{Lavoie2009OE}.
Recently, several strategies for removing artifacts have been proposed and contributed to the development of QOCT \cite{Kolenderska2020OE,Kolenderska2021SR,Krzysztof2021}.
In 2020, Ibarra-Borja et al. have demonstrated full field QOCT, achieving an impressive depth resolution of 6 $\mu$m \cite{Ibarra-Borja2020PR}.

In particular, in 2022, Hayama et al. achieved a depth resolution of 2.5 $\mu$m, setting a new record as the highest value ever achieved in QOCT imaging \cite{Hayama2022OL}.
Figure\,\ref{HOMimaging}(b) depicts the detailed structure of the aluminum-coated glass sample.
Figure\,\ref{HOMimaging}(c1-c2) shows the OCT and QOCT images without the insertion of a dispersion material, while Figure\,\ref{HOMimaging}(d1-d2) displays the OCT and QOCT images with the insertion of a dispersion material (ZnSe window). Notably, the QOCT images clearly demonstrate the ability to cancel out the effects of dispersion.
This implies that QOCT offers a promising solution to overcome the resolution limitations imposed by dispersion in OCT systems.

Another important example of quantum imaging is the HOM interferometer-based microscopy.
In 2022,  Ndagano et al. reported a scan-free quantum imaging technique that utilizes HOM interference to reconstruct the surface depth profile of transparent samples \cite{Ndagano2022NP}.
They demonstrated the capability to retrieve images with micrometer-scale depth features using a low photon flux of only seven photon pairs per frame.
By employing a single-photon avalanche diode (SPAD) camera, they measured both bunched and anti-bunched photon-pair distributions at the output of an HOM interferometer. These distributions were combined to generate a lower-noise image of the sample, improving the image quality.
Figure\,\ref{HOMimaging}(e) displays the experimental setup.
 %
The two outputs from the BS were depicted in Fig.\,\ref{HOMimaging}(f,g), where pixel positions A (B) and A' (B') correspond to the photon paths in the two outputs of the BS.
 The sample was fabricated by etching the letters `UofG' onto a glass substrate to a depth of 8.36 $\mu m$. Fig.\,\ref{HOMimaging}(h) displays the profile along the yellow dashed line, which was measured using a profilometer.
Figure\,\ref{HOMimaging}(i) presents the intensity image, which does not reveal any details about the shape of the etched sample.
This is primarily due to the sample's transparency and the fact that the image is acquired through direct illumination, rather than through interference.
Figure\,\ref{HOMimaging}(j) displays the anti-bunching coincidence image at the camera’s native resolution of 32$\times$32 pixels. 
The `UofG' pattern is visible but strongly under-resolved in this image.
By performing a 2$\times$2 raster scan to increase the pixel resolution by a factor of four, achieving 64$\times$64 pixels, the sample becomes clearly visible in the anti-bunching HOM image shown in Fig.\,\ref{HOMimaging}(k).
Figure\,\ref{HOMimaging}(l) presents the retrieved depth image from Fig.\,\ref{HOMimaging}(k). The combination of weighted bunching and anti-bunching images produces a normalized image with reduced noise, as depicted in Fig.\,\ref{HOMimaging}(m).
This experiment demonstrates the potential of HOM microscopy as a tool for label-free imaging of transparent samples in the very low photon regime.

\section{N00N State Interferometer}

N00N state is an entangled state with $N$ photons that occupy either one of two optical modes (e.g., polarization modes or path modes). The N00N state could be written as $\frac{1}{{\sqrt 2 }}(\left| {N0} \right\rangle  + \left| {0N} \right\rangle )$. N00N state interferometry is a powerful tool for improving phase detection precision to the Heisenberg limit of $\Delta \phi=1/N$, much higher than the shot-noise limit of $\Delta \phi=1/\sqrt{N}$, which is the precision limit of a classical interferometer \cite{Giovannetti2004Science}. N00N state interferometer (N00N-SI) has been widely used in quantum measurements \cite{Mitchell2004Nature,Walther2004,Nagata2007Science,Afek2010Science,Slussarenko2017NP,Zhou2017PRAppl}, quantum imaging \cite{Ono2013NC}, quantum spectroscopy \cite{Jin2018Optica}, and error correction \cite{Bergmann2016PRA}. In this section, we will first discuss the concept of N00N-SI, then move on to the generation of the N00N state, and finally explore the applications of N00N-SI.

\subsection{Principles of N00N state interferometer}
\subsubsection{Single-mode theory of N00N state interferometer}
\begin{figure}[htbp]
\centering
\includegraphics[width=0.85\textwidth]{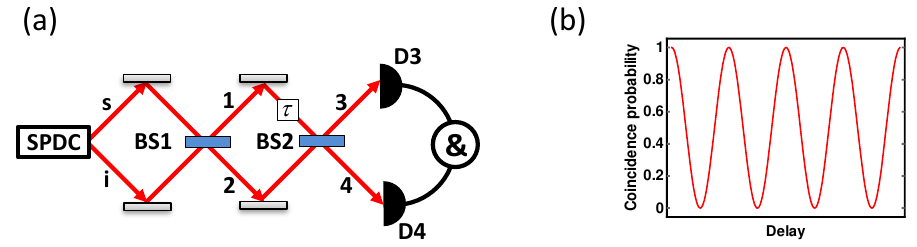}
\caption{(a) The setup of single-mode N00N state interferometer and (b) the corresponding interference pattern. Reprinted from Ref.\,\cite{Jin2018Optica}. }
\label{N00Nsetup1}
\end{figure}
The setup of single-mode N00N-SI is shown in Fig.\,\ref{N00Nsetup1} (a).
The input state is
${\left| \psi  \right\rangle } = {\left| {11} \right\rangle _{s,i}} = \hat a_s^\dag \hat a_i^\dag \left| {00} \right\rangle$, where $\hat a_s^\dag $ and $\hat a_i^\dag $ are the creation operators of the signal and the idler after SPDC. Follow the notation in Fig. \ref{N00Nsetup1}(a), the annihilation operators $\hat a_1  $, $\hat a_2  $, $\hat a_3  $, and $\hat a_4  $ are assigned to the optical paths 1, 2, 3, and 4, respectively. Phase shift $\omega \tau $ is added to arm 1, by scanning an optical path delay $\tau$.  The operators evolve as follow
\begin{equation}\label{Eq:2-1-2}
\begin{aligned}
\hat{a}_1 &=\sqrt{\frac{1}{2}}\left(\hat{a}_s +\hat{a}_i \right),
\hat{a}_2 =\sqrt{\frac{1}{2}}\left(\hat{a}_s -\hat{a}_i \right),\\
\hat{a}_3 &=\sqrt{\frac{1}{2}}\left(\hat{a}_1  e^{i\omega \tau}+\hat{a}_2 \right),
\hat{a}_4 =\sqrt{\frac{1}{2}}\left(\hat{a}_1 e^{i\omega \tau}-\hat{a}_2 \right).
\end{aligned}
\end{equation}
Therefore,
\begin{equation}\label{Eq:2-1-4}
\begin{aligned}
\hat{a}_3&=\frac{1}{2}\left[(1+e^{i\omega \tau})\hat{a}_s  +(-1+e^{i\omega \tau})\hat{a}_i \right],\\
\hat{a}_4 &=\frac{1}{2}\left[(-1+e^{i\omega \tau})\hat{a}_s  +(1+e^{i\omega \tau})\hat{a}_i \right].
\end{aligned}
\end{equation}

The coincidence probability $P(\tau)$ is determined by the second order correlation function, i.e.,
$P(\tau) = \left\langle \psi  \right| \hat a_4^\dag \hat a_3^\dag  {{\hat a}_3}{{\hat a}_4}\left| \psi  \right\rangle$.
First, we consider
${{\hat a}_3}{{\hat a}_4}\left| \psi  \right\rangle$,

\begin{equation}\label{Eq:2-1-5}
\hat{a}_3  \hat{a}_4 \left| \psi  \right\rangle =\frac{1}{4}\left[(1+e^{i\omega \tau})^2\hat{a}_s \hat{a}_i  +(-1+e^{i\omega \tau})^2\hat{a}_i \hat{a}_s \right] \left| {11} \right\rangle _{s,i} =\frac{1}{2}(1+e^{i 2\omega \tau})  \left| {00} \right\rangle _{s,i}.
\end{equation}
Note in the above calculation, $ \hat{a}_s \hat{a}_s\left| {00} \right\rangle _{s,i}=0$ and  $ \hat{a}_i \hat{a}_i\left| {00} \right\rangle _{s,i}=0$.
Therefore, 
\begin{equation}\label{Eq:2-1-6}
 P(\tau) = \left\langle \psi  \right| \hat a_4^\dag \hat a_3^\dag  {{\hat a}_3}{{\hat a}_4}\left| \psi  \right\rangle
 =\frac{1}{2}[1+\cos{(2\omega \tau)}]. 
\end{equation}
The corresponding interference pattern is shown in Fig.\,\ref{N00Nsetup1}(b).

\subsubsection{The multi-frequency theory of N00N state interferometer for biphotons with no spectral correlation}
%
\begin{figure}[htbp]
\centering
\includegraphics[width= 0.85\textwidth]{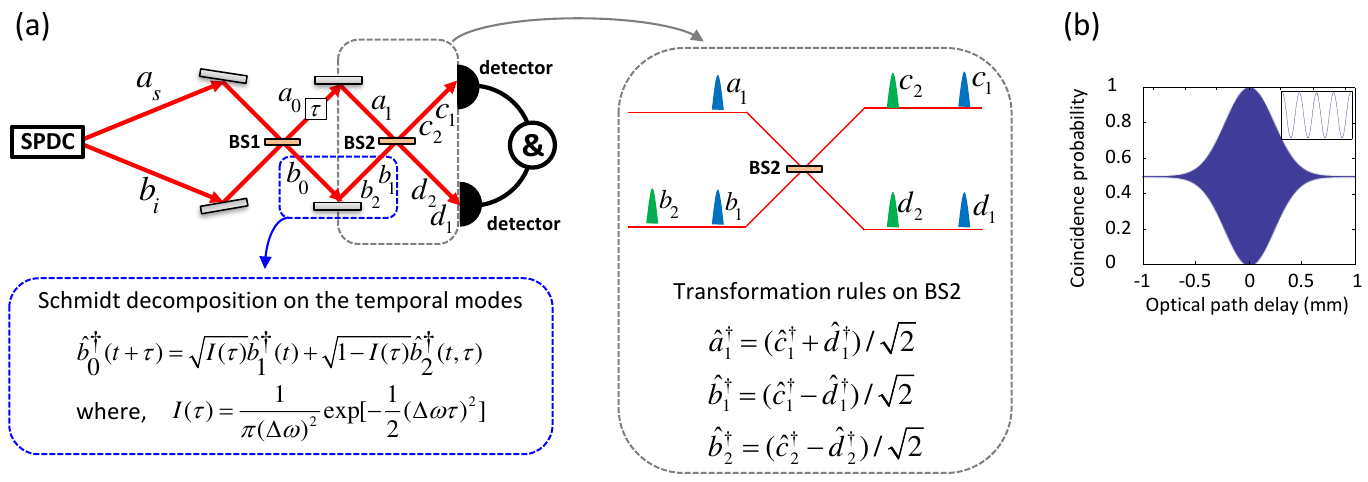}
\caption{(a) The setup of multi-frequency N00N state interference. (b) The multi-frequency N00N-state interference pattern in a large delay, and inset is the fine pattern in two periods. Reprinted from Ref.\,\cite{Jin2016SR}.}
\label{N00Nsetup2}
\end{figure}
%
The theory of multi-frequency N00N-SI with uncorrelated biphotons was introduced in Ref.\,\cite{Jin2016SR}. The setup of the multi-frequency N00N-SI is shown in Fig.\,\ref{N00Nsetup2}(a). The input state is given by $\left| {11} \right\rangle _{s,i}  = \hat a_s^\dag  \hat b_i^\dag  \left| 0 \right\rangle$. After BS1
%
and a phase shift $\omega \tau$ in the arm $a_0$, the state evolves to
\begin{equation}\label{Eq:2-2-3}
\frac{1}{2}[(e^{i\omega \tau } \hat a_0^\dag  (t))^2  - \hat b_0^\dag  (t + \tau )^2 ]\left| 0 \right\rangle.
\end{equation}
In the above equation, due to the optical path delay between the arm $a_0$ and $b_0$, the time information for $\hat a_0^\dag $ is $t$, and the time information for $\hat b_0^\dag $ is $t + \tau$.
We can apply a Schmidt decomposition on the temporal modes for arm $b_0$:
\begin{equation}\label{Eq:2-2-4}
\hat b_0^\dag  (t + \tau ) = \sqrt {I(\tau )} \hat b_1^\dag  (t) + \sqrt {1 - I(\tau )} \hat b_2^\dag  (t,\tau ),
\end{equation}
and we obtain the state
\begin{equation}\label{Eq:2-2-5}
\frac{1}{2}[e^{i2\omega \tau } \hat a_1^\dag  (t)^2  - (\sqrt {I(\tau )} \hat b_1^\dag  (t) + \sqrt {1 - I(\tau )} \hat b_2^\dag  (t,\tau ))^2 ]\left| 0 \right\rangle,
\end{equation}
where $\hat a_0^\dag  (t)=\hat a_1^\dag  (t)$ and $I(\tau ) = \frac{1}{{\pi (\Delta \omega )^2 }}\rm{exp}[ - \frac{1}{2}(\Delta \omega \tau )^2 ]$ is the indistinguishability. 

The state after BS2 is given by
\begin{equation}\label{Eq:2-2-10}
\frac{1}{4}\{ e^{i2\omega \tau } (\hat c_1^\dag   + \hat d_1^\dag  )^2  - [\sqrt I(\tau) (\hat c_1^\dag   - \hat d_1^\dag  ) + \sqrt {1 - I(\tau)} (\hat c_2^\dag   - \hat d_2^\dag  )]^2 \} \left| 0 \right\rangle,
\end{equation}
where the time label for each operator is omitted for simplicity. For $\left| {11} \right\rangle$ state detection at ports $c$ and $d$, the corresponding projected state is
\begin{equation}\label{Eq:2-2-11}
\frac{1}{2}\{e^{i2\omega \tau } \left| {11} \right\rangle _{c_1 d_1 }  + I(\tau )\left| {11} \right\rangle _{c_1 d_1 }  + [1 - I(\tau )]\left| {11} \right\rangle _{c_2 d_2 }  + \sqrt {I(\tau )[1 - I(\tau )]} (\left| {11} \right\rangle _{c_1 d_2 }  + \left| {11} \right\rangle _{c_2 d_1 } )\}.
\end{equation}
The detection probability of $\left| {11} \right\rangle$ state is
\begin{equation}\label{Eq:2-2-12}
P_{11}  = \frac{1}{4}\{{\rm{|}}e^{i2\omega \tau } + I(\tau ){\rm{|}}^2  + {\rm{|}}1 - I(\tau ){\rm{|}}^2  + 2{\rm{|}}\sqrt {I(\tau )[1 - I(\tau )]} {\rm{|}}^2 \}
=\frac{1}{2}[1 + I(\tau )\rm{cos}(2\omega \tau )].
\end{equation}
The corresponding interference pattern is shown in Fig.\,\ref{N00Nsetup2}(b).

This theoretical model can be readily extended to investigate multi-photon N00N state interference cases, as demonstrated in Ref.\,\cite{Jin2016SR}.
However, in this model, the biphotons generated from SPDC exhibit no spectral correlation. Consequently, this theory cannot account for scenarios where biphotons exhibit spectral correlation, which is a common characteristic observed in quantum sources. Therefore, in the next section, we will delve into the multi-frequency theory of N00N state interferometers for biphotons with arbitrary spectral distributions.

\subsubsection{The multi-frequency theory of N00N state interferometer for biphotons with arbitrary distribution}
\begin{figure}[!h]
\centering
\includegraphics[width= 0.95\textwidth]{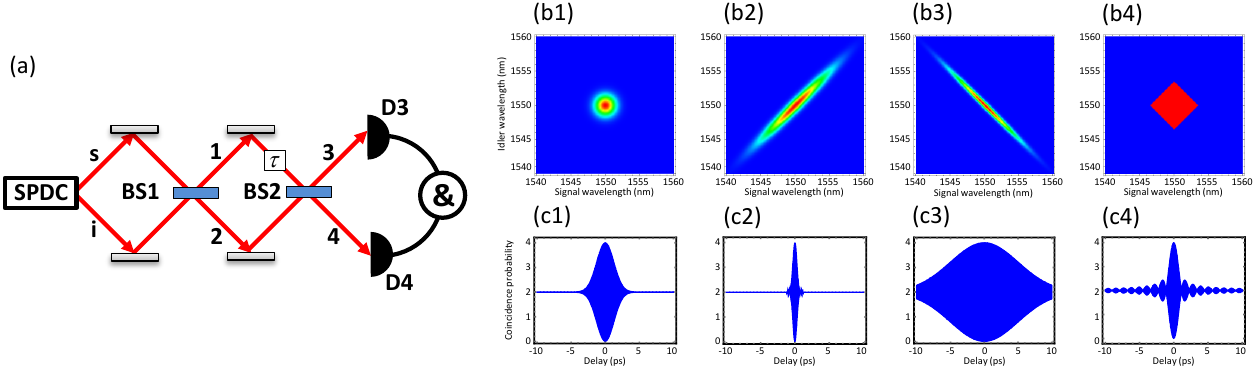}
\caption{(a) Setup of N00N state interference. (b1-c4) The joint spectral amplitudes of the N00N-state and the corresponding interference patterns.
}
\label{N00Narbitrary}
\end{figure}

In this section, we consider the multi-frequency model with arbitrary joint spectral amplitudes, with the setup shown in Fig.\,\ref{N00Narbitrary}(a) \cite{Jin2018Optica}. Similar to the notation in the previous section, the two-photon state from a SPDC process can be described as $\left| \psi  \right\rangle  = \int_0^\infty  {\int_0^\infty  {d\omega _s d\omega _i } } f(\omega _s ,\omega _i )\hat a_s^\dag  (\omega _s )\hat a_i^\dag  (\omega _i )\left| {00} \right\rangle.$ The field operators of detector 3 (D3) and detector 4 (D4) are $\hat E_3^{( + )} (t_3 ) = \frac{1}{{\sqrt {2\pi } }}\int_0^\infty  {d\omega _3 } \hat a_3 (\omega _3 )e^{ - i\omega _3 t_3 } $ and $\hat E_4^{( + )} (t_4) = \frac{1}{{\sqrt {2\pi } }}\int_0^\infty  {d\omega _4 \hat a_4 (\omega _4 )} e^{ - i\omega _4 t_4 }$, respectively. After BS1 and BS2, we can rewrite the detection field operators as
\begin{equation}\label{Eq:2-3-2}
\begin{aligned}
\hat E_3^{( + )} (t_3 ) &= \frac{1}{{2\sqrt {2\pi } }}\int_0^\infty  d \omega _3 [\hat a_s (\omega _3 )(e^{ - i\omega _3 \tau }  + 1)e^{ - i\omega _3 t_3 }  + \hat a_i (\omega _3 )(e^{ - i\omega _3 \tau }  - 1)e^{ - i\omega _3 t_3 } ],\\
\hat E_4^{( + )} (t_4 ) &= \frac{1}{{2\sqrt {2\pi } }}\int_0^\infty  d \omega _4 [\hat a_s (\omega _4 )(e^{ - i\omega _4 \tau }  - 1)e^{ - i\omega _4 t_4 }  + \hat a_i (\omega _4 )(e^{ - i\omega _4 \tau }  + 1)e^{ - i\omega _4 t_4 } ].
 \end{aligned}
 \end{equation}
The probability of coincidence detection $P(\tau )$ can be expressed as
\begin{equation}\label{Eq:2-3-4}
\begin{aligned}
  P(\tau )  =& \int {\int {dt_3 dt_4 } } \left\langle {\psi \left| {\hat E_3^{( - )} \hat E_4^{( - )} \hat E_4^{( + )} \hat E_3^{( + )} } \right|\psi } \right\rangle\\
   =& \frac{1}{{16}}\int_0^\infty  {\int_0^\infty  {d\omega _3 } d\omega _4 } \left| {[f(\omega _3 ,\omega _4 )(e^{ - i\omega _3 \tau }  + 1)(e^{ - i\omega _4 \tau }  + 1) + f(\omega _4 ,\omega _3 )(e^{ - i\omega _3 \tau }  - 1)(e^{ - i\omega _4 \tau }  - 1)]} \right|^2. 
\end{aligned}   
\end{equation}
Using the above equation, we can simulate the N00N state interference with joint spectral amplitudes featuring arbitrary distributions, as illustrated in Fig.\,\ref{N00Narbitrary}(b1-c4) \cite{Jin2018Optica}. Assuming $f$ is symmetric and normalized, $ P(\tau )$ can be further simplified as 
\begin{equation}\label{Eq:2-3-12}
P(\tau ) = \frac{1}{2}[1 + \int_0^\infty  {\int_0^\infty  {d\omega _s } d\omega _i } {\rm{|}}f(\omega _s ,\omega _i ){\rm{|}}^2 {\rm{cos(}}\omega _s {\rm{ + }}\omega _i {\rm{)}}\tau].
\end{equation}
This is the simplified equation for the N00N state interference. Using this equation, it is also possible to demonstrate the spectrally resolved N00N state interference \cite{Jin2021arXiv}.

Now, we use the parameters $\omega _+= (\omega _s  + \omega _i ) $ and
$\omega _- = (\omega _s  - \omega _i ) $, so $\omega _s = \frac{1}{2} (\omega _+ + \omega _-)$ and $\omega _i = \frac{1}{2} (\omega _+ - \omega _-)$. $P(\tau)$ can be rewritten as
\begin{equation}\label{Eq:2-3-13}
 P(\tau)
  = \frac{1}{2}[1 + \frac{1}{2}\int_0^\infty \int_{-\infty}^\infty  {d\omega _+ } d\omega _- |f(\omega _s ,\omega _i ) |^2  \cos(  \omega _+ \tau)]
  =\frac{1}{2}[1 + \int_{0}^\infty d\omega _+ F_2(\omega _+)  \cos(\omega _+ \tau)], \\
\end{equation}
where $F_2(\omega _+) \equiv  \frac{1}{2} \int_{-\infty}^\infty  d\omega _-  {\rm{|}} f(\omega _s ,\omega _i ) {\rm{|}}^{\rm{2}}$ is the projection of ${\rm{|}} f(\omega _s ,\omega _i ) {\rm{|}}^{\rm{2}} $ on to the anti-diagonal axis. Therefore,
\begin{equation}\label{Eq:2-3-15}
P(\tau )  =   \frac{1}{2}[1 +  \int_{0}^\infty d\omega _+ F_2(\omega _+)  \cos( \omega _+ \tau)]. \\
\end{equation}
Omitting the constant component (``direct current'' component) and the coefficients, we can define the second-order correlation function $G_2(\tau)$ in the N00N-state interference
\begin{equation}\label{Eq:2-3-16}
G_2(\tau)  \equiv \int_{0}^\infty d\omega _+ F_2(\omega _+)  e^{-i \omega _+ \tau}, \\
\end{equation}
where $ P_{2}(\tau)= \frac{1}{2}[1 +  \rm{Re} \{G_2(\tau)\}]$.
The inverse Fourier transform of $G_2(\tau)$ is
\begin{equation}\label{Eq:2-3-17}
F_2(\omega _+) = \frac{1}{2\pi} \int_{0}^\infty d\tau G_2(\tau) e^{i \omega _+ \tau}. \\
\end{equation}
This is the quantum version of the Wiener-Khinchin theorem (WKT), which will be introduced in detail in the next section.

\subsubsection{The quantum Wiener-Khinchin theorem}
%
%
%
\begin{figure}[htbp]
\centering
\includegraphics[width= 0.85\textwidth]{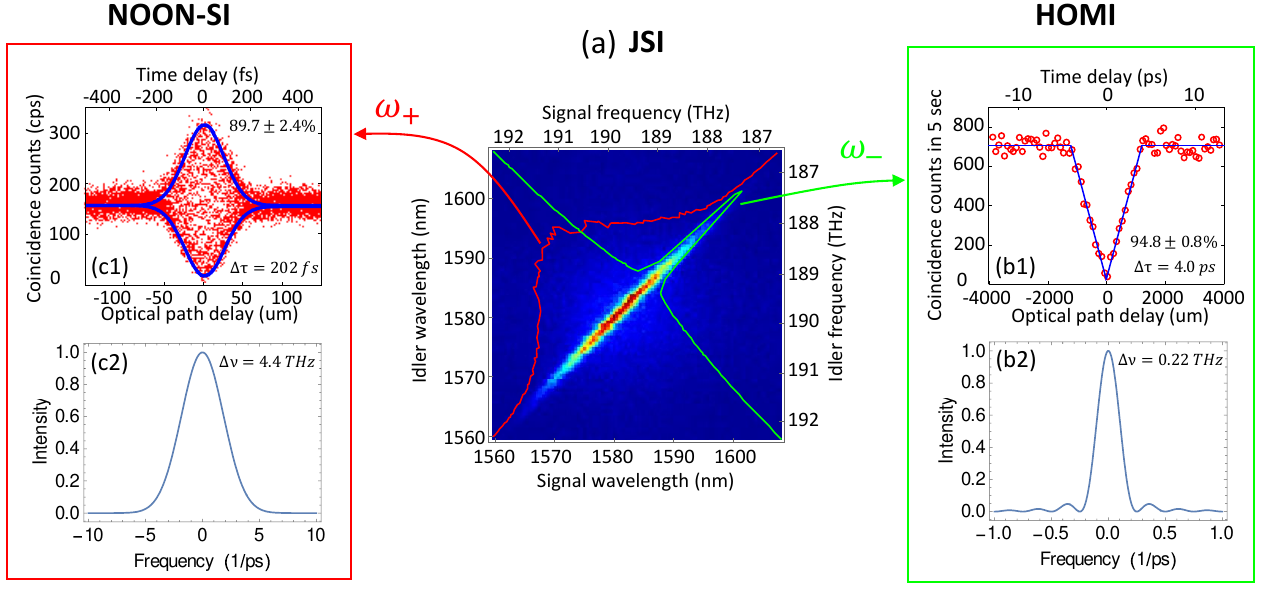}
\caption{(a) The measured JSI of the biphoton: Red curve: The projection of JSI on the $\omega_+$ direction. Green curve: The projection of JSI on the $\omega_-$ direction. (b1-b2) The HOM interference pattern and its Fourier transformation.
(c1-c2) The N00N-state interference pattern and its Fourier transformation. Reprinted from Ref.\,\cite{Jin2018Optica}.
}
\label{qwkt}
\end{figure}
%
%
%
The  WKT is an expression that relates the power spectrum to the autocorrelation function through Fourier transformation. In classical optics, the WKT is a fundamental theorem and serves as the basis for Fourier transform infrared spectroscopy, which finds commercial applications in chemical analysis, polymer testing, and pharmaceutical analysis \cite{Griffiths2007book}. The WKT is constructed based on classical interferometers, such as the Michelson interferometer or Mach-Zehnder interferometer.
In addition to the classical version, a quantum version of the WKT can be constructed. This quantum version connects correlated biphoton spectral information using quantum interferometers, such as the HOM interferometer and N00N state interferometer \cite{Jin2018Optica}.

The input state is $\left| \psi_2  \right\rangle  = \int_{-\infty}^\infty  {\int_{-\infty}^\infty  {d\omega _s d\omega _i } } f_2(\omega _s ,\omega _i )\hat a_s^\dag  (\omega _s )\hat a_i^\dag  (\omega _i )\left| {00} \right\rangle$,
where $f_2(\omega _s ,\omega _i )$ is the two-photon spectral amplitude. The corresponding two-photon detection probability $P_2^{\pm}(\tau)$ is given by
\begin{equation}\label{Eq:2-4-2}
 P_2^{\pm}(\tau)
  = \frac{1}{2}[1 \pm \int_{-\infty}^\infty \int_{-\infty}^\infty  {d\omega _s } d\omega _i {\rm{|}} f_2(\omega _s ,\omega _i ) {\rm{|}}^2 \cos(\omega _s  \pm \omega _i )\tau],
\end{equation}
where   $P_2^{+}$ is for N00N-state interference, while  $P_2^{-}$ is for HOM interference.

After the transformation of variables: $\omega _{\pm}= \omega _s  \pm \omega _i $,
\begin{equation}\label{Eq:2-4-3}
 P_2^{\pm}(\tau)
  = \frac{1}{2}[1 \pm \frac{1}{2}\int_{-\infty}^\infty \int_{-\infty}^\infty  {d\omega _+ } d\omega _- {\rm{|}} f_2(\omega _s ,\omega _i ) {\rm{|}}^{\rm{2}} \cos(\omega _ \pm  \tau)]. 
\end{equation}
Using the definition of the sum- or difference-frequency spectrum of the two-photon state, $F_2^{\pm}(\omega _ \pm) \equiv \frac{1}{2}\int_{-\infty}^\infty d\omega _\mp | f_2(\omega _s ,\omega _i ) |^2$, 
$P_2^{\pm}(\tau)$ can be further simplified as
\begin{equation}\label{Eq:2-4-5}
 P_2^{\pm}(\tau)
  = \frac{1}{2}[1 \pm \int_{-\infty}^\infty d\omega_{\pm} F_2^{\pm}(\omega _ \pm)  \cos(\omega _ \pm  \tau)]. 
\end{equation}

Omitting the constant component (``direct current'' component) and the coefficients, we can define the second-order correlation function $G_2^{\pm}(\tau)$ as
\begin{equation}\label{Eq:2-4-6}
G_2^{\pm}(\tau)  \equiv  \int_{-\infty}^\infty d\omega_{\pm} F_2(\omega _{\pm})  e^{-i \omega _{\pm} \tau}. \\
\end{equation}
The connection between $P_{2}^{\pm}$ and $G_2^{\pm}$ is,
\begin{equation}\label{Eq:2-4-7}
 P_{2}^{\pm}(\tau)= \frac{1}{2}[1 \pm \rm{Re} \{G_2^{\pm}(\tau)\}],
\end{equation}
where $G_2^{+}$ is for N00N-state interference, while  $G_2^{-}$ is for HOM interference.
The inverse Fourier transform of $G_2^{\pm}(\tau)$ is
\begin{equation}\label{Eq:2-4-8}
F_2^{\pm}(\omega _{\pm}) = \frac{1}{2\pi} \int_{-\infty}^\infty d\tau G_2^{\pm}(\tau) e^{i \omega _{\pm} \tau}. \\
\end{equation}
This is the unified form of quantum WKT.

According to the quantum WKT in Eq.\,(\ref{Eq:2-4-8}), the difference-frequency distribution of the biphoton wavefunctions can be extracted by applying a Fourier transform on the time-domain HOM interference patterns, while the sum-frequency distribution can be obtained by performing a Fourier transform on the time-domain N00N state interference patterns, as depicted in Fig.\,\ref{qwkt} \cite{Jin2018Optica}. Alternatively, both difference-frequency and sum-frequency distributions can be obtained simultaneously by performing a Fourier transform on the time-domain combination interference patterns \cite{Li2023PRA}.
Additionally, another form of the quantum WKT for continuous-wave (CW) laser pumped HOM interferometer can be found in Ref.\,\cite{Chen2022PRAppl-2}.

\subsubsection{Multi-mode HOM interferometer and N00N state interferometer}
%
%
%
\begin{figure}[htbp]
\centering
\includegraphics[width= 0.85\textwidth]{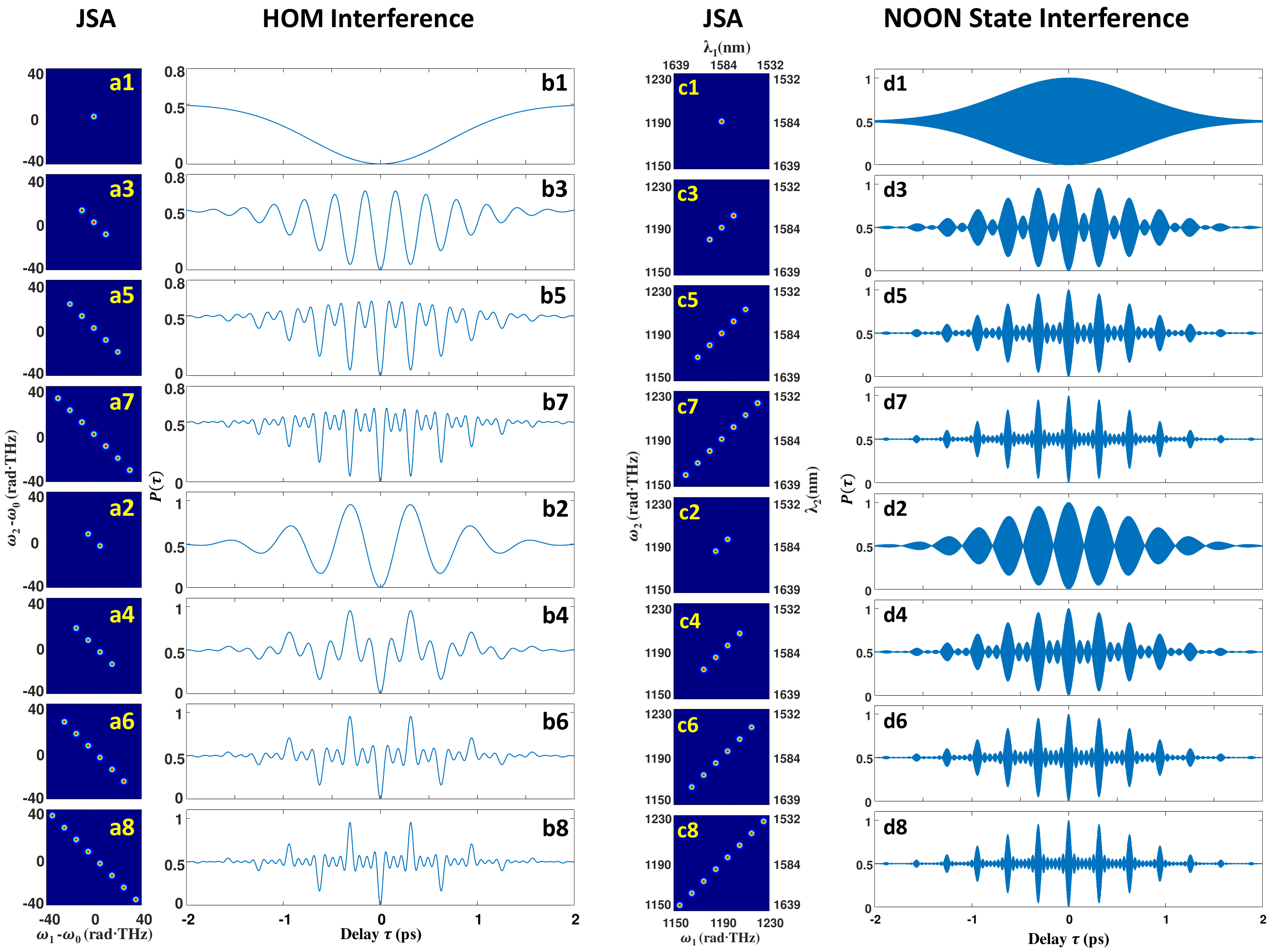}
\caption{(a1-a8) and (b1-b8) are the simulated JSAs and multi-mode HOM interference patterns. (c1-c8) and (d1-d8) are the simulated JSAs and multi-mode N00N state interference patterns. Reprinted from Ref.\,\cite{Guo2023OE}.
}
\label{MMHOMNOONI}
\end{figure}
%
%
%
In traditional HOM interferometer and N00N state interferometer, the biphotons are usually correlated in a single discrete spectral mode. However, the biphotons can be correlated in multiple discrete spectral modes, and this two-body high-dimensional entangled state can be referred to as entangled qudits \cite{Cozzolino2019AQT, Erhard2020NRP, Yang2023OL}. The multi-mode HOM interferometer or multi-mode N00N state interferometer can be defined as HOM or N00N state interferometer using frequency entangled qudits.
Figure \ref{MMHOMNOONI}(a1-a8) shows the simulated joint spectral amplitudes (JSAs), which represent entangled qudits (for mode numbers greater than 3). Figure \ref{MMHOMNOONI}(b1-b8) displays the corresponding multi-mode HOM interference patterns \cite{Guo2023OE}.
One important characteristic of multi-mode HOMI is that its interference patterns are significantly narrower compared to those in single-mode HOM interference. These narrow interference fringes provide increased Fisher information for phase estimation \cite{Xiang2013SR, Lyons2018SA, Chen2019npj, Guo2023OE}.
Figures \ref{MMHOMNOONI}(c1-c8) and (d1-d8) illustrate the simulated JSAs and the corresponding multi-mode N00N  state interference patterns . Both  multi-mode HOM interference and multi-mode N00N  state interference are determined by two factors: the envelope factor and the details factor. The envelope factor is contributed by the single spectral mode, while the details factor is determined by the mode number $N$ \cite{Guo2023OE}.

\subsection{Generation of N00N states from different systems}
\begin{figure}[htbp]
\centering
\includegraphics[width= 0.92\textwidth]{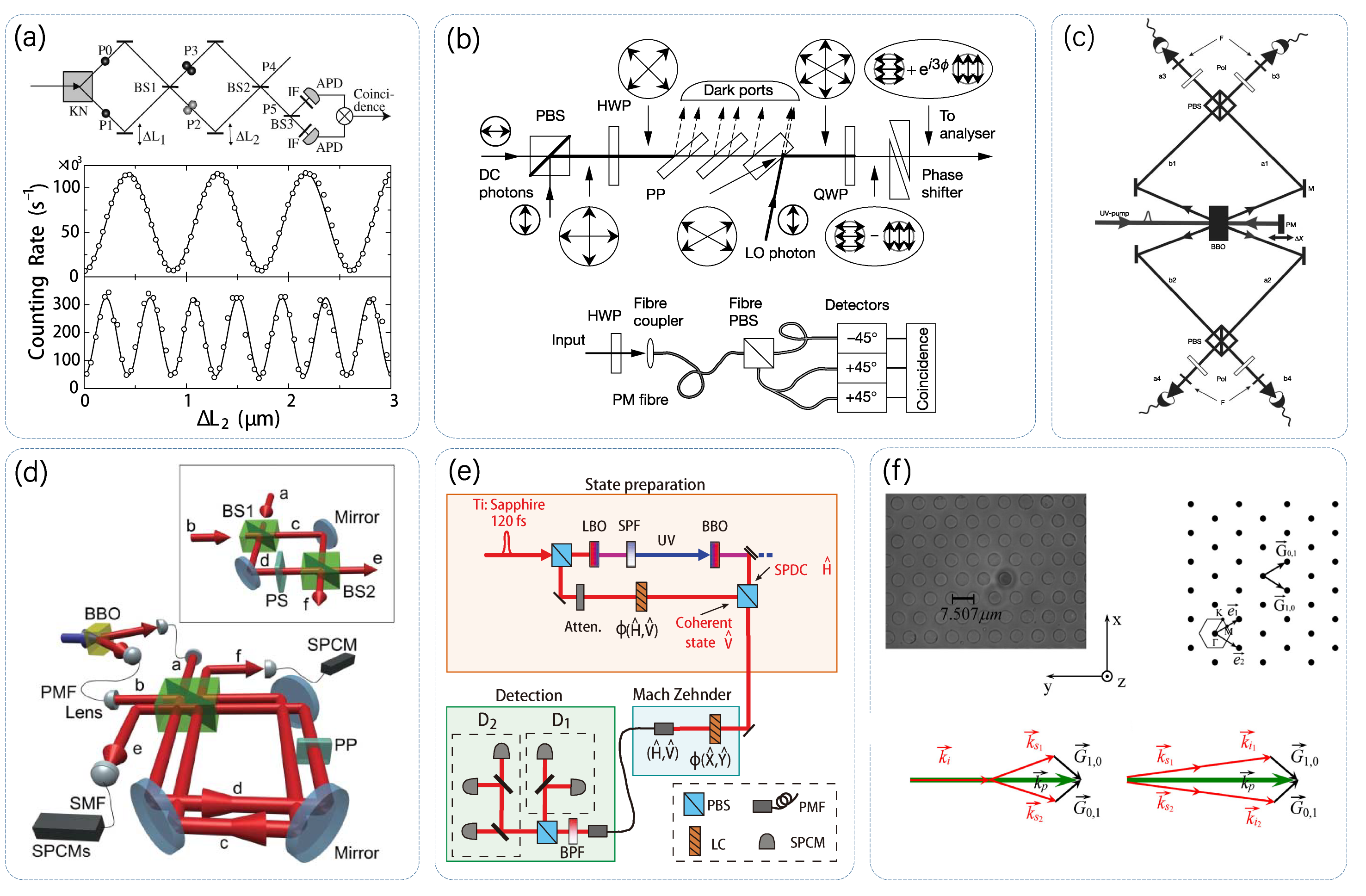}
\caption{The generation of N00N states from different setups. (a) The experimental generation of the two-photon N00N state (upper) and interference patterns in the one photon and biphotons (lower). Reprinted from Ref.\,\cite{Edamatsu2002PRL}. (b) Experimental generation of the three-photon N00N state. Reprinted from Ref.\,\cite{Mitchell2004Nature}. (c, d) The experimental generation of the four-photon N00N state. Reprinted from Refs. \cite{Walther2004, Nagata2007Science}. (e) The experimental generation of the five-photon N00N state by mixing a weak coherent light source and an SPDC source. Reprinted from Ref.\,\cite{Afek2010Science}. (f) Experimental generation of the two-photon N00N state from a hexagonally poled lithium tantalate. Reprinted from Ref.\,\cite{Jin2014PRL}. }
\label{N00Ngeneration}
\end{figure}

The theoretical framework for the N00N state interferometer was initially proposed by Boto et al. in 2000 \cite{Boto2000PRL}. In this work, the authors demonstrated that by using the entangled N00N state instead of the classical coherent state, it is possible to surpass the classical diffraction limit in optical lithography.
The term ``N00N state" first appeared in print as a footnote in Ref.\,\cite{Lee2002JMO} in 2002. Following this pioneering work in Ref.\,\cite{Boto2000PRL}, numerous theoretical and experimental works have been dedicated to the implementation and characterization of N00N state interferometers. On the theoretical side, Giovannetti et al. proposed the generation of the N00N state by combing parametric down-conversion and HOM interference in 2002 \cite{Giovannetti2002PRL}. Kok et al. proposed the general method based on post-selection via photodetection in 2002 \cite{Kok2002PRA}. Pryde and White subsequently introduced a method using intensity-symmetric multiport beam splitters, single photon inputs, and either heralded or conditional measurement \cite{Pryde2003PRA}. Cable and Dowling proposed an efficient method using only linear optics and feed-forward \cite{Cable2007PRL}. In 2008, Dowling reviewed the theoretical and experimental advances of quantum optical metrology using N00N state \cite{Dowling2008CP}.

On the experimental side, researchers have explored different physical platforms, including photonic systems (bulk nonlinear crystals, waveguides, integrated chips), quantum dots, plasmons, trapped ions, and atoms (Bose-Einstein condensates), superconducting transmons, and optomechanics, to generate and manipulate the N00N state for interferometric measurements, as listed in Tab.\,\ref{Tab:N00N1}.
\begin{table}[h]
\caption{Generation of N00N states from different systems}
\label{Tab:N00N1}
\begin{tabular}{l|l|l}
\hline \hline
 Physical system         & Process                  &  References                                        \\ \hline   
 Bulk crystal           & SPDC                     & \begin{tabular}[c]{@{}l@{}} \cite{Edamatsu2002PRL,Walther2004,Mitchell2004Nature,Eisenberg2005PRL,Sun2006PRA,Nagata2007Science,Sciarrino2008PRA,Vitelli2009JOSAB,Kim2009OE,Afek2010Science,Yabuno2012PRA,Kim2015SR,Ra2015OE,Ming2015IEEE,Zhou2017PRAppl,Liu2018CPL,Domenico2022OE} \end{tabular} \\  \hline 
 Waveguide, chip      & SPDC or SFWM                      & \begin{tabular}[c]{@{}l@{}} \cite{Longhi2011PRA,Jin2014PRL,Kruse2015PRA,Mohanty2017NC,Eguibar2019AnnPhys, Pryde2003PRA,Jin2013PRL,Silverstone2013NP,Preble2015PRAppl,Feng2016NC,Georgi2019LSA,Hernandez2022LPR}    \end{tabular} \\    \hline                                 
 Quantaum dots           & Cascaded  emission, Quantum dot-cavity coupling                 & \begin{tabular}[c]{@{}l@{}}    \cite{Mueller2017PRL,Kamide2017PRA} \end{tabular}  \\  \hline                  
Plasmon                 & Coherent light matter interaction & \begin{tabular}[c]{@{}l@{}}  \cite{Vest2018NJP,Chen2018Optica,Li2015NL}    \end{tabular}  \\  \hline 
Trapped-ion             & Coupling between vibrational states and internal states                  & \begin{tabular}[c]{@{}l@{}}  \cite{Ivanov2013NJP,Zhang2018PRL}   \end{tabular}  \\  \hline 
Atoms                   & Atom-cavity coupling, quantum walk in optical lattice                    & \begin{tabular}[c]{@{}l@{}}  \cite{Nikoghosyan2012PRL,Liu2013QIP,Liu2014QIP,Li2016PRL,Opanchuk2016PRA,Song2016QIP,Compagno2017PRA,Bychek2018PRA,Grun2022CP} \end{tabular} \\  \hline 
Circuit or waveguide QED & Transmon-microwave resonator coupling                      & \begin{tabular}[c]{@{}l@{}}  \cite{Su2014SR,Chen2017PRA,Kannan2020SA} \end{tabular}  \\  \hline 
 Optomechanics           & Photon-phonon entanglement, optomechanical ultrastrong coupling                     & \begin{tabular}[c]{@{}l@{}} \cite{Ren2013PRA,Macri2016PRA} \end{tabular}  \\     \hline                
 Atoms       &  Fock-state Bose-Einstein condensates      &\begin{tabular}[c]{@{}l@{}}    \cite{Cable2011PRA,Cable2011PRA,Vanhaele2021PRA}   \end{tabular} \\ 
\hline \hline      
\end{tabular}
\end{table}

Photonic systems are the most widely used platform for the demonstration of N00N-SI.
Figure\,\ref{N00Ngeneration} show several noteworthy experiments for the generation of the N00N state and demonstrates its application in quantum metrology.
In 2002, Edamatsu et al. demonstrated 2-photon N00N state interference patterns using entangled photon pairs generated by spontaneous parametric down-conversion, as shown in Fig.\,\ref{N00Ngeneration}(a)  \cite{Edamatsu2002PRL}. This experiment can also be interpreted as the measurement of the photonic de Broglie wavelength. In 2004, Mitchell et al. demonstrated 3-photon N00N state interference using linear optical elements and post-selection, as shown in Fig.\,\ref{N00Ngeneration}(b) \cite{Mitchell2004Nature}. In the same year, Walther et al. presented the 4-photon N00N state interferometer with the setup shown in Fig.\,\ref{N00Ngeneration}(c) \cite{Walther2004}. The result confirms the theoretical expectation that the de Broglie wavelength of a four-photon state is one-quarter that of a single photon. In 2006, Xiang et al. used the N00N-state projection measurement technique to quantitatively characterize the temporal distinguishability of a four- or six-photon state \cite{Xiang2006PRL}. In 2007, Natata et al. demonstrated an optical phase measurement with a 4 photon N00N state and the visibility was greater than the threshold to beat the standard quantum limit, where the setup is shown in Fig.\,\ref{N00Ngeneration}(d) \cite{Nagata2007Science}. In 2010, Afek et al. generated ``high-N00N" states (N = 5) by multiphoton interference of quantum down-converted light with a classical coherent state, which is shown in Fig.\,\ref{N00Ngeneration}(e) \cite{Afek2010Science}.
Integrated waveguides are also a good platform for the generation of N00N states. As shown in Fig.\,\ref{N00Ngeneration}(f), Jin et al. reported the compact engineering of steerable photonic path-entangled states from a monolithic quadratic nonlinear photonic crystal \cite{Jin2013PRL}.

There are many other investigations on the characteristics of the N00N state interferometers as listed in Tab.\,\ref{Tab:N00N2}.
\begin{table}[h]
\caption{Studies on the characteristics of the N00N state interferometer}
\label{Tab:N00N2}
\begin{tabular}{l|l}
\hline \hline
Characteristic of N00N-SI                      & References                                  \\ \hline   
 Nonlinear behavior of geometric phases         &  \cite{Kobayashi2011PRA}    \\
 Role of pump coherence for                       &  \cite{Liang2011PRA}      \\                         
 Wigner function of   N00N state                &  \cite{Xu2013APS}      \\
Entanglement concentration of less-entangled N00N state    &  \cite{Zhou2012QIP}      \\
 Entanglement and phase properties of noisy N00N state      &   \cite{Bohmann2015PRA}    \\   
 Nonlocality of N00N state                     &  \cite{Teh2016PRA}  \\ 
 Spectrally resolved N00N-SI                    &  \cite{Jin2021arXiv,Triggiani2023PRAppl} \\
 N00N-SI in frequency domain& \cite{Lee2024LSA}
\\                                                                                                     \hline \hline      
\end{tabular}
\end{table}

\subsection{Holland-Burnett state: The experimental approximation of N00N state }
\begin{figure}[htbp]
\centering
\includegraphics[width= 0.92\textwidth]{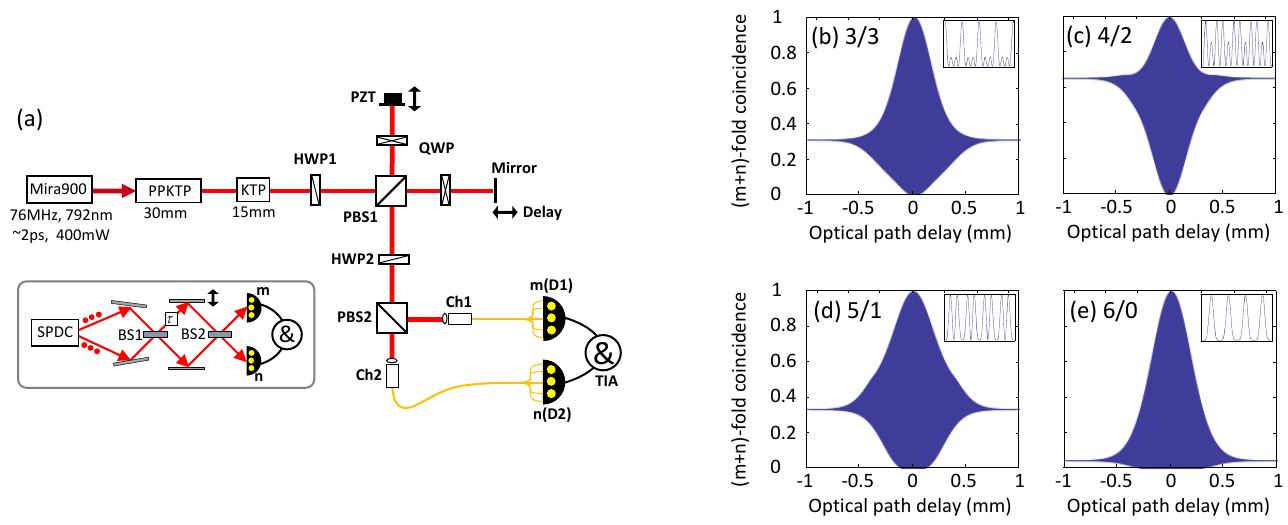}
\caption{Experimental testing of the detection-dependent properties of the HB state. (a) The experimental setup, where the inset depicts a standard configuration of the HB state interference using path-mode. (b-e) Numerical simulations of the HB interference patterns with different detection schemes. Insets are the fine interference patterns near the zero delay position. Reprinted from Ref.\,\cite{Jin2016SR}. }
\label{N00NHBstate}
\end{figure}

The ideal N00N state is challenging to prepare in experiments, therefore, the Holland-Burnett (HB) state is traditionally used to approximate the N00N state \cite{Holland1993PRL,Xiang2013SR}.
The HB state can be easily prepared in experiments using photon pairs from SPDC. By passing a $\left| N/2, N/2  \right\rangle$ initial state through a BS and introducing a phase shift of $\phi$,
the HB state can be generated in the form of
$\left| \psi  \right\rangle  = \sum\limits_{n = 0}^{N/2} {c_n } \left| {2n,N - 2n} \right\rangle$, with
$c_n  = \frac{{\sqrt {(2n)!(N - 2n)!} }}{{2^{N/2} n!(N/2 - n)!}}e^{i2n\phi }$ \cite{Xiang2013SR}. The HB state approximates the N00N state such that they can attain the Heisenberg limit for phase estimation, but the photon number exhibits a quadratic variance in $N$ \cite{Braunstein1994PRL}. The HB state has been shown to be almost optimally robust to imperfect state preparation and detection, and losses \cite{Datta2011PRA}.

The previous HB state interferometer (HBSI) schemes mainly considered period-based applications and only measured the single-mode interference pattern, which comprises solely the portion of the interference patterns around zero optical path length difference. As a result, an analysis of the full-range properties of the HBSI, e.g., the overall envelope shape and the coherence time, was omitted.
Therefore, although detailed in their analysis of the single-mode interference patterns, the previous experiments are not enough to fully characterize the HBSI, especially for higher photon numbers ($N$$>$2).
As shown in Fig.\,\ref{N00NHBstate}, Jin et al. considered the properties of the HBSI pattern over a wide range of optical path length differences and experimentally measured HBSI patterns up to six photons \cite{Jin2016SR}.
It was found that the shape, the coherence time and the visibility of the interference patterns strongly depend on the detection schemes. This experiment can be used for applications that are based on the envelope of the HBSI pattern, such as quantum spectroscopy and quantum metrology.

\subsection{Applications of N00N state interferometer}
\subsubsection{Precision phase measurement}
\begin{figure}[htbp]
\centering
\includegraphics[width= 0.92\textwidth]{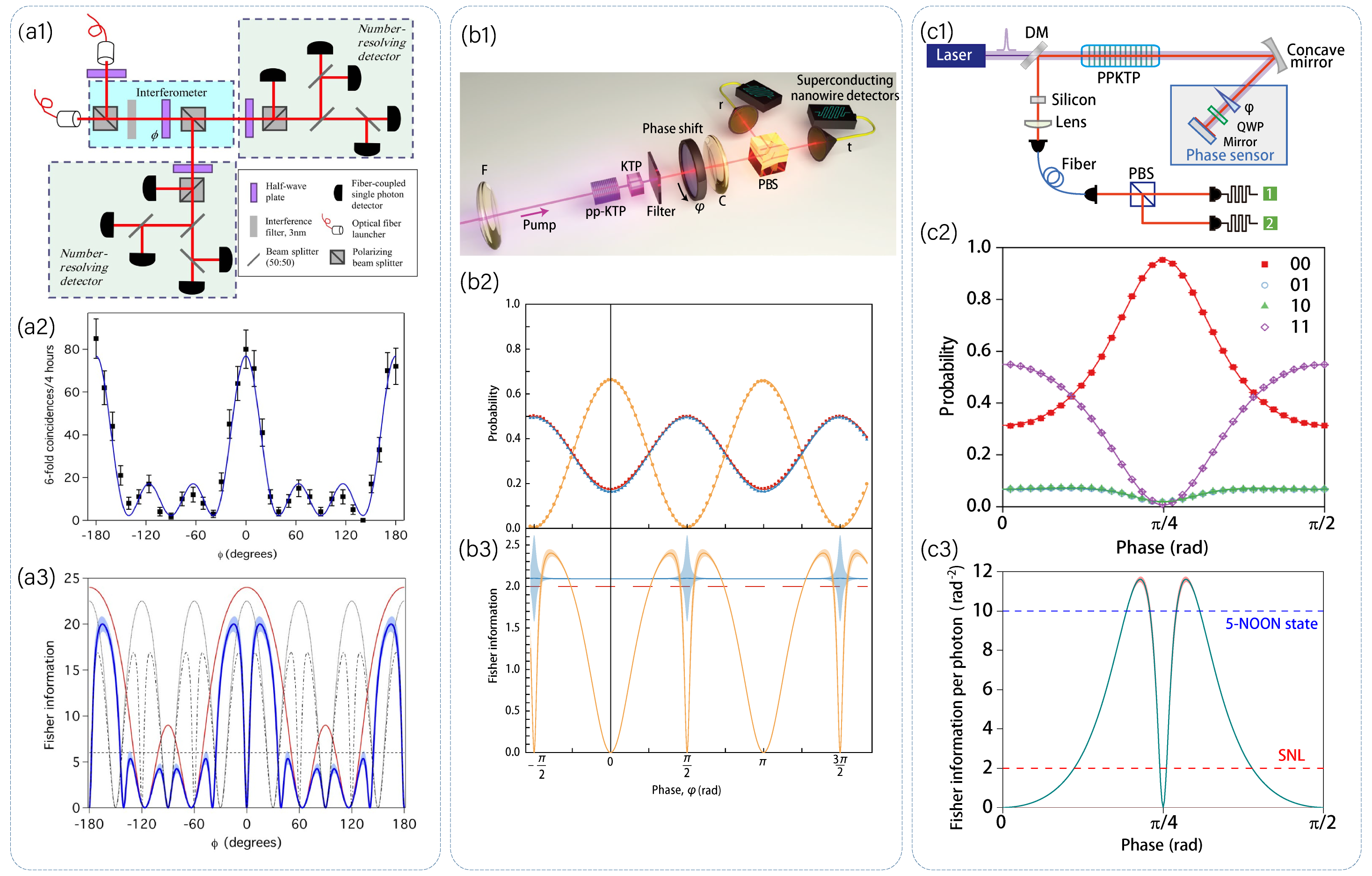}
\caption{Applications of N00N-SI in high precision phase measurement. (a1-a3) Experimental setup, interference patterns, and Fisher information of the optimal multi-photon phase sensing. Reprinted from Ref.\,\cite{Xiang2013SR}.  (b1-b3) The experimental setup for the N=2 N00N state interferometer, the output detection probability measured experimentally and the corresponding Fisher information. Reprinted from Ref.\,\cite{Slussarenko2017NP}. (c1-c3) The experimental setup for the N=2 N00N state interferometer; Experimentally measured output detection probability and the corresponding Fisher information. Reprinted from Ref.\,\cite{Qin2023PRL}. }
\label{N00NphaseM}
\end{figure}
Precision measurements are crucial across all scientific fields. Optical phase measurements, in particular, are utilized to measure distance, position, displacement, acceleration, and optical path length. Quantum entanglement allows for higher precision beyond what would otherwise be unachievable.
In particular, N00N-SI can help increase the precision of optical phase measurements beyond the shot-noise limit (SNL) and reach the ultimate Heisenberg limit.
Numerous theoretical and experimental works have demonstrated the applications of N00N-SI in high-precision phase measurements \cite{Giovannetti2004Science,Kawabe2007OE, Rubin2007PRA,Sun2008EPL, Okamoto2008NJP, Berry2009PRA, Kim2011OE, Jin2013SR,Slussarenko2017NP, Hong2021NC,Qin2023PRL,You2021APL,Zhou2018PRL,Li2022LSA}.
Xiang et al. showed that the 6-photon HB state is more optimal than the 6-photon N00N state to realize enhanced sensitivity \cite{Xiang2013SR}. Figure\,\ref{N00NphaseM}(a1) shows the corresponding experimental setup. The input state is $\left| 33 \right\rangle$, which is generated from an SPDC source. The state is incident on a polarization-based interferometer and measured by photon number-resolved detectors, which are, in fact, single-photon counting module (SPCM) arrays. Figure\,\ref{N00NphaseM}(a2) is the interference pattern of the $\left| 33 \right\rangle$ state. Figure\,\ref{N00NphaseM}(a3) presents the calculated Fisher information. It is clear that the 6-photon HB state (solid blue) has a higher Fisher information than the 6-photon N00N state (dotted and dot-dash curve).

Although quantum-enhanced metrology holds the promise of improved sensitivity, no experiment using photonic quantum states has truly surpassed the SNL.
All previous experiments have considered only a subset of the photons used due to photon loss, detector inefficiency, or other imperfections.
In 2017, Slussarenko et al. performed unconditional entanglement-enhanced photonic interferometry using an ultra-high-efficiency photon source and detectors \cite{Slussarenko2017NP}. By sampling a birefringent phase shift, they demonstrated precision beyond the SNL without artificially correcting for loss and imperfections.
Figure\,\ref{N00NphaseM}(b) shows their experimental setup and measured interference patterns and the corresponding Fisher information. In Fig.\,\ref{N00NphaseM}(b3), the orange curve is the Fisher information determined from the probability fringes in Fig.\,\ref{N00NphaseM}(b2). The dashed red line is the Fisher information at the SNL, the solid blue line is the SNL when taking into account the inefficiency and multi-photon emission. Their results show a clear violation of the adjusted SNL bound that takes into account the information in unrecorded trials. 

While multiphoton-entangled N00N states can in principle beat the SNL and reach the Heisenberg limit, high N00N states are difficult to prepare and fragile to photon loss, which hinders them from reaching unconditional quantum metrological advantages. In 2023, Qin et al. combined the ideas of unconventional nonlinear interferometers and stimulated emission of squeezed light to demonstrate a new scheme that achieves a quantum metrological advantage \cite{Qin2023PRL}. They observe a 5.8(1)-fold enhancement above the SNL in the Fisher information extracted per photon, without discounting for photon loss and imperfections, which outperforms the ideal 5-N00N state. Figure\,\ref{N00NphaseM}(c1) shows their experimental setup for the stimulated emission of the two-mode squeezed state. Figure\,\ref{N00NphaseM}(c2-c3) shows the experimentally measured output detection probability and the Fisher information per photon as a function of the phase. The cyan curve corresponds to the experimental results, while the red lines correspond to the SNL.

\subsubsection{Fiber optical gyroscope}
\begin{figure}[htbp]
\centering
\includegraphics[width= 0.85\textwidth]{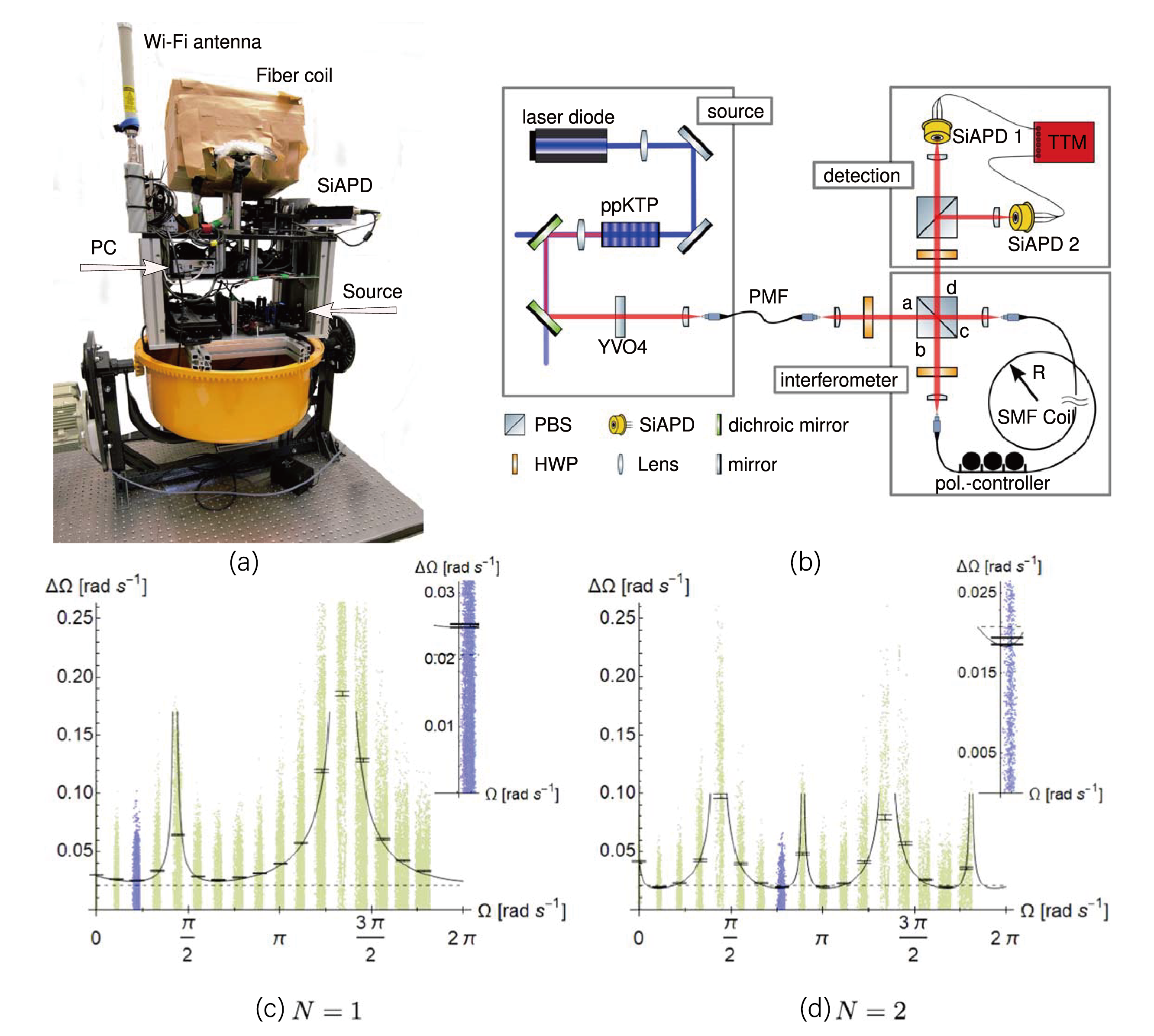}
\caption{(a-b) Experimental arrangement and optical setup of the entangled-photon based optical gyroscope. (c) The precision measurement of the rotational velocity, measured with the one photon state. (d) Precision measurement of the rotational velocity, measured with the two-photon state. Reprinted from Ref.\,\cite{Fink2019NJP}. }
\label{N00Ngyroscope}
\end{figure}

Fiber optic gyroscopes (FOG), which utilize the Sagnac interferometer, are highly valuable for sensing and navigation purposes, providing precise measurements across a wide range of applications \cite{Lefevre2022BOOK}. Similar to conventional optical sensors, the performance of these devices is ultimately limited by the SNL.
Quantum-enhanced interferometry provides a solution to surpass this limitation by utilizing nonclassical light states. In Ref.\,\cite{Fink2019NJP}, Fink et al. report on an entangled-photon gyroscope that employs two-photon N00N-state interferometers, enabling super-resolution and phase sensitivity that surpass the SNL. Figure\,\ref{N00Ngyroscope} shows the experimental setup and the measured results of the N00N-SI based FOG. Although quantum-enhanced FOG technology is not yet competitive with classical FOG, it still possesses distinct advantages of its own. Laser-driven FOG uses an optical power of approximately 20 $\mu$W , which corresponds to a rate of $156 \times 10^{12}$ photons per second (at $\lambda$ = 1550 nm). In contrast, the detected photon rate of the N00N state is $100\times10^3$ in this experiment. This relatively low photon rate was mainly limited by the detectors they used. This work is an important step towards reaching the utmost sensitivity limits in Sagnac interferometry using N00N states.

\subsubsection{Precision angular measurement}
\begin{figure}[htbp]
\centering
\includegraphics[width= 0.85\textwidth]{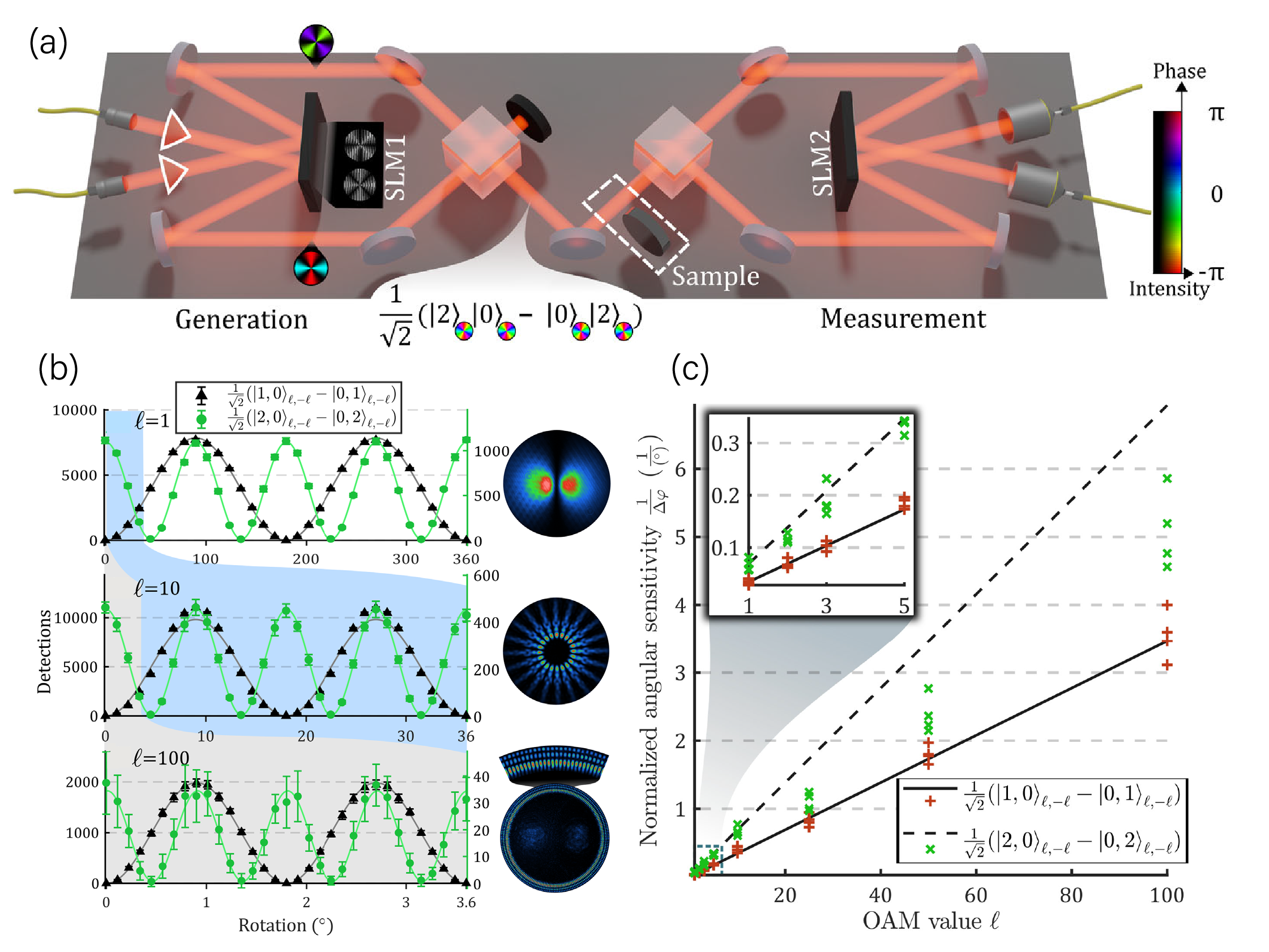}
\caption{Applications of N00N-SI in high-precision angular measurement. (a) The experiment setup. (b) The rotation measurements with l= {$\pm$ 1; $\pm$ 10; $\pm$ 100}. (c) Measurement sensitivities of single-photon and two-photon N00N states. Reprinted from Ref.\,\cite{Hiekkamaeki2021PRL}.}
\label{N00NangularM}
\end{figure}
The N00N state can increase the phase sensitivity of the photon not only in paths or polarization, but also in orbital angular momentum (OAM). In 2021, Hiekkam$\ddot{a}$ki et al. experimentally combined the phase sensitivity of the N00N state with the OAM of photons up to 100$\hbar$, to resolve rotations of a light field around its optical axis \cite{Hiekkamaeki2021PRL, Hiekkamaeki2022NP}. The experimental setup is shown in Fig.\,\ref{N00NangularM}(a).
Photon pairs generated from an SPDC source were guided into single-mode fibers and then impinged onto two separate regions of a spatial light modulator (SLM). Then the photons are structured using holographic phase and amplitude modulation. After that, the structured photons are combined on a BS to prepare the state of $\frac{1}{{\sqrt 2 }}({\left| {2,0} \right\rangle _{l, - l}} - {\left| {0,2} \right\rangle _{l, - l}})$. To measure the two-photon state, a second beam splitter probabilistically separates the photons, and a second SLM (SLM2) is used in conjunction with two single-mode fibers to filter the spatial structures of the photons independently. Figure\,\ref{N00NangularM}(b) shows the detected single photons and two-photon coincidences as a function of the rotation angle. The single photons were prepared in the modes shown in the insets, and the corresponding two-photon N00N states were created by imprinting the same structure on one photon and its orthogonal stricture on the other. Figure\,\ref{N00NangularM}(c) shows that the measured two-photon states are more sensitive than their one-photon counterparts, although the theoretical scaling was not reached with two-photon states with large OAM.

\subsubsection{Quantum microscope and quantum imaging }
\begin{figure}[htbp]
\centering
\includegraphics[width= 0.92\textwidth]{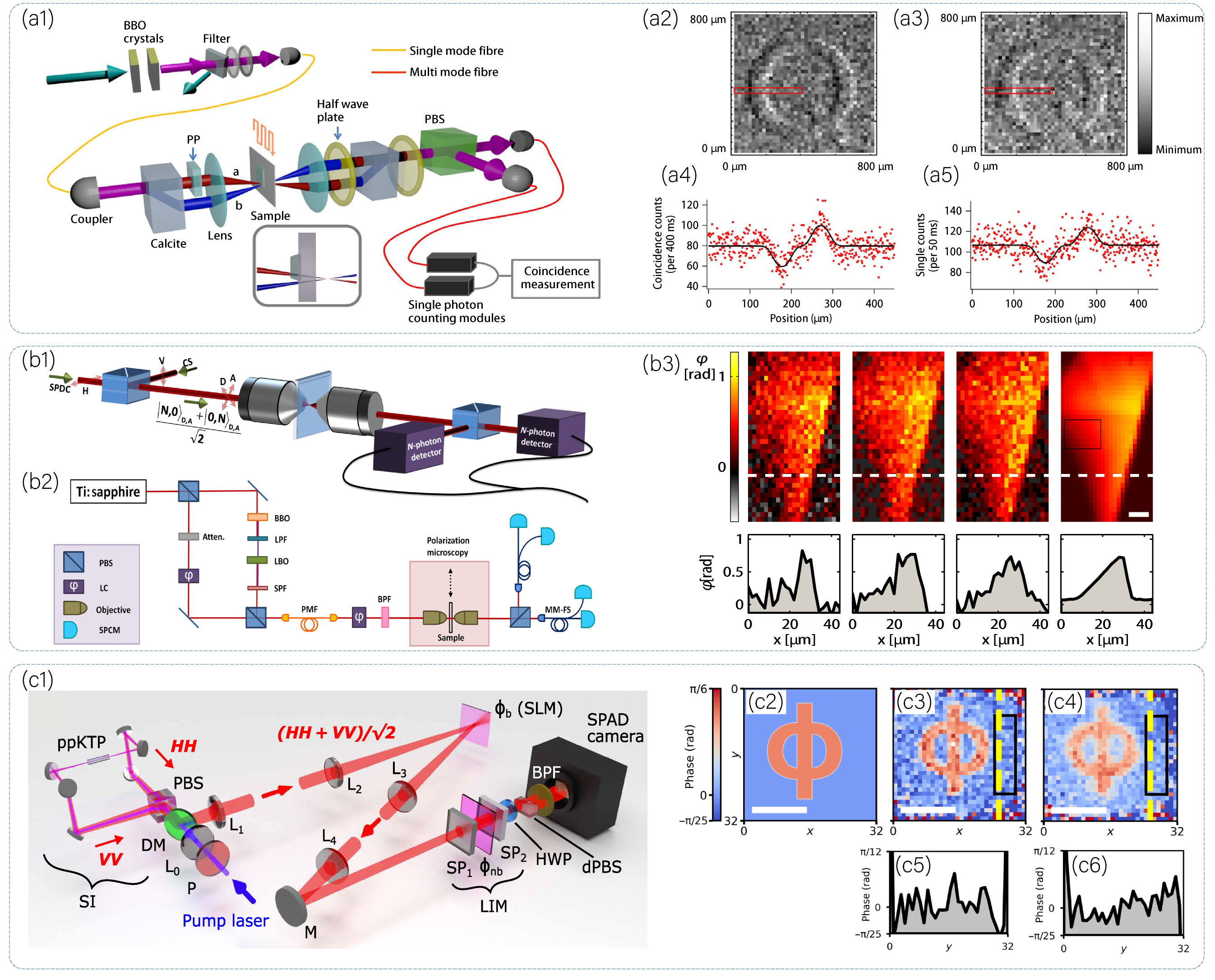}
\caption{Applications of N00N-SI in quantum microscope and quantum imaging. (a1-a5) The setup and results of a two-photon-N00N-state based microscope. Reprinted from Ref.\,\cite{Ono2013NC}. (b1-b3) The setup and results of  three-photon-N00N-state based polarization microscopy. Reprinted from Ref.\,\cite{Israel2014PRL}. (c1-c6) The setup and  the retrieved phase images of a birefringent sample. Reprinted from Ref.\,\cite{Camphausen2021SA}.  }
\label{N00Nimaging}
\end{figure}
An entanglement-enhanced microscope is a type of microscope that utilizes quantum entanglement to enhance its imaging capabilities. Traditional microscopes rely on classical light sources and detectors to capture images. In contrast, entanglement-enhanced microscopes leverage the quantum properties of light to achieve higher resolution and increased sensitivity. Entanglement-enhanced microscopes have the potential to revolutionize imaging and microscopy by providing the ability to detect faint signals that would be challenging to observe using classical microscopy techniques \cite{Bhusal2022npj}. 
In 2013, Ono et al. reported the demonstration of an entanglement-enhanced microscope, which is a confocal-type differential interference contrast microscope where an entangled photon pair source ($N=2$) is used for illumination \cite{Ono2013NC}. As shown in Fig.\,\ref{N00Nimaging}(a1), an image of a Q shape carved in relief on the glass surface is obtained with better visibility than with a classical light source. Specifically, Fig.\,\ref{N00Nimaging}(a2) is the image of the sample using an entanglement-enhanced microscope where two-photon entangled state is used to illuminate the sample. Figure\,\ref{N00Nimaging}(a3) is the image of the sample using a classical light source. 
Figure\,\ref{N00Nimaging}(a4) and (a5) are the one-dimensional fine scan data for
the area outlined in red in (a2) and (a3). The signal-to-noise ratio of the results with two-photon entangled state is 1.35$\pm$0.12 times better than that with the classical light source. 

In 2014, Israel et al. demonstrated a quantum polarized light microscope using entangled N00N states with $N=2$ and $N=3$ \cite{Israel2014PRL}. In this experiment, the N00N state is created by combining a coherent state with an SPDC source. They successfully captured images of birefringent objects under extremely low light conditions, with only 50 photons per pixel. In such challenging scenarios, classical imaging is greatly hindered by the shot-noise. Figure\,\ref{N00Nimaging}(b1-b2) are the conceptual and detailed layout of the setup. Figure\,\ref{N00Nimaging}(b3) shows the polarization microscopy images of a sample of a single quartz crystal. The upper four figures show the images using weak coherent light only (50 single photons), the N00N state with $N=2$ (25 pairs), the N00N state with $N=3$ (17 triples), and strong coherent light. The lower four figures show the phase distribution of the line scans in the upper four figures. We can clearly observe that strong coherent light exhibits the best signal-to-noise ratio. However, when considering the same intensity level of 50 photons, the N00N states outperform the weak coherent light.

Finally, N00N-SI can be used to enhance the signal-to-noise ratio not only in quantum microscopy but also in general optical imaging. Utilizing the latest advancements in single-photon avalanche diode array cameras and multiphoton detection techniques, Camphausen et al. have introduced a supersensitive phase imager \cite{Camphausen2021SA}. This innovative imager utilizes space-polarization hyperentanglement, allowing it to operate over a large field of view without the need for scanning operations. This technology can easily be adjusted to capture high-resolution images, making it an important step towards practical quantum-enhanced imaging. Figure\,\ref{N00Nimaging}(c1) show the corresponding experimental setup. The entangled photon pairs are imaged onto the SLM and then reimaged again into the lens-free interferometric microscope (LIM) and single-photon avalanche diode (SPAD) camera. Figure\,\ref{N00Nimaging}(c2-c4) are the phase profiles applied to SLM, the classical phase image, and the entanglement-enhanced phase image, respectively. Figure\,\ref{N00Nimaging}(c5) and (c6) shows the cross section of the phase profile along the yellow dashed line in (c3) and (c4). The sensitivity enhancement can be quantified by computing the local uncertainty of the images, which is calculated as 0.091 for the phase image in (c5) and 0.065 for the phase image in (c6). These experimental results confirmed the supersensitivity of the two-photon N00N-SI based imaging.

\subsubsection{Quantum spectroscopy}
\begin{figure}[htbp]
\centering
\includegraphics[width= 0.92\textwidth]{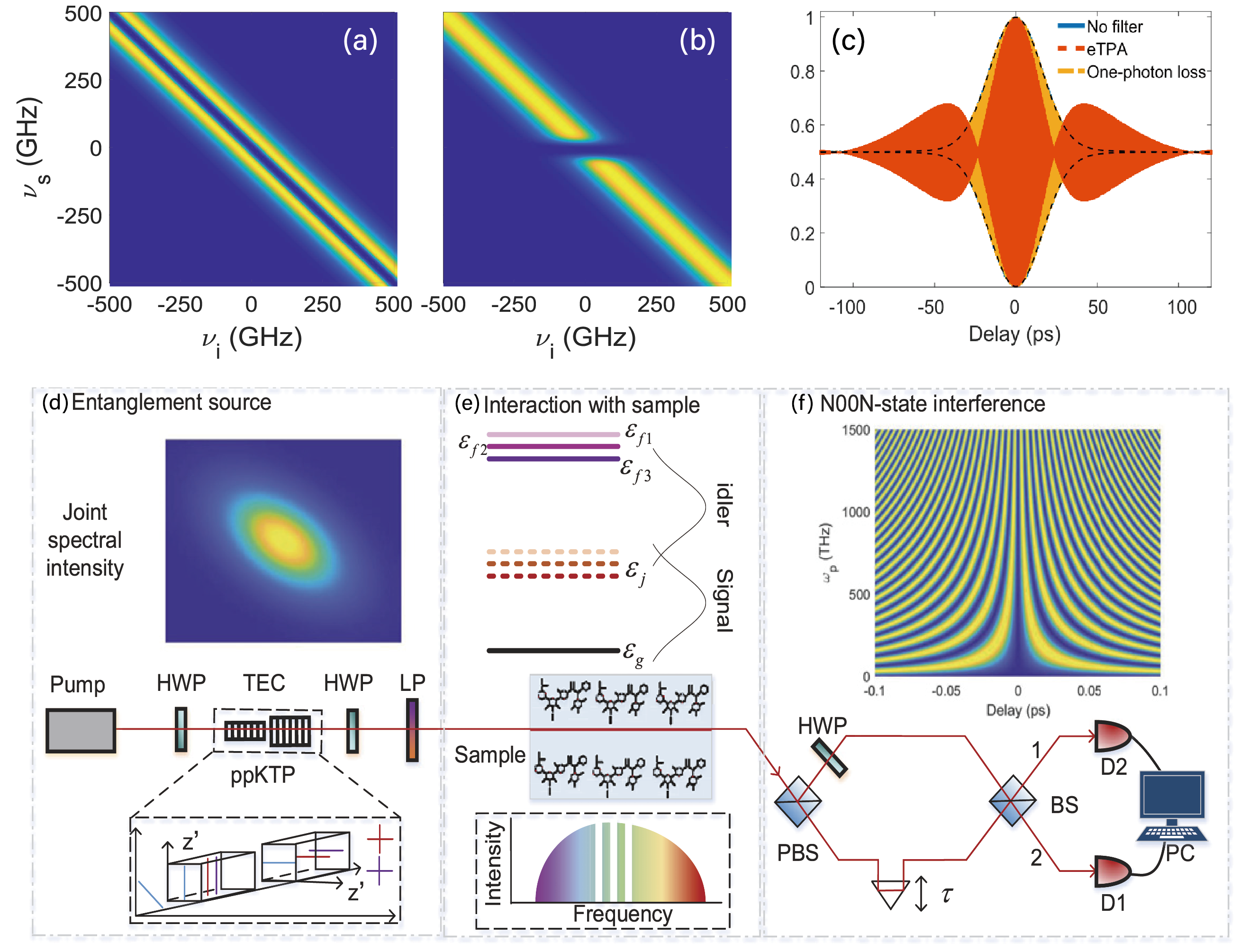}
\caption{Applications of N00N state interferometer in quantum spectroscopy. (a) JSI after entangled two-photon absorption. (b) JSI after single-photon loss. (c) N00N state interference patterns after entangled two-photon absorption and single-photon loss. Reprinted from Ref.\,\cite{MartinezTapia2023}. (d) Entangled photon source. (e) The interaction with the sample. (f) N00N state interference. Reprinted from Ref.\,\cite{Chen2022NJP}. }
\label{N00Netpa}
\end{figure}

Nonlinear spectroscopy has proven to be a powerful technique to extract valuable information on the energy dynamics and chemical structure of new substances and molecules \cite{Mukamel1995}.
Recent research has suggested that the utilization of entangled photon pairs has the potential to unlock new possibilities in experimental two-photon absorption spectroscopy \cite{Mukamel2020JPB}.
Despite the numerous experimental studies conducted on entangled two-photon absorption (eTPA), there is still ongoing debate as to whether eTPA has truly been observed \cite{Raymer2021JCP}.

In 2022, Martinez-Tapia et al. theoretically demonstrated that the N00N state interferometer is a powerful method to experimentally certify true eTPA, since it is insensitive to linear (single-photon) losses \cite{MartinezTapia2023}. Figure\,\ref{N00Netpa}(a) and (b) show the simulated joint spectral intensities with two-photon absorption and one-photon absorption and (c) shows the N00N state interference patterns. It is clear that the pattern resulting from one-photon absorption exhibits a single peak with a Gaussian profile, whereas the pattern associated with eTPA displays three distinct peaks. These distinctive characteristics make it easy to differentiate between the two processes. In addition, the N00N-SI can also be used to extract information about levels in the two-photon excitation band. Chen et al. theoretically demonstrated such a two-photon excitation spectroscopy scheme using a N00N state interferometer \cite{Chen2022NJP}. The two-photon excitation spectrum of the sample can be extracted by performing a Fourier transform on the second-order correlation function of the interference pattern. Figure\,\ref{N00Netpa}(d) shows the entangled photon source, (e) shows the interaction between the samples and biphotons, and (f) shows the interference of the N00N state. Different samples have different excitation bands and therefore will have different N00N state interference patterns. This approach enables the extraction of information regarding the electronic structure of atoms or molecules' two-photon excited-state.

\subsubsection{Chemical and biological sensing }
\begin{figure}[htbp]
\centering
\includegraphics[width= 0.92\textwidth]{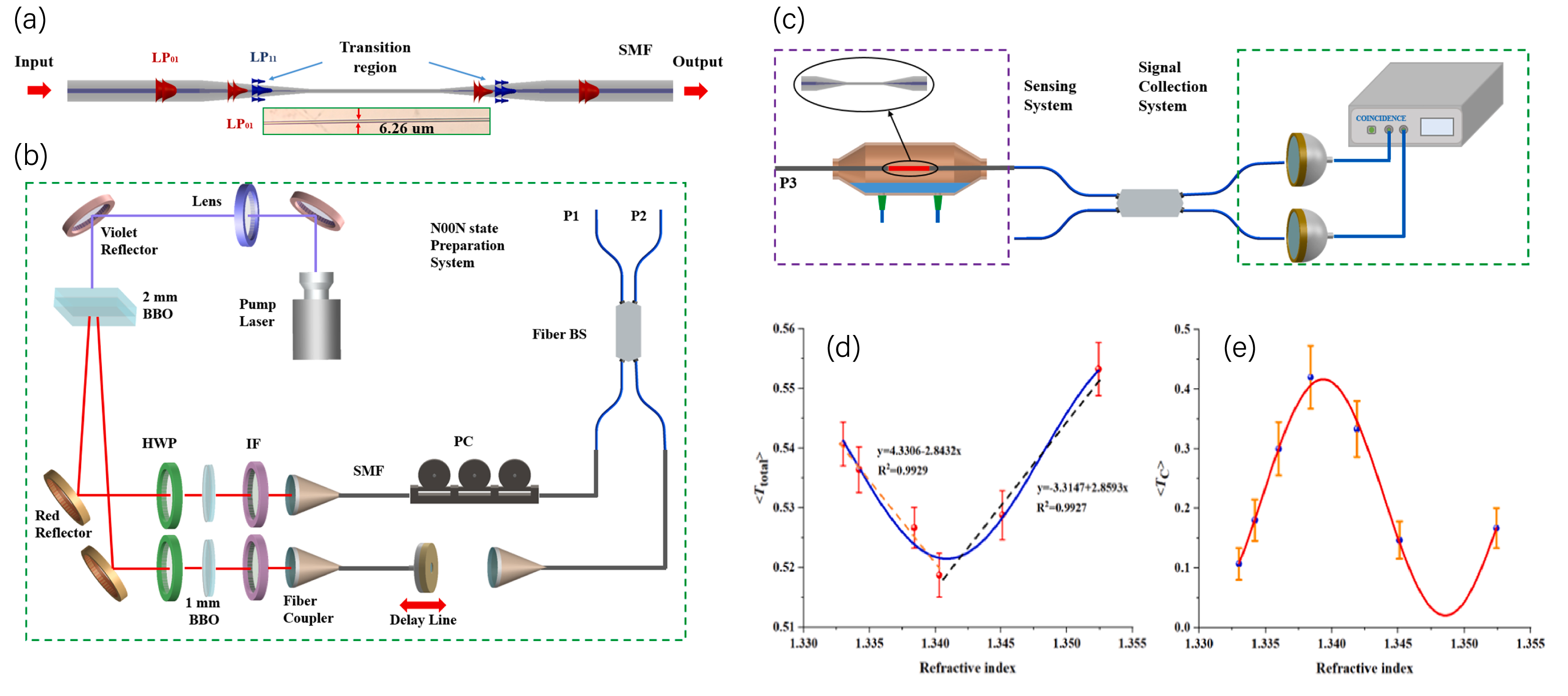}
\caption{Applications of N00N-SI in chemical and biological sensing. (a) Schematic of the proposed microfiber MZI. (b) Schematic of the two-photon N00N state preparation system. (c) Schematic of the sensing system and the signal collection system. (d-e) Comparison diagram of single-photon transmittance and two-photon N00N state transmittance based on the microfiber MZI sensing system. Reprinted from Ref.\,\cite{ Peng2023}. }
\label{N00Nsensing}
\end{figure}
The N00N-SI can also be used in chemical and biological sensing \cite{Haas2018, Peng2023}. In 2023, Peng et al. demonstrated a microfiber quantum sensor for protein measurement with the two-photon N00N-SI, which can eliminate the disturbance of classical interference fringes. The concentration of biological samples can be measured by monitoring changes in the coincidence count. Experimental results reveal that the quantum sensing system exhibits an ultra-high refractive index (RI) sensitivity. Compared to an RI sensor employing single-photon ``classical" interference, the proposed sensor achieves a nearly 21-fold improvement in RI sensitivity. By achieving remarkable sensitivity and precision at low photon levels, this research introduces a novel approach for label-free detection of photosensitive biomolecules. Figure\,\ref{N00Nsensing}(a) and (b) shows the schematic and experiment of the proposed microfiber Mach-Zehnder interferometer (MZI) for high-precision protein measurement with a two-photon N00N state. The MZI consists of a microfiber with a uniform waist region and two transition regions. The microfiber serves as the sensing element, and the waist region is where the two-photon N00N state is injected. The two transition regions are used to activate higher-order modes in the microfiber. By connecting P2 and P3 in Fig.\,\ref{N00Nsensing}(c), the two-photon N00N state can be introduced into the sensing system, and the coincidence counter is used to measure and record the number of two-photon N00N states.
To measure the RI, Fig.\,\ref{N00Nsensing}(d-e) shows the single photon transmittance and two-photon N00N state transmittance based on the microfiber MZI sensing system.
Obviously, the period of interference fringes using two-photon N00N state quantum interference is about half of the period of single photon classical interference fringes. This means that the sensitivity of the quantum sensor can be improved by using the N00N state.

\section{Franson Interferometer}
Franson interferometer was first proposed in 1989 to verify the Bell inequality in the degrees of freedom of energy and time \cite{Franson1989PRL, Jogenfors2014JPA}. For different aims, several distinct configurations of Franson interferometer are invented, including the original version \cite{Franson1989PRL},  the hugged version \cite{Cabello2009PRL,Lima2010PRA}, the conjugate version \cite{Zhang2014PRL,Chen2021PRL}, and the three-photon version as well \cite{Agne2017PRL,OBrien2003Nature}. As a general setup, the Franson interferometer is effective for photons from several physical processes, including SPDC in $\chi^{(2)}$ medium \cite{Khan2006PRA,Lima2010PRA,Vallone2011PRA,Barreiro2005PRL,Shi2006NJP,Bessire2014NJP}, spontaneous four-wave mixing (SFWM) in $\chi^{(3)}$ medium \cite{Ma2018QST,Mittal2021NP,Oser2020npj,Kumar2015OE,Suo2015OE,Mazeas2016OE}, biexciton-excition emission in quantum dots \cite{Jayakumar2014NC,Sun2017OE,Peiris2017PRL}, etc. The Franson interference also has been applied in a variety of fields, from characterizing entanglement \cite{Kwiat1993PRA,Horne1989PRL,Jayakumar2014NC,Chen2018APL} to testing fundamental physical principles \cite{Cuevas2013NC,Stefanov2002PRL}, quantum cryptography \cite{Tittel2000PRL,Kim2022APL,AliKhan2007PRL}, quantum network \cite{Halder2007NP,Saglamyurek2015NP}, quantum spectroscopy \cite{Tabakaev2021PRA,Franson1989Springer}, and imaging \cite{Gao2019APL}. In the next sections, we will review the Franson interferometer in detail.

\subsection{Principles of Franson Interferometer}
\subsubsection{Single-mode model of Franson interferometer}

\begin{figure}[htbp]
\centering
\includegraphics[width= 0.8\textwidth]{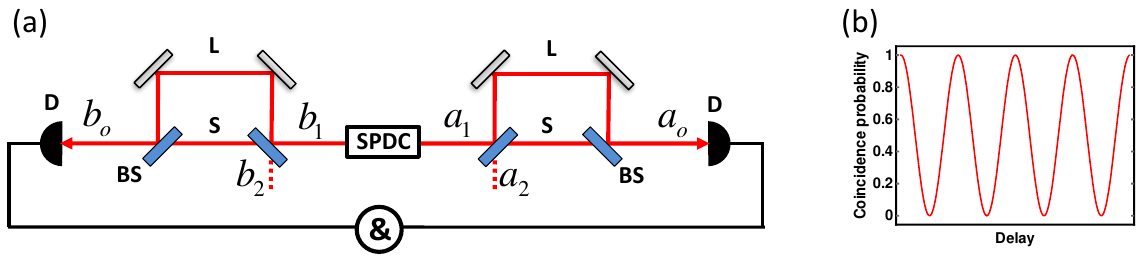}
\caption{The model of Franson interferometer. (a) The setup. (b) The interference patterns.}
\label{FransonI}
\end{figure}

In the Franson interferometer, as shown in Fig. \ref{FransonI}(a), two photons from an SPDC source pass through two unequal-arm MZI independently. The long arm and the short arm are labeled as $L$ and $S$, respectively. The path-length difference is given by $L-S=\Delta L$, which satisfies $l_{c1}  < \Delta L < l_{c2}$, where $l_{c1}$ and $l_{c2}$ are the coherent lengths of a single photon and two photons. The first condition $l_{c1}  < \Delta L$ is used to ensure that there is no single-photon interference. The second condition $\Delta L < l_{c2}$ makes it impossible to distinguish both photons along the long arm and the short arm.

The initial state is given by
$\left| \psi  \right\rangle  = \hat a_1^\dag \hat b_1^\dag {\left| 0 \right\rangle _{a1}}{\left| 0 \right\rangle _{b_1}}{\left| 0 \right\rangle _{a_2}}{\left| 0 \right\rangle _{b_2}} = {\left| 1 \right\rangle _{a_1}}{\left| 1 \right\rangle _{b_1}}{\left| 0 \right\rangle _{a_2}}{\left| 0 \right\rangle _{b_2}}$, where $a_1$ and $b_1$ represent the input ports, while $a_2$ and $b_2$ correspond to the vacuum input ports, as illustrated in Fig. \ref{FransonI}(a). The output field operator at port $a_o$ is
${{\hat a}_o} = \frac{1}{{\sqrt 2 }}({{\hat a}_L}{e^{i\phi }} + {{\hat a}_S})$, where $\phi$ is the relative phase difference between the long arm and the short.
Considering ${{\hat a}_L} = \frac{1}{{\sqrt 2 }}({{\hat a}_1} + {{\hat a}_2})$ and ${{\hat a}_S} = \frac{1}{{\sqrt 2 }}({{\hat a}_1} - {{\hat a}_2})$, we can obtain
${{\hat a}_o} = \frac{1}{2}[({{\hat a}_1} + {{\hat a}_2}){e^{i\phi }} + ({{\hat a}_1} - {{\hat a}_2})] = \frac{1}{2}[{{\hat a}_1}({e^{i\phi }} + 1) + {{\hat a}_2}({e^{i\phi }} - 1)]$.
Similarly, we can obtain the field operator at port $b_0$ as
${{\hat b}_o} = \frac{1}{2}[{{\hat b}_1}({e^{i\phi }} + 1) + {{\hat b}_2}({e^{i\phi }} - 1)]$.
The coincidence probability $P$ is determined by the second order correlation function, i.e.,
\begin{equation}\label{Eq:3-1-4}
 {P} = \left\langle \psi  \right| \hat b_o^\dag \hat a_o^\dag  {{\hat a}_o}{{\hat b}_o}\left| \psi  \right\rangle= {\left| {\frac{1}{4}({e^{i\phi }} + 1)({e^{i\phi }} + 1)} \right|^2} = \frac{1}{4}(1 + \cos \phi )^2
\end{equation}
With this equation, we can plot the interference patterns of Franson interference, as shown in Fig. \ref{FransonI}(b).
The single counts probability at port  $a_o$  can be calculated as follow:
\begin{equation}\label{Eq:3-1-5}
{P_{sc}} = {\left\langle 0 \right|_{a2}}{\left\langle 1 \right|_{a1}}\hat a_o^\dag {{\hat a}_o}{\left| 1 \right\rangle _{a1}}{\left| 0 \right\rangle _{a2}} = {\left| {\frac{1}{2}({e^{i\phi }} + 1)} \right|^2} = \frac{1}{2}(1 + \cos \phi ).
\end{equation}

\subsubsection{Multi-frequency model of Franson interferometer}

\begin{figure}[htbp]
\centering
\includegraphics[width= 0.8\textwidth]{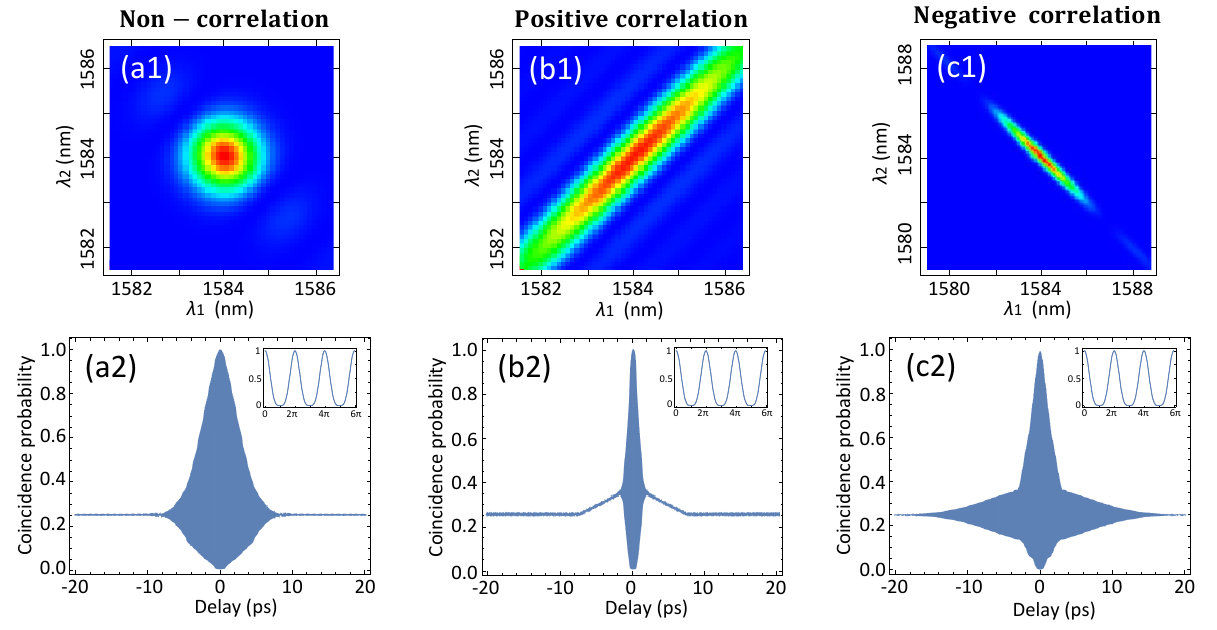}
\caption{Franson interference using biphotons with different spectral correlations: (a1-a2) non-correlation,  (b1-b2) positive correlation, and (c1-c2) negative correlation. Reprinted from Ref. \cite{Jin2024SCPMA}.  The joint spectral amplitudes and the corresponding HOM interference patterns.}
\label{multifmFranson}
\end{figure}

In this section, we deduce the equations for the Franson interference using multi-mode theory \cite{Jin2024SCPMA}. Similar to previous sections, the two-photon state from a SPDC process can be described as
$\left| \psi  \right\rangle  = \int_0^{\infty}{\int_0^\infty  {d\omega _s d\omega _i } } f(\omega _s ,\omega _i )\hat a_s^\dag  (\omega _s )\hat a_i^\dag  (\omega _i )\left| {00} \right\rangle$. The transformation rule of the operators after the delay time $T_1$ and $T_2$ is $\hat{a}_1\left(\omega_1\right)=\frac{1}{{2}}\left[\hat{a}_s\left(\omega_1\right)+\hat{a}_s\left(\omega_1\right) e^{-i \omega_1 T_1}\right],$ and $\hat{a}_2\left(\omega_2\right)=\frac{1}{{2}}\left[\hat{a}_i\left(\omega_2\right)+\hat{a}_i\left(\omega_2\right) e^{-i \omega_2 T_2}\right].$ Therefore, we can rewrite the field operators at the detectors as
\begin{equation}
\begin{aligned}
    \hat{E}_1^{(+)}\left(t_1\right)&=\frac{1}{2 \sqrt{2 \pi}} \int_0^{\infty} d \omega_1\left[\hat{a}_s\left(\omega_1\right)+\hat{a}_s\left(\omega_1\right) e^{-i \omega_1 T_1}\right] e^{-i \omega_1 t_1},\\
    \hat{E}_2^{(+)}\left(t_2\right)&=\frac{1}{2 \sqrt{2 \pi}} \int_0^{\infty} d \omega_2\left[\hat{a}_i\left(\omega_2\right)+\hat{a}_i\left(\omega_2\right) e^{-i \omega_2 T_2}\right] e^{-i \omega_2 t_2}.
\end{aligned}
\end{equation}
The coincidence probability $P(\tau)$ can be expressed as
\begin{equation}\label{Eq:3-2-8}
\begin{aligned}
    P(\tau )   &= \int_{-\infty}^{\infty}  {\int_{-\infty}^{\infty} {dt_1 dt_2 } } \left\langle {\psi \left| {\hat E_1^{( - )} \hat E_2^{( - )} \hat E_2^{( + )} \hat E_1^{( + )} } \right|\psi } \right\rangle.\\
& =\frac{1}{4} \int_0^{\infty} \int_0^{\infty} d \omega_1 d \omega_2\left|f\left(\omega_1, \omega_2\right)\right|^2\left[1+\operatorname{cos}\left(\omega_1 T_1\right)\right]\left[1+\operatorname{cos}\left(\omega_2 T_2\right)\right].
\end{aligned}
\end{equation}
If the delay of the two paths is equal, i.e., $T_1=T_2=T$, then
\begin{equation}\label{Eq:3-2-14}
\begin{array}{lll}
\begin{aligned}
P(\tau)=\frac{1}{4} \int_0^{\infty} \int_0^{\infty} d \omega_1 d \omega_2\left|f\left(\omega_1, \omega_2\right)\right|^2\left[1+\operatorname{cos}\left(\omega_1 T\right)\right]\left[1+\operatorname{cos}\left(\omega_2 T\right)\right]  \\
\end{aligned}
 \end{array}.
\end{equation}
With this equation, it is possible to plot the Franson interference with different joint spectral distribution, as shown in Fig.\,\ref{multifmFranson}. We note that it is also possible to demonstrate the spectrally resolved Franson interference \cite{Jin2024SCPMA}.

 In a Franson interferometer, it is crucial to eliminate any single-photon interference. Therefore, we will now examine the probability of a single count in the multi-mode form, taking detector 1 as an illustrative example.
For simplicity in calculations, we make the assumption that the two-photon states produced by the SPDC process are separable.
Under this assumption, the state of the signal is given by:
 $\left| \psi  \right\rangle  = \int_0^{\infty}    {d{\omega _s}} f({\omega _s})\hat a_s^\dag ({\omega _s})\left| 0 \right\rangle $.  
The  field operator is given by $\hat E_1^{( + )}({t_1}) = \frac{1}{{2\sqrt {2\pi } }}\int_0^{\infty}  {d{\omega _1}} [{\hat a_s}({\omega _1}) + {\hat a_s}({\omega _1}){e^{ - i{\omega _1}{T}}}]{e^{ - i{\omega _1}{t_1}}}.$
The single count probability $P_{sc}(\tau)$, can be expressed as
\begin{equation}\label{Eq:3-2-19}
    P_{sc}(\tau ) = {\int_{-\infty}^{\infty} {d{t_1}} } \left\langle {\psi \left| {\hat E_1^{( - )}\hat E_1^{( + )}} \right|\psi } \right\rangle
 = \frac{1}{2}\int_0^{\infty}  {  {d{\omega _1}} }  {\left| {f({\omega _1})} \right|^2}[1 + \cos({\omega _1}{T})].\\
\end{equation}

\subsection{Franson interferometers based on different setups}

\subsubsection{The original and the revised setup}
\label{IV.B.1}
\begin{figure}[htbp]
\centering
\includegraphics[width= 0.82\textwidth]{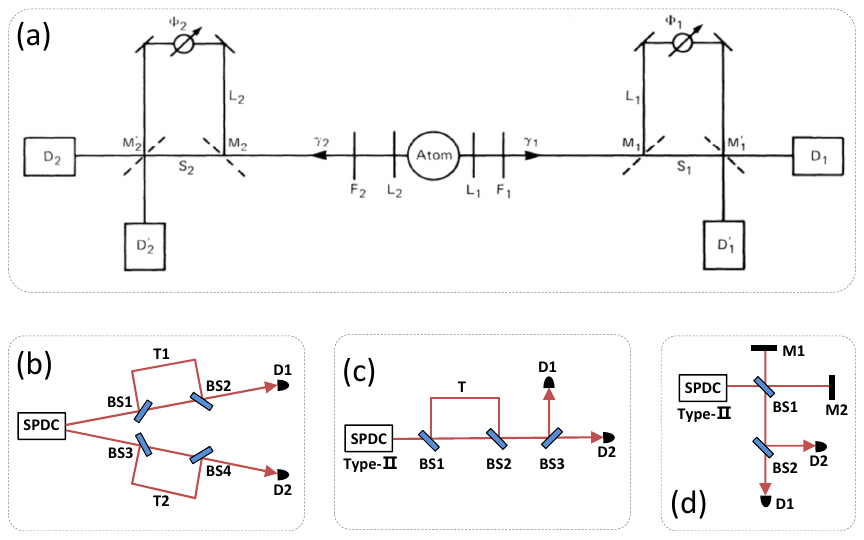}
\caption{(a) The original version of Franson interferometer. Reprinted from Ref.\,\cite{Franson1989PRL}. (b) The unfolded Franson interferometer with two detectors on the same side. (c) The Mach–Zehnder-type folded Franson interferometer. (d) The Michelson-type folded Franson interferometer. Reprinted from Ref.\,\cite{Jin2024SCPMA}.}
\label{Franson/origin.pdf}
\end{figure}
In 1989, Franson first proposed an interferometer to test the Bell inequality about time and energy \cite{Franson1989PRL}. The original configuration of this interferometer is shown in Fig.\,\ref{Franson/origin.pdf}(a). In this configuration, the two photons cascadingly emitted from a three-level atomic structure are separated by their wavelengths and each one of them is sent to an unbalanced Mach–Zehnder interferometer (UMZI). In each UMZI, a variable phase is introduced in one arm. The photons from the two output ports of each UMZI are detected by two single-photon detectors. Under different phases in UMZI, the single counts in each output port and the coincidence counts between two arbitrary output ports of the two UMZIs are recorded. The three-level structure in Franson's original scheme is composed of an upper level, a lower intermediate level, and a lowest ground state, in which the upper one has a lifetime significantly longer than the intermediate one. Therefore, the probability amplitude of the two-photon cascade emission is distributed over a large range of time, defining a long two-photon coherence time. Meanwhile, the single-photon coherence time, i.e., the coherence time of the photon emitted from the upper level or the photon emitted from the intermediate level, is considerably shorter.
The time delay due to the path length difference in each UMZI can be set to be much longer than the single-photon coherence time, but much shorter than the two-photon coherence time. The former condition can suppress single-photon interference, i.e. second-order interference, in each UMZI. However, calculations show that a cosine oscillation exists in the coincidence counts when the phase in either UMZI is varied. These calculations clearly indicate that such an oscillation is caused by nonlocal fourth-order interference, later called Franson interference. This interference originates from the correlation between the emission times of the two photons and their energy constraint. To intuitively understand this interference, the pathway of photons can be considered. In each UMZI, the input photon would select either the short or long arms. It is obvious that the events in which the two photons select pathways of different lengths are temporally distinguishable. However, it is impossible to distinguish the two events in which the two photons both select short or long pathways if they are identical in all other degrees of freedom. Thus, the probability amplitudes of the latter two events can interfere with each other and lead to the Franson interference fringes. Experimentally, several types of UMZI have been used to construct Franson interferometers, such as fiber-based UMZI \cite{Brendel1991PRL}, free-space UMZI \cite{Kwiat1993PRA}, and integrated UMZI \cite{Honjo2007OE}.

The configuration of the UMZI used can be standard as shown in Fig.\,\ref{Franson/origin.pdf}(a).
In fact, this setup can also be revised.  Figure\,\ref{Franson/origin.pdf}(b) illustrates an alternative setup where two photons are on the same side.
Additionally, Fig.\,\ref{Franson/origin.pdf}(c) and (d)  present folded forms of the UMZI based on a Mach-Zehnder interferometer \cite{Jin2024SCPMA} or a Michelson interferometer \cite{Jayakumar2014NC}. 
In these cases, the nonlinear crystals in the SPDC sources are type-II phase matched, and the signal and idler photons have orthogonal polarizations. These modifications allow for different interference configurations and can be tailored to specific experimental requirements.

\subsubsection{The hug Franson interferometer}
\label{IV.B.2}
\begin{figure}[!h]
\centering
\includegraphics[width= 0.6\textwidth]{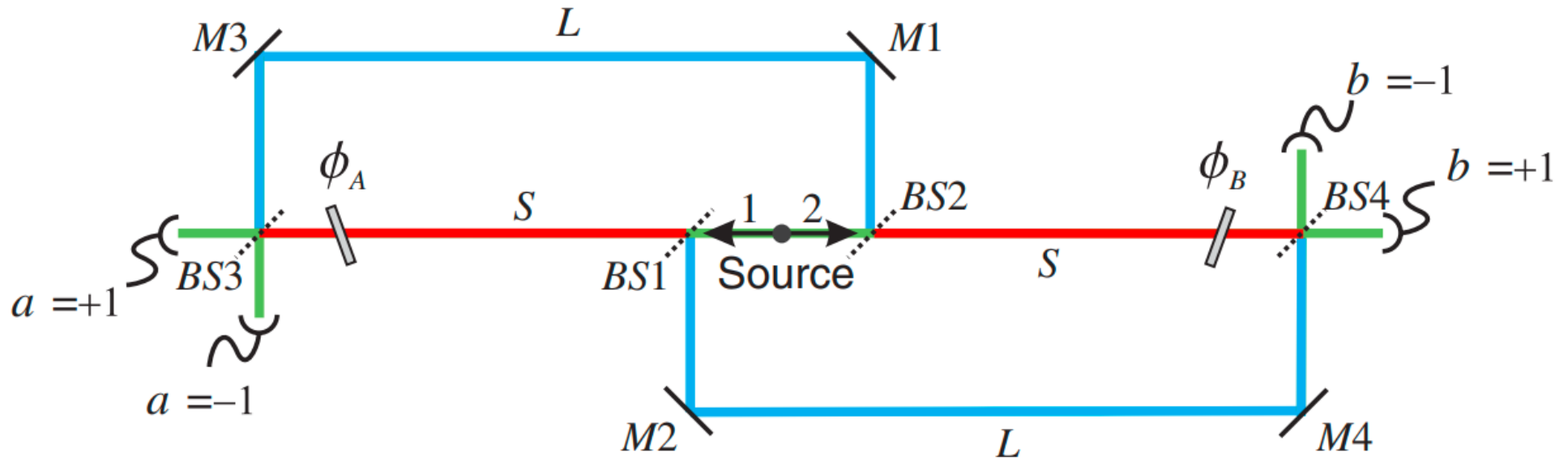}
\caption{The hug configuration of Franson interferometer. Reprinted from Ref.\,\cite{Cabello2009PRL}. 
}
\label{Franson}
\end{figure}
The original Franson interferometer was proposed to demonstrate the violation of local realism using energy-time entanglement \cite{Franson1989PRL}. Researchers had believed that this criterion was undoubtedly correct until a model constructed in 1999 demonstrated that the measured Franson interference by the original interferometer does not necessarily violate local realism, due to the post-selection loophole \cite{Aerts1999PRL}. To close this loophole, a modified interferometer, called hug Franson interferometer, is proposed by Cabello et al. in 2009 \cite{Cabello2009PRL}. The structure of the hug Franson interferometer is shown in Fig.\,\ref{Franson}, where $BS_{i}$ and $M_{i}$ ($i=1, 2, 3, 4$) are beam splitters and mirrors, respectively. Alice on the left and Bob on the right are the two terminals that receive photons from the source after going through the interferometer. $\phi _{A}$ ($\phi _{B}$) represents the phase shift in the short arms of the interferometer on the path to Alice (Bob). Compared with the original Franson interferometer, this revised setup has the distinct feature that a photon will go to different sides, i.e., Alice and Bob, if it selects the short and long paths when encountering the first $BS$ after its emission. This modification ensures that all the coincidence detections between Alice and Bob result from the events in which the two photons arrive at their terminals through paths of the same length. Events with one photon taking a short path and its twin taking a long path will lead to both photons reaching the same side. In this case, the temporal postselection (which introduced the loophole) is no longer necessary to obtain the interference curves. It should be noted that a prototype of a hug Franson interferometer was proposed and experimentally constructed in 2008 to measure the time-bin entanglement without postselection \cite{Rossi2008PRA}. However, that experimental setup could introduce additional losses, and its usage in Bell experiments was not discussed.

Later, Lima et al. experimentally realized the hug Franson interferometer to perform a post-selection loophole-free energy-time Bell experiment \cite{Lima2010PRA}. Subsequently, Vallone et al. used the same interferometer to implement a Hardy-type test of quantum nonlocality free from the post-selection loophole \cite{Vallone2011PRA}. The long-distance fiber-based distribution of genuine energy-time entanglement has also been achieved \cite{Cuevas2013NC} and deployed in an optical fiber network \cite{Carvacho2015PRL} with the help of hug Franson interferometer. In 2019, a novel application of this interferometer was proposed by Gianani \cite{Gianani2019PRR}  for reconstructing the joint spectral phase of photon pairs in a noisy environment. In 2021, Lum et al. reported an experiment in which the interferometer was used to witness the survival of energy-time entanglement after the photons propagated through biological tissue and scattering media \cite{Lum2021BOE}. Recently, the hug Franson interferometer has been realized on a chip and used to certify genuine energy-time and time-bin  entanglement \cite{Santagiustina2023arXiv}. Dispite of these progresses, most works involving energy-time or time-bin entanglement, even conducted much recently, still employ the original version of Franson interferometer as the tool to implement Bell test \cite{Kim2022APL,Hohn2023PRR}. It is argued that we cannot simply claim whether Franson's original setup can or cannot be used for the violation of local realism without the details of the experiment and goal. For more discussions, interested reader should refer to Ref. \cite {Jogenfors2014JPA}.

\subsubsection{The conjugate-Franson interferometer}
\label{IV.B.3}
\begin{figure}[htbp]
\centering
\includegraphics[width= 0.92\textwidth]{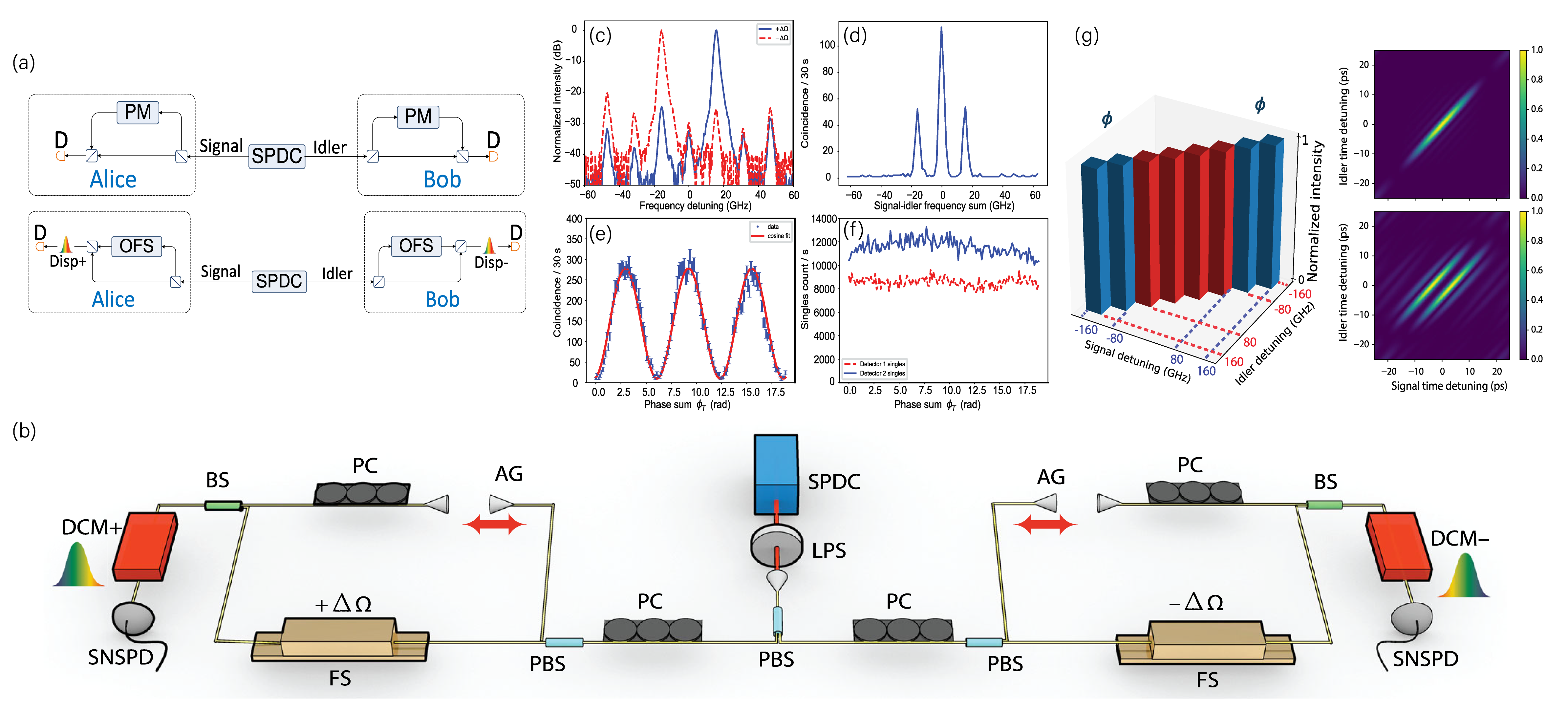}
\caption{The principle, experiment setup and some results of the conjugate-Franson interferometer. (a) Comparison between the original (top panel) and conjugate-Franson interferometers (bottom panel). Reprinted from Ref.\,\cite{Zhang2014PRL}. (b) Experimental setup of the conjugate-Franson interferometer. (c) Measured output spectra of the frequency shifter in (a). (d) Frequency-resolved coincidence counts versus signal-idler frequency sum. (e) Coincidences after frequency post-selection versus MZI phase sum $\phi_T$. (f) Singles count rates for signal and idler photons versus $\phi_T$. (g) Calculated JSI and JTI normalized to one. Reprinted from Ref.\,\cite{Chen2021PRL}. }
\label{Conjugate-Franson.pdf}
\end{figure}
A complete characterization of biphotons should reveal the structures of nonclassical correlations in the frequency and time domains. It has been concluded that the original Franson interferometer only provides the correlation information in the frequency domain, while that in the time domain is still absent \cite{Zhang2014PRL}. This limitation becomes particularly noticeable in the studies of quantum frequency combs, where two frequency combs that differ only in their spectral phase structures cannot be distinguished from each other \cite{Lingaraju2019OE}. Chang et al. \cite{Chang2021npj} pointed out that the interference curves measured by the original Franson interferometers only relate to the joint spectral intensity (JSI), while the joint temporal intensity (JTI) could be obtained through conjugate-Franson interferometry. This interferometry has been first proposed by Zhang et al. \cite{Zhang2014PRL} to provide unconditional security against collective attacks in energy-time entanglement-based QKD.

Figure\,\ref{Conjugate-Franson.pdf}(a) illustrates the comparison between the original Franson interferometer (upper panel) and its conjugate version (lower panel). The conjugate version can be considered as a transformation of the original version according to the frequency-time conjugation relationship. In the upper panel, the time delay between the two arms is replaced by the frequency shift in the lower panel. Additionally, the time-resolved detection after the beam combiner in the upper panel is replaced by the frequency-resolved detection in the lower one, which can be achieved through dispersion. In 2021, Chen et al. demonstrated the conjugate-Franson interferometer experimentally, as shown in Fig.\,\ref{Conjugate-Franson.pdf}(b) \cite{Chen2021PRL}. In this setup, the frequency shift in Fig.\,\ref{Conjugate-Franson.pdf}(a) is realized using dual-drive quadrature phase-shift keying modulators operating in single sideband generation mode. In the experiment, a CW laser pumps a type-II phase-matched SPDC process and generates frequency-nondegenerate signal and idler photons with orthogonal polarizations. After being separated by a PBS, the signal and idler photons pass through the conjugate-Franson interferometer and finally are detected by the SNSPDs. The performance of the frequency shifter is shown in Fig.\,\ref{Conjugate-Franson.pdf}(c). It is obvious that both the positive and negative frequency shifters shift the spectra of the signal and idler photons successfully.

Figure\,\ref{Conjugate-Franson.pdf}(d) presents a typical frequency-resolved coincidence between photons after the conjugate-Franson interferometer. Similar to the time-resolved coincidence in original Franson interferometer \cite{Brendel1991PRL}, there are three peaks. The middle peak results from the events in which both the signal and idler photons select the path with or without frequency shift, whereas the two side peaks can be attributed to the events in which one photon experiences a frequency shift but its twin does not. The two types of events contributing to the middle peak are spectrally indistinguishable and, therefore, can interfere with each other. A sample of the measured interference fringe is shown in Fig.\,\ref{Conjugate-Franson.pdf}(e). To eliminate the possibility of single-photon interference, the single count rates of the signal and idler photons after the two interferometers are measured against the phase sum, and the results are shown in  Fig.\,\ref{Conjugate-Franson.pdf}(f), which exhibit no oscillatory signature. Consequently, the interference fringe in Fig.\,\ref{Conjugate-Franson.pdf}(e) unambiguously validates that the energy-time entanglement can be certified by conjugate-Franson interferometer.

It has been shown that the visibility of the interference fringe in Fig.\,\ref{Conjugate-Franson.pdf}(e) would change under different JTI of the biphoton states \cite{Chen2021PRL}. Therefore, this visibility can be utilized to distinguish different biphoton states with the same JSIs but different spectral phase contents, which could result in different JTIs. In Ref.\,\cite{Chen2021PRL}, this prediction is experimentally verified by measuring the variation of the interference fringe when frequency-dependent phase modulation is applied to a biphoton state without intrinsic spectral phase variation. The theoretically calculated JSI and JTI of states with and without phase modulation are shown in Fig.\,\ref{Conjugate-Franson.pdf}(g), clearly demonstrating that JTI changes significantly after spectral phase modulation $\phi$ is introduced to the signal and idler photons. The experimental results indicate a significant reduction in the measured visibility of the interference fringe when $\phi$ is varied from 0 to $\pi$. This unequivocally verifies the capability of conjugate-Franson interferometry to reveal the temporal correlation in energy-time entanglement. In fact, a simple duality can be summarized, i.e., the frequency correlation can be revealed if time shift is introduced into one arm in the interferometer, while the time correlation can be demonstrated if frequency shift is introduced. These rules could be extended to verify the entanglement in other pairs of dual variables, such as position and momentum, OAM and azimuth.

\subsubsection{The three-photon Franson interferometer}
\begin{figure}[htbp]
\centering
\includegraphics[width= 0.92\textwidth]{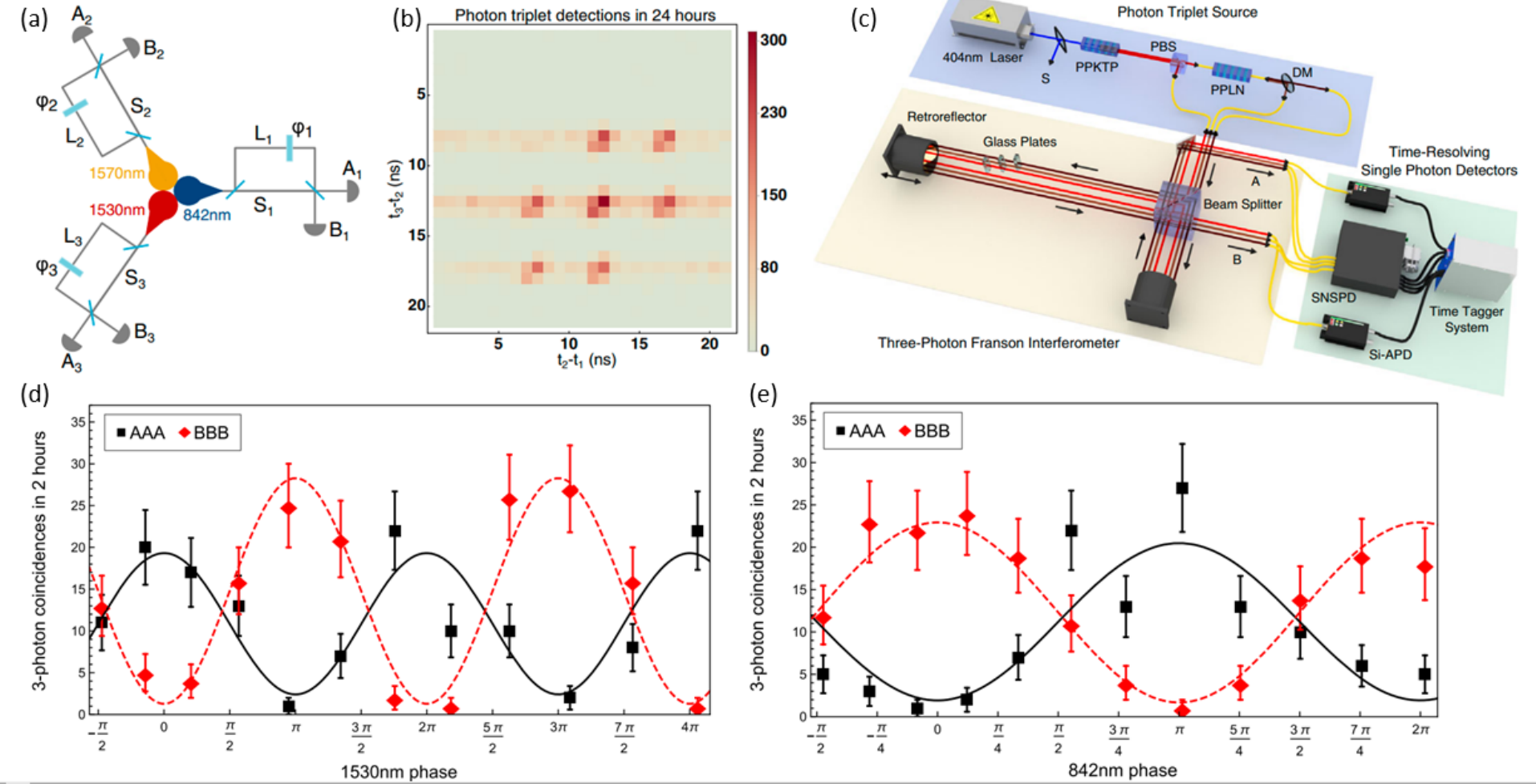}
\caption{The schematic, experimental setup and results of the three-photon Franson interferometer. (a) Schematic of the three-photon Franson interferometer. (b) Triplet coincidence histogram after a three-photon Franson interferometer. (c) Experimental setup for observing three-photon Franson interference. (d) and (e) are the measured interference fringes when the phases for photons with different wavelengths are scanned. Reprinted from Ref.\,\cite{Agne2017PRL}. 
}
\label{Franson/3-photon.pdf}
\end{figure}
Multiparticle quantum interference plays a critical role in our understanding and exploitation of quantum information \cite{OBrien2003Nature,Pan2012RMP}, and in fundamental tests of quantum mechanics \cite{Menssen2017PRL}. The extension of the original Franson interferometer to three photons was first theoretically proposed by Greenberger et al. to investigate the Einstein-Podolsky-Rosen paradox using the three-photon correlation \cite{Greenberger1990AJP}. This three-photon interferometer was then mentioned and studied in several theoretical works \cite{Barnett1998JMO,Greenberger1993PT}, while the three-photon polarization entanglement has been experimentally created and measured by conventional polarization analyzers \cite{Bouwmeester1999PRL,Hamel2014NP}. An illustration of a three-photon Franson interferometer is shown in Fig.\,\ref{Franson/3-photon.pdf}(a), in which three correlated photons are individually input into the three UMZIs and coincidences between arbitrary two ports from two different UMZIs are measured.

The main experimental challenge in realizing the three-photon Franson interference is the extremely low coincidence counts. Agne et al. demonstrated a high-rate source of photon triplets and constructed a three-photon Franson interferometer to measure the energy-time entanglement \cite{Agne2017PRL}. The experimental setup is depicted in Fig.\,\ref{Franson/3-photon.pdf}(c), in which a CW laser is used to pump the SPDC process to generate correlated photons with different wavelengths. One photon in each photon pair is then used to pump another SPDC process, generating frequency-nondegenerate correlated photons in the telecom-band. The generated photons are all incident on a common unbalanced Michelson interferometer (UMI). At the output ports of the interferometer, photons with different wavelengths are separated and detected. Figure\,\ref{Franson/3-photon.pdf}(b) shows the measured three-fold coincidence versus the arrival time difference between photons in different wavelength channels. Similar to the three peaks observed in two-photon Franson interference, the seven peaks shown in Fig.\,\ref{Franson/3-photon.pdf}(b) can be attributed to the events where the photons select short or long paths in the UMI. The central peak results from the event in which all three photons go through either the short arms or the long arms. The measured three-photon Franson interference fringes are displayed in Fig.\,\ref{Franson/3-photon.pdf}(d) and (e), in which the phases of photons at two different wavelengths are scanned, respectively. An averaged visibility of 84.6$\%$ is obtained for the two cases, revealing genuine three-photon Franson interference due to energy-time entanglement. In this experiment, a single Michelson interferometer acts as the unbalanced interferometer for all the three entangled photons. However, independent unbalanced interferometers are required if the nonlocality is tested for multiple photons distributed to different nodes. Such test has high requirement on the uniformity of the path difference in all the unbalanced interferometers. When more and more entangled photons are involved, this requirement will be challenging. Utilizing the state-of-art integrated interferometer might be a promising way to cope with this challenge \cite{ Zhang2018OL}.

\subsection{Franson interferometers with photons from different systems}

Franson interference occurs in energy-time entanglement or time-bin entanglement. The generation of such entanglement can be achieved through various physical mechanisms in different media. The following sections will summarize most of these mechanisms, including SPDC in bulk crystals or waveguiding structures with $\chi^{(2)}$ nonlinearity, SFWM in silica fiber, integrated waveguides or microresonators with $\chi^{(3)}$ nonlinearity, SFWM in warm atomic ensembles, and two-photon emission in quantum dots (QDs). 

\subsubsection{SPDC in bulk crystals and waveguides}
\begin{table}[h]
\caption{The $\chi^{(2)}$ media for creating energy-time and time-bin entanglement}
\label{Tab:Franson1}
\begin{tabular}{l|l|l}
\hline \hline
Nonlinear medium & Types of entanglement  &  References                                        \\ \hline   
\multirow{2}{*}{BBO crystal}& Energy-time     &  \cite{Khan2006PRA,Lima2010PRA,Vallone2011PRA,Barreiro2005PRL,AliKhan2007PRL,Broadbent2008PRL}  \\  \cline{2-3} 
&Time-bin& \cite{Kwon2013OE} \\  \hline
LBO crystal& Time-bin     & \cite{Thew2004PRL}  \\  \hline
\multirow{2}{*}{KNbO$_3$ crystal}& Energy-time     &  \cite{Fasel2004EPJD,Tittel1998PRA,Tittel1999PRA,Tittel1998PRL,Fasel2005PRL}                   \\  \cline{2-3}
&Time-bin& \cite{Marcikic2002PRA,Riedmatten2002QIC} \\  \hline
LiIO$_3$ crystal& Energy-time     & \cite{Ou1990PRL}  \\  \hline
LiNbO$_3$ crystal &Time-bin &\cite{Brendel1999PRL} \\  \hline
PPKTP crystal& Energy-time     & \cite{Bulla2023PRA,Mann2023arXiv,Steinlechner2017NC,Shalm2013NP,Shi2006NJP,Bessire2014NJP,Xu2020OE,Ecker2019PRX,Bulla2023PRX,Tiranov2017PRA,Agne2017PRL,Ecker2021PRL}  \\  \hline
\multirow{2}{*}{PPLN crystal} & Energy-time& \cite{Rakonjac2021PRL,Achatz2023npj} \\  \cline{2-3}
&Time-bin& \cite{Martin2017PRL} \\  \hline
\multirow{2}{*}{PPKTP waveguide} & Energy-time& \cite{Clausen2011Nature,Zhong2012OE,Tiranov2015Optica,Xu2016Optica} \\  \cline{2-3}
&Time-bin& \cite{Ma2009OE} \\  \hline
\multirow{2}{*}{PPLN waveguide} & Energy-time& \cite{Thew2004QIC,Kaiser2016APL,Liu2023COL,Lefebvre2021JOSAB,Aktas2016LPR,Kaiser2017LSA,Shalm2013NP,Clausen2011Nature,Pomarico2009NJP,Huang2022npj,Zhang2021npj,Honjo2007OEwaveguide,Zhang2008OE,Tiranov2015Optica,Stefanov2003PRA,Huang2022PRAppl,Thew2004PRL,Agne2017PRL,Sanaka2001PRL,Zhao2020PRL,Vergyris2019QST,Xue2021PRAppl,Xia2023PRAppl} \\  \cline{2-3}
&Time-bin& \cite{Zhang2021npj,Honjo2007OE,Zhang2008OE-10GHz,Thew2002PRA,Martin2013PRA} \\  \hline
AlGaAs waveguide& Energy-time     & \cite{Autebert2016Optica} \\  \hline
Silicon nitride waveguide& Energy-time     & \cite{Dalidet2022OE} \\  \hline
PPLN microring resonator &Energy-time     & \cite{Ma2020PRL} 
\\ 
\hline \hline      
\end{tabular}
\end{table}

\begin{figure}[htbp]
\centering
\includegraphics[width= 0.92\textwidth]{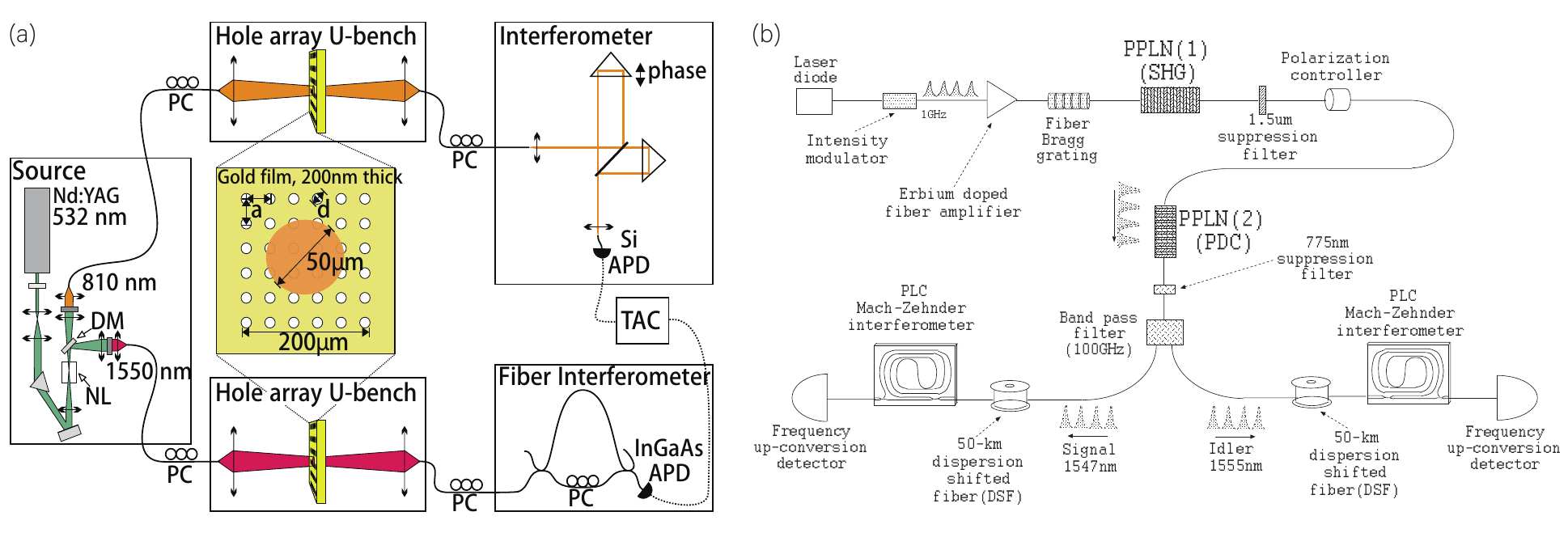}
\caption{Two experimental setups involving the observations of Franson interference of energy-time and time-bin entanglement generated via $\chi^{(2)}$ media. (a) The setups used to observe the survival of energy-time entanglement in plasmon-assisted light transmission. Reprinted from Ref.\,\cite{Fasel2005PRL}. (b) The setup used for distributing time-bin entanglement. Reprinted from Ref.\,\cite{Honjo2007OE}.
}
\label{SPDC}
\end{figure}

SPDC is one of the earliest processes used to generate correlated photons \cite{Brendel1991PRL}. In fact, energy-time or time-bin entangled biphoton states are superpositions of correlated biphoton states in a wide temporal continuous domain or two discrete time bins. Thus, it is natural to employ SPDC to create photon pairs for Franson interference. For crystals with $\chi^{(2)}$ nonlinearity, SPDC processes can occur when the phase matching condition is satisfied, which requires proper synergy between the refractive indices and the propagation directions of the pump field and the fields created. After energy-time entangled photon pairs around 700 nm were first generated by pumping bulk LiNbO$_3$ crystal with a laser of 351.1 nm \cite{Ou1990PRL}, several $\chi^{(2)}$ materials have been used to prepare biphoton states characterized by Franson interference. Over the last two decades, straight waveguides and microresonators made from $\chi^{(2)}$ materials have been increasingly employed to perform this task, thanks to advancements in nanofabrication technology. The main experimental results are summarized in Tab.\,\ref{Tab:Franson1}.

Figure\,\ref{SPDC} illustrates two experiments that utilize SPDC processes to generate energy-time and time-bin entangled photon pairs, respectively, and measure them using Franson interferometers \cite{Fasel2005PRL}. In Fig.\,\ref{SPDC}(a), a CW laser is used to pump a KNbO$_3$ nonlinear crystal, generating energy-time entangled photon pairs through a type-I SPDC process. The entangled photons have distinct wavelengths but possess identical polarization. One photon from the pair goes through a gold film with a hole array, which supports surface plasmon polaritons (SPPs). In that experiment, the gold film can be inserted in the path of either photon in each pair. After coupling to the SPPs on the gold film and transforming back into free-space propagating photons again, the photons enter a free-space or fiber-based unbalanced interferometer, and are subsequently detected by a detector. Franson interference is observed in the coincidences, and the results demonstrate that the visibility of the interference fringe is nearly reserved after the gold film is inserted into the propagating paths of the photons. This indicates that the photon-plasmon-photon transformation could be coherent and energy-time entanglement can survive during this transformation.

Figure\,\ref{SPDC}(b) illustrates a setup used to distribute time-bin entangled photon pairs over 100 km of fiber \cite{Honjo2007OE}, which are generated by the SPDC process in a PPLN waveguide. Paired laser pulses are employed to pump this process. The coherence between the pair of pump pulses leads to a coherent superposition between the two correlated biphotons located in the two time bins loaded with the pump pulses, i.e., resulting in the formation of time-bin entanglement. The time-bin entangled photons subsequently propagate through 50 km of dispersion-shifted fiber and are finally coupled into two UMZIs of a planar lightwave circuit (PLC) for the measurement of Franson interference. The arm length difference in the UMZIs is set to match the spacing between the two time bins. After the long-distance distribution, an interference fringe with a visibility of 81$\%$ is obtained, indicating the survival of time-bin entanglement.

In 2023, correlated photon pairs has been generated via the SPDC process in a novel 2D layered materials, niobium oxide dichloride (NbOCl2) \cite{Guo2023Nature}. In this work, the coincidence peaks have been observed for the generated photons. Considering the natural direct link between photon correlation and energy-time entanglement, we can expect the Franson interference of photons generated in this ultrathin quantum light source. With the increasing interests in enhancing second-order nonlinearity in low-dimensional materials \cite{Chen2017LSA,Wang2017NL}, it is expected that more  SPDC sources can be realized in ultrathin materials. The extreme relaxation of phase matching of SPDC in such materials could make the bandwidth of the generated photons much broader than that in bulk materials. This might shorten the length requirement of the path difference in interferometers for Franson interference. In this case, integrated interferometer would be more suitable because the uniformity of the path difference is relatively easy to realize in fabrication, while long path difference is difficult.

\subsubsection{SFWM in fibers and integrated optics}
\begin{table}[h]
\caption{The $\chi^{(3)}$ media for creating energy-time and time-bin entanglement}
\label{Tab:Franson2}
\begin{tabular}{l|l|l}
\hline \hline
Nonlinear medium  & Types of entanglement  &  References                                        \\ \hline  
\multirow{2}{*}{Silica optical fiber}& Energy-time &\cite{Dong2014OE,Dong2015SR}  \\  \cline{2-3} 
&Time-bin &\cite{Dong2017ACS,Takesue2006OE,Takesue2009OE,Agarwal2014PRX}\\ \hline 
\multirow{2}{*}{Silicon wire waveguide} &Energy-time &\cite{Liu2023PR,Yao2018PRA,Ren2023PhotoniX,Villegas2017Optica} \\  \cline{2-3} 
&Time-bin&\cite{Takesue2007APL,Li2017PRAppl,Fang2018OE,Harada2008OE} \\ \hline
Silicon photonic crystal waveguide &Time-bin &\cite{Takesue2014SR} \\ \hline
AlGaAs waveguide& Time-bin     & \cite{Chen2018APL} \\ \hline
\multirow{2}{*}{Silicon microring resonator}&Energy-time &\cite{Ma2018QST,Mittal2021NP,Oser2020npj,Kumar2015OE,Suo2015OE,Mazeas2016OE,Ma2017OE,Grassani2015Optica,Feng2023Optica} \\  \cline{2-3} 
& Time-bin&\cite{Wakabayashi2015OE} \\ \hline
\multirow{2}{*}{Silicon nitride microring resonator}&Energy-time &\cite{Wen2022PRA,Samara2021QST} \\  \cline{2-3}
&Time-bin&\cite{Samara2019OE,Zhang2018OL,Xiong2015Optica}\\ \hline
AlGaAs microring resonator& Energy-time &\cite{Steiner2021PRXQuantum} \\ \hline
InP microring resonator&Time-bin&\cite{Kumar2019APL} \\ \hline
Silicon carbide microring resonator&Energy-time &\cite{Anouar2023} \\ \hline
High refractive index glass&Time-bin&\cite{Reimer2018NP}\\ 
\hline \hline      
\end{tabular}
\end{table}

Four-wave mixing (FWM) is a typical third-order nonlinear optical phenomenon that finds wide applications in wavelength conversion \cite{Foster2006Nature}, ultrashort pulse generation \cite{Kippenberg2018Science}, microscopy \cite{Frischwasser2021NP}. In some applications, both pump and seeded fields are injected into the $\chi^{(3)}$ medium, where the FWM process coherently amplifies the seeded fields to a desired power by consuming the pump power \cite{Foster2006Nature}. In other applications, only the pump field is injected into a medium placed in a resonator \cite{Razzari2009NP}. The FWM process first amplifies the vacuum fluctuations, and the resulting creations, acting as seeds, are further coherently amplified to a significant power level via gain and feedback in the resonator. In these classical applications, the created optical fields are generated by the stimulation and coherent amplification of the seeded fields, which results in considerable power outputs. In 2002, an experiment was conducted for the first time to investigate a completely spontaneous FWM process, where the generated fields arise solely from the parametric amplification of vacuum fluctuations, resulting in power at the single-photon level \cite{Fiorentino2002PTL}. This process, later known as spontaneous FWM (SFWM), exhibits significant photon correlations. Subsequently, numerous studies have been conducted to create correlated or entangled photons through the SFWM process, utilizing nonresonant $\chi^{(3)}$ nonlinearities in silica fiber \cite{Li2005PRL}, silicon waveguides \cite{Du2007PRL}, silicon nitride waveguides \cite{Sharping2006OE}, and resonant $\chi^{(3)}$ in atomic \cite{Zhang2016JO} or molecular ensembles \cite{Shi2017NC}. Advancements in integrated optics have enabled SFWM on a chip, making it an effective approach for generating non-classical optical fields. Tab.\,\ref{Tab:Franson2} presents some prominent non-resonant $\chi^{(3)}$ materials or structures used for generating energy-time and time-bin entanglement.

\begin{figure}[htbp]
\centering
\includegraphics[width= 0.92\textwidth]{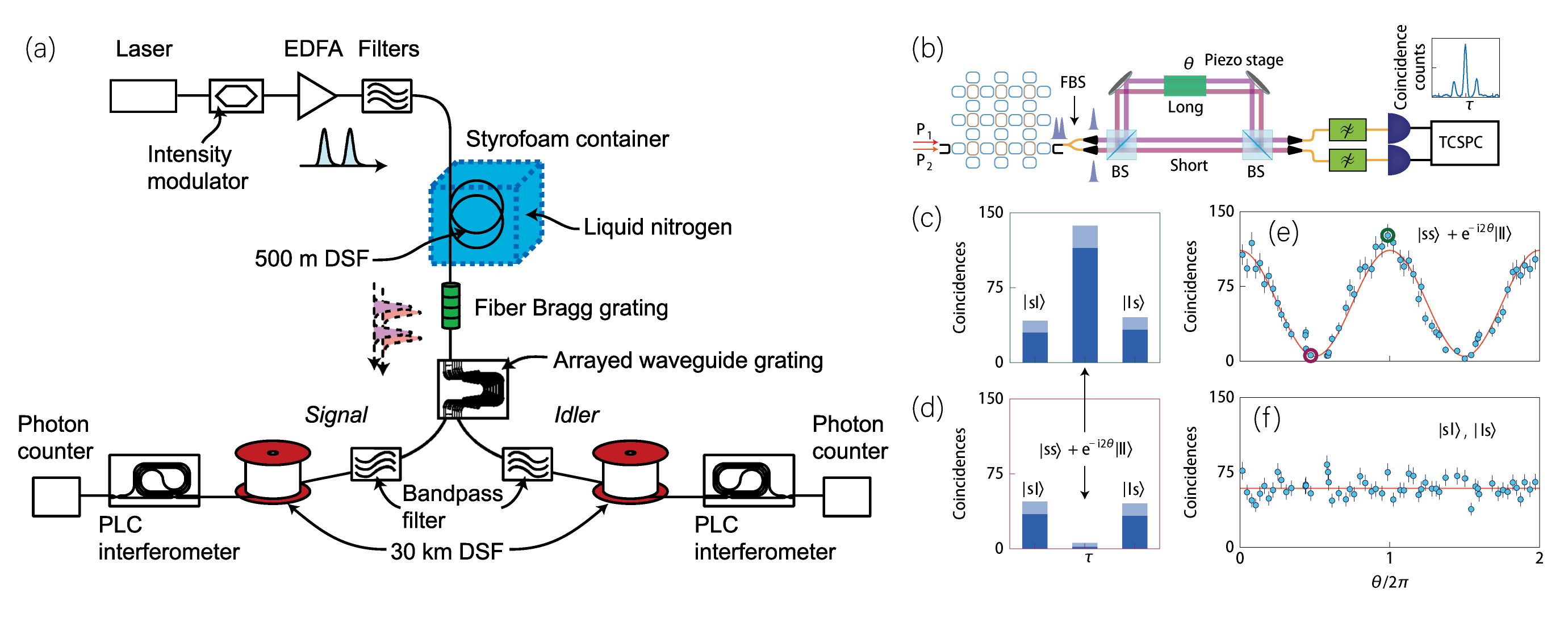}
\caption{Two experimental setups involving the generation of time-bin and energy-time entangled photon pairs by SFWM in $\chi^{(3)}$ media. (a) Experimental setup for generating time-bin entanglement in a silicon wire waveguide. Reprinted from Ref.\,\cite{Takesue2006OE}. (b) Franson interferometer is used to measure the energy-time entanglement of photon pairs generated in a topological system comprising a SOI-based two-dimensional array of resonators. (c) and (d) are the measured coincidence histograms with three-peak structure, corresponding to the constructive and destructive cases of Franson interference, respectively. (e) The measured Franson interference of the energy-time entangled photons. (f) The measured side peak height in (c) versus phase. Reprinted from Ref.\,\cite{Mittal2021NP}. 
}
\label{Fiberandintegratedoptics}
\end{figure}

The first generation of energy-time entangled photons in optical fiber was carried out by Dong et al. in 2015 \cite{Dong2015SR}. The researchers cooled a piece of commercial DSF in a liquid nitrogen bath to suppress the Raman noise, following the approach described in Ref.\,\cite{Takesue2006OE}. Figure\,\ref{Fiberandintegratedoptics}(a) depicts the experimental setup used by Takesue to prepare time-bin entangled photon pairs in cooled DSF and distribute them over 60 km of optical fiber \cite{Takesue2006OE}. In Fig.\,\ref{Fiberandintegratedoptics}(a), paired laser pulses modulated from a CW laser were used to pump the SFWM in DSF, generating time-bin entangled photon pairs. The signal and idler photons are both transmitted over 30 km DSFs, and finally input into two UMZIs with PLC technologies. The Franson interference was measured by the coincident detections of photons output from the two UMZIs. A visibility of 75.8$\%$ was obtained in the interference fringe, indicating the successful distribution of the time-bin entanglement generated from DSF over a long-distance fiber.

In 2007, Takesue et al. created time-bin entangled photon pairs via SFWM in a silicon wire waveguide \cite{Takesue2007APL}. The setup is similar to that in Fig.\,\ref{Fiberandintegratedoptics}(a), except that the cooled DSF is replaced by a silicon wire waveguide at room temperature. Since then, several studies have been conducted to utilize integrated silicon photonics for generating entangled photons suitable for Franson interference. A recent work, depicted in Fig.\,\ref{Fiberandintegratedoptics}(b), achieved energy-time entanglement using a topological system comprising a two-dimensional array of resonators fabricated on a silicon-on-insulator (SOI) platform \cite{Mittal2021NP}. In this configuration, the non-degenerate pumped SFWM produces correlated photons in topological edge modes, and the spectral correlations are topologically protected from fabrication disorders. The output signal and idler photons are separated and injected into a Franson interferometer. The constructive and destructive interference of the middle peak in the coincidence histogram between signal and idler photons is shown in Fig.\,\ref{Fiberandintegratedoptics}(c) and (d), respectively. In Fig.\,\ref{Fiberandintegratedoptics}(e), a complete Franson interference fringe is shown by post-selecting the middle peak, while Fig.\,\ref{Fiberandintegratedoptics}(f) shows constant values for the two side peaks. 
It is well known that the third-order nonlinearity is much weaker than the second-order one. However, there are massive instances employing SFWM in generating photon pairs. One reason is that the wavelengths involved in the system can all be in the telecom band, which is easy to manage. The other reason is that CMOS-compatible manufacturing technology can be used to fabricate integrated devices used in third-order sources. The latter advantage is also beneficial to the fabrication of Franson inteferometers integrated in complex quantum photonic circuit \cite{Kim2008OL}.

\subsubsection{SFWM in warm atomic ensembles}

\begin{figure}[htbp]
\centering
\includegraphics[width= 0.92\textwidth]{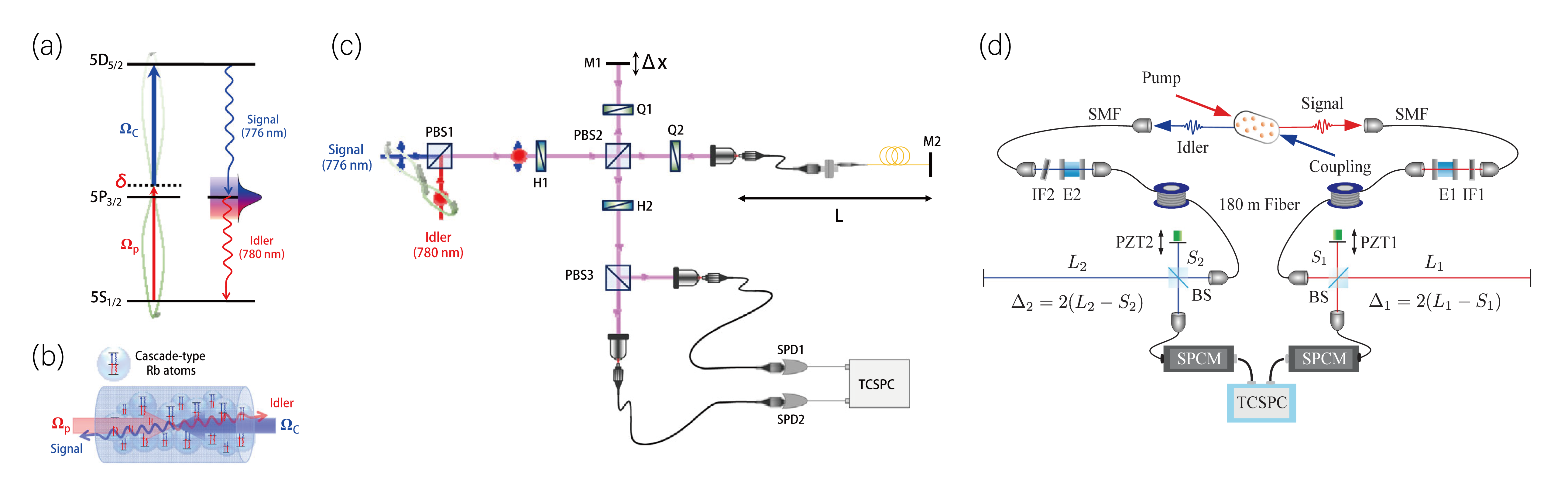}
\caption{The schematics and experimental setups used to generate energy-time entanglement via SFWM in warm atomic ensembles and implement the Franson interference. (a) The energy-level diagram for SFWM. (b) The geometric configuration of the beams in SFWM. (c) The setup used for observing the Franson interference of energy-time entangled photons. Reprinted from Ref.\,\cite{Park2018PRL}. (d) The setup used for implementing nonlocal Franson interference. Reprinted from Ref.\,\cite{Lee2019PRA}.
}
\label{atom}
\end{figure}
As mentioned in the previous section, SFWM can occur in an atomic ensemble where $\chi^{(3)}$ could be enhanced by atomic resonance. The first creation of entangled photons in an atomic gas, however, did not utilize this mechanism but the cascaded emission \cite{Freedman1972PRL}. In 2003, Kaindl et al. successfully employed SFWM in cold Cs atoms to generate correlated photon pairs \cite{Kuzmich2003Nature}. In 2018, Park et al. generated energy-time entanglement through the SFWM process in a three-level ladder-type system using warm $^{87}$Rb atoms \cite{Park2018PRL}. The energy level system and the optical fields involved in the transitions are depicted in Fig.\,\ref{atom}(a), while the geometric configuration of the optical fields is shown in Fig.\,\ref{atom}(b). In this system, the annihilation of a pump photon and a coupling photon can lead to the creation of a pair of correlated signal and idler photons. However, the intrinsic ratio between the two-photon coherence time and the single-photon coherence time is not sufficient to suppress the single-photon interference in Franson interference. The effect of collective two-photon coherence is introduced to overcome this problem, which can enlarge that ratio by shortening the single-photon coherence time. As shown in Fig.\,\ref{atom}(a), the signal and idler photons have opposite propagating directions and orthogonal polarizations. Thus, the Franson interferometer in Fig.\,\ref{atom}(b) utilizes the polarization degree of freedom to split and combine beams, and a common interferometer is used for both signal and idler photons to make the entire setup compact. A piece of fiber is inserted into the long arm of the interferometer to ensure the elimination of single-photon interference. With this method, a Franson interference visibility of 97 $\%$ is obtained, confirming the successful creation of energy-time entanglement in a warm atomic ensemble.

In 2019, with the same entanglement generation scheme in Fig.\,\ref{atom}(a), but with separate interferometers for the signal and idler photons, the nonlocal Franson interference was achieved for photon pairs emitted in a ladder-type atomic system \cite{Lee2019PRA}, as originally proposed by Franson \cite{Franson1989PRL}. The setup in Ref.\,\cite{Lee2019PRA} is depicted in Fig.\,\ref{atom}(d), where two spools of fibers are placed in the pathways of the signal and idler photons to ensure the nonlocality of interference. Unlike the configurations in Fig.\,\ref{atom}(b), the Franson interferometer in Fig.\,\ref{atom}(d) does not split and combine optical fields based on polarization but directly manipulates the spatial pathways. One advantage of the energy-time entangled photons generated from atomic ensemble is the narrow bandwidth which matches with the storage bandwidth of atomic ensemble-based quantum memory \cite{Lee2019PRA}. The interferometers used for this type of photons should have a long path difference, and therefore the requirement on the stability of the interferometer is stringent.

\subsubsection{Photons from quantum dots}

Quantum dots (QDs) are artificial atoms that play important roles in quantum communication \cite{Vajner2022AQT} and quantum information processing \cite{Zhang2018NSR}. In addition to their widespread use in the generation of single photons, increasing attention is paid to the utilization of QDs to create entangled photons \cite{Dousse2010Nature,Wang2019PRL}. The main mechanism used for this purpose is biexciton-exciton cascaded emission, which can avoid the generation of multiple photon pairs in SPDC or SFWM when appropriate energy levels structures and pump configurations are used \cite{Huber2018JO}. Several studies have been conducted to generate polarization-entangled photons in QDs, but the presence of fine structure splitting (FSS) in QDs can degrade polarization entanglement \cite{Hafenbrak2007NJP}. In the past decade, efforts have also been made to generate time-bin or energy-time entanglement in QDs. The motivation behind these works is not only for practical long-distance quantum communication but also based on the fact that these two types of entanglement are insensitive to the FSS.

\begin{figure}[htbp]
\centering
\includegraphics[width= 0.92\textwidth]{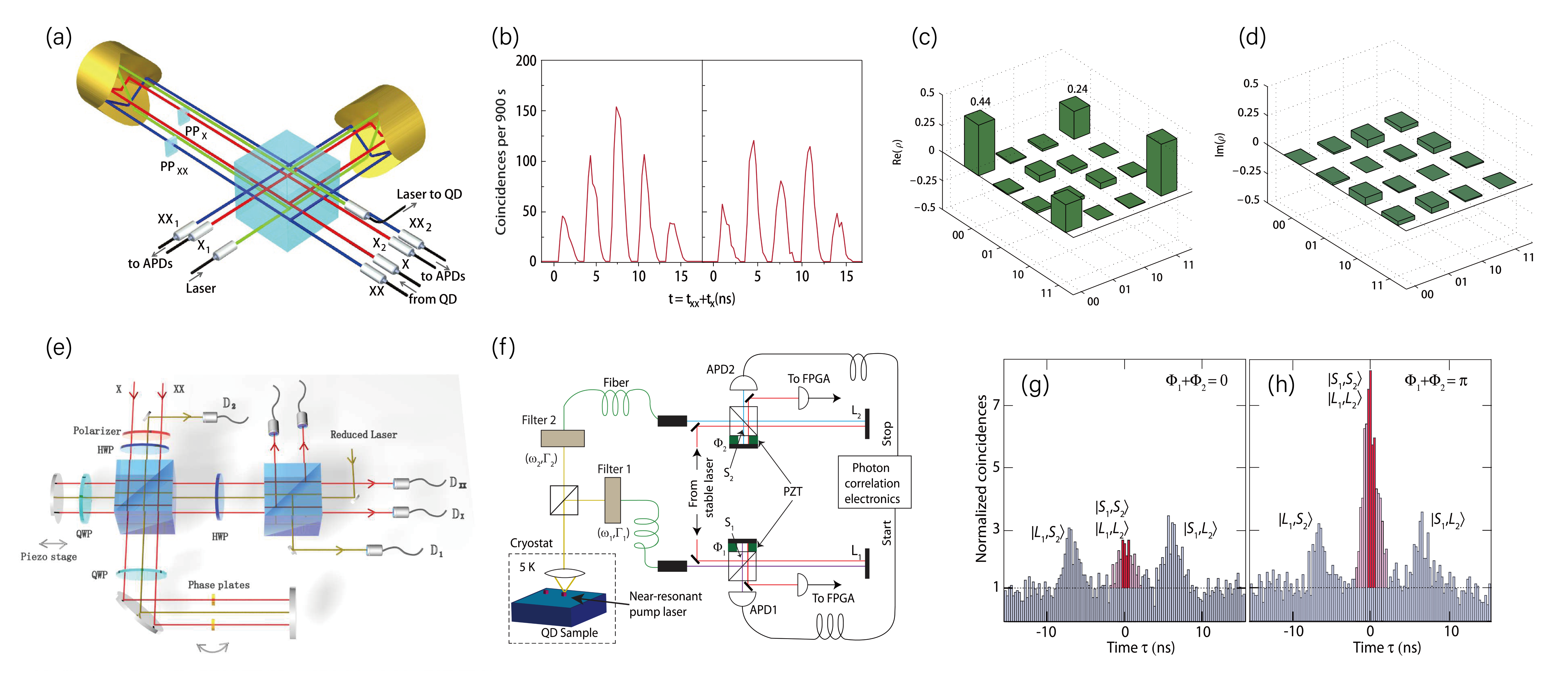}
\caption{The experiment setups used for generating and characterizing the time-bin or energy-time entangled photons from QDs. (a) A single bulk interferometer is used to prepare paired pumping pulses and to implement Franson interference. (b) Triplet coincidences of XX photon, X photon, and a synchronous signal, corresponding to the constructive (left panel) and destructive (right panel) cases in Franson interference. (c) and (d) are the real and imaginary parts of the reconstructed density matrix via quantum state tomography with a Franson interferometer. Reprinted from Ref.\,\cite{Jayakumar2014NC}. (e) A Franson interferometer used to measure the inhomogeneous broadening of a biexciton state by Franson interference of cascaded emission in QDs. Reprinted from Ref.\,\cite{Sun2017OE}. (f) Experimental setup for generating photon pairs in a two-level QD and observing the Franson interference. (g) and (h) are the coincidence histograms with three-peak structures corresponding to the destructive and constructive Franson interference, respectively. Reprinted from Ref.\,\cite{Peiris2017PRL}.}
\label{quantumdotsandbiexciton}
\end{figure}

In 2014, Jayakumar et al. first prepared time-bin entangled photons in quantum dots \cite{Jayakumar2014NC}. In this experiment, a doubly pulsed laser was used to pump self-assembled InAs QDs embedded in a distributed Bragg reflector microcavity. The pump laser excited the QDs to the biexciton state via two-photon excitation. The subsequent cascaded decays from the biexciton and exciton states emitted the XX and X photons with different wavelengths, respectively. The XX and X photons were then separated by wavelength and analyzed by an unbalanced interferometer. In Ref.\,\cite{Jayakumar2014NC}, a folded bulk interferometer was used to construct the double pumping pulses and observe the Franson interference between XX and X photons, as shown in Fig.\,\ref{quantumdotsandbiexciton}(a). Figure\,\ref{quantumdotsandbiexciton}(b) shows the three-fold coincidence histogram among XX, X photons and a pump laser trigger. It is evident that five peaks are observed, resulting from different combinations of early (late) time-bin and long (short) transmission arms of XX and X photons \cite{Jayakumar2014NC}. Here, the height of the central peak changes significantly when the phase in the interferometer is changed, indicating that the Franson interference can be observed by scanning the phase. Using this setup, time-bin tomography of the two-photon state of XX and X photons was implemented, and the results of the state ${\left| {\Phi^+} \right\rangle}$ are shown in Fig.\,\ref{quantumdotsandbiexciton}(c) and (d) with a fidelity of 69$\%$. In 2015, Versteegh et al. proposed and performed an experimental scheme to eliminate the residual probability of emission of two pairs \cite{Versteegh2015PRA}. In this case, a single pulse train was used to pump the QDs and generate polarization-entangled photons. An UMZI with a PBS was used to transform the orthogonal polarizations into different time bins. Subsequently, all the photons passed through a 45° polarizer to erase the polarization distinguishability and output time-bin entangled photons, which were then measured by a Franson interferometer established with a common UMZI for XX and X photons. In 2016, using the same interferometer shown in Fig.\,\ref{quantumdotsandbiexciton}(a), Huber et al. studied the relationship between the degree of time-bin entanglement and the pump pulse in QDs \cite{Huber2016PRB}.

In 2017, Sun et al. measured the inhomogeneous broadening of a biexciton state in an InAs QD using Franson interference. The Franson interferometer used in this work is shown in Fig.\,\ref{quantumdotsandbiexciton}(e), where the XX and X photons generated in biexciton and exciton decays pass through the same unbalanced interferometer, but are completely isolated in spatial pathways. When the difference between the time delays in the long and short arms of the interferometer increases, the visibility of the Franson interference fringe significantly decreases. By exponentially fitting this reduction, we can obtain the total broadening of the biexciton state, which is further compared with the homogeneous broadening measured by correlation to calculate the inhomogeneous broadening. It should be noted that a visibility of approximately 35$\%$ is obtained with a time delay difference of 7 ns, which is attributed to the incoherent excitation process. In the same year, Peiris et al. observed a Franson interference with visibility exceeding 66$\%$ for photons from a two-level QD under coherent excitation \cite{Peiris2017PRL}. The experimental setup is shown in Fig.\,\ref{quantumdotsandbiexciton}(f), in which a near-resonant pump laser is used to excite the QDs from the ground state to the neutral exciton state and resonance fluorescence is collected in a perpendicular direction. After being split into two equal parts, the resonance fluorescence photons in the two parts are filtered and input into two unbalanced interferometers to observe the nonlocal Franson interference. Figure\,\ref{quantumdotsandbiexciton}(g) and (h) show the measured destructive and constructive interference, respectively, in the central peak in the coincidence histogram. The measured Franson interference visibility reaches 66$\%$, which is higher than the classical limit of 50$\%$ and approaches the limit to violate the Bell inequality. In 2023, Hohn et al. improved the visibility of the Franson interference fringe to (73 $\pm$ 2)$\%$ for a resonantly driven biexciton and exciton cascaded emission, thus providing the first verification of the generation of energy-time entanglement in QDs \cite{Hohn2023PRR}. QD-based photon pair sources could realize deterministic emission with high brightness \cite{Liu2019NN}. In preparing practical time-bin entanglement, the realization of this advantage needs pulsed pumping lasers with high repetition rate, and therefore the shorter path difference in Franson interferomter is preferred, which is suitable for integrated realizations \cite{Zhang2021PRL}.

\subsection{Applications of Franson interferometer}

\subsubsection{Characterizing energy-time entanglement and time-bin entanglement }
\begin{figure}[htbp]
\centering
\includegraphics[width= 0.92\textwidth]{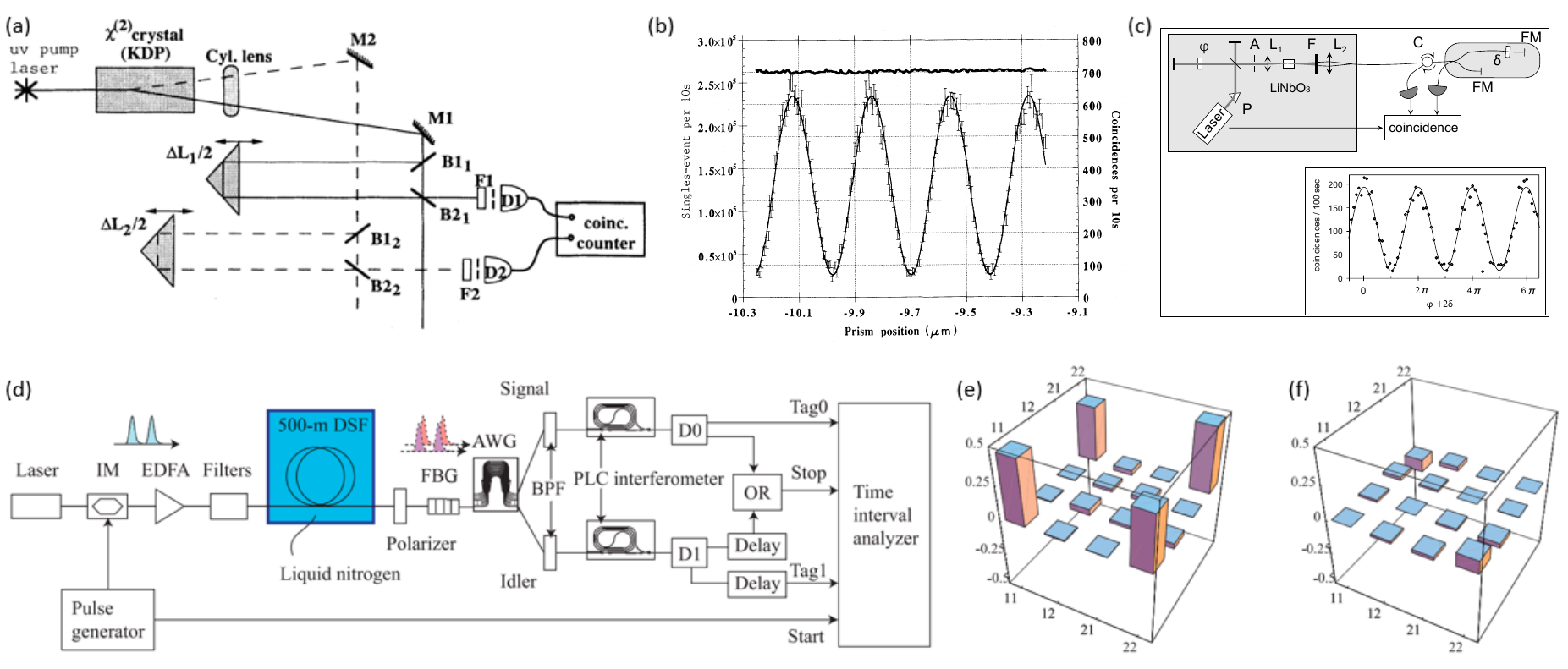}
\caption{The setups and results of three experiments involving the characterization of energy-time or time-bin entanglement. (a) The first experiment demonstrating the nonlocal interference of energy-time entangled states. (b) Measured single counts and coincidence counts in (a). Reprinted from Ref.\,\cite{Kwiat1993PRA}. (c) The first experiment setup used for generating and characterizing time-bin entanglement, with the inset showing a measured Franson interference fringe. Reprinted from Ref.\,\cite{Brendel1991PRL}. (d) Experimental setup for QST of time-bin entangled photons by using a Franson interferometer. (e) and (f) are the real and imaginary parts of the reconstructed density matrix via QST, respectively. Reprinted from Ref.\,\cite{Takesue2009OE}. 
}
\label{Franson/E-TandTime-bin}
\end{figure}
The original motivation for proposing the Franson interferometer in 1989 was to test the Bell inequality \cite{Franson1989PRL}. Similarly to the scheme demonstrated in the earliest test of CHSH equality, where the polarizations of the entangled photons are measured \cite{Freedman1972PRL}, Franson's scheme involves measuring the arrival times of photons. As discussed in the previous section, in the original scheme, the two photons are generated by cascaded emission in a ladder-type three-level atomic system, with the lifetime of the top (intermediate) level being much longer (shorter) than the time-delay difference in the UMZIs. Later studies demonstrated that two photons generated in SPDC can also be used to observe Franson interference \cite{Horne1989PRL,Ou1990PRL,Brendel1991PRL,Kwiat1990PRA}. In both methods, the generation times of the two photons are nearly the same with some uncertainty, and their frequencies are constrained by the energy conversation. Thus, this type of two-photon state used to test Bell inequality can be referred to as an energy-time entangled state \cite{Strekalov1996PRA}, although this term was not introduced until several years after Franson's scheme with significant theoretical and experimental progress in this field. In some early experiments, the visibilities of the interference fringe were limited to being less than 50$\%$ due to the low temporal resolution in coincidence measurements \cite{Ou1990PRL,Kwiat1990PRA}. By improving this resolution, temporal post-selection can be achieved, and visibility higher than 71$\%$ has been observed, verifying the existence of energy-time entanglement and violating the Bell inequality \cite{Brendel1991PRL,Kwiat1993PRA}. Subsequently, numerous studies have been conducted to create energy-time entangled states characterized by different types of Franson interferometers. Currently, the visibility of Franson interference has improved to be higher than 99$\%$ \cite{Park2019OL}.

The nonlocal Franson interference of energy-time entangled photons was first achieved by Kwait et al. in 1993, with the setup shown in Fig.\,\ref{Franson/E-TandTime-bin}(a) and the corresponding Franson interference fringe and interference-free single counts shown in Fig.\,\ref{Franson/E-TandTime-bin}(b) \cite{Kwiat1993PRA}. In 1999, time-bin entanglement was proposed and realized as a method to avoid decoherence in polarization-encoded quantum communication over long-distance fiber \cite{Brendel1999PRL}. Figure\,\ref{Franson/E-TandTime-bin}(c) shows the experimental setup used in this work, where a fiber-based single Michelson configuration is used as the Franson interferometer. A notable difference between the observations of Franson interference in CW laser-pumped energy-time entanglement and pulsed laser-pumped time-bin entanglement is that only the two-fold coincidence counts are measured in the former, while an additional synchronous signal is required for the latter in coincidence measurement. In time-bin entanglement, if only the two-fold coincidence counts are measured, a three-peak histogram will be obtained in which the central peak is composed of four types of events: (i) two photons generated in the early bin both going through short arms; (ii) two photons generated in the late bin both going through long arms; (iii) two photons generated in the early bin both going through long arms; (iv) two photons generated in the late bin both going through short arms. Among these cases, only the third and fourth cases are indistinguishable in any degree of freedom and are able to interfere with each other. The maximum visibility is 50$\%$ in the interference of two-fold coincidence counts. However, by introducing a synchronous signal, the central peak in two-fold coincidence counts can be split into three peaks. The two side peaks correspond to the first and second cases mentioned above, while the middle peak results from the indistinguishable third and fourth cases. In this way, the upper bound of the visibility of Franson interference is increased to 100$\%$. An interference fringe measured in the Ref.\,\cite{Brendel1999PRL} is also shown in Fig.\,\ref{Franson/E-TandTime-bin}(c).

Quantum state tomography (QST) provides a more direct description of quantum states and has been successfully applied to various types of entangled states \cite{White1999PRL,Peters2004PRL}. In 2009, Takesue and Noguchi first demonstrated the QST of a time-bin entangled two-photon state by using the Franson interferometer \cite{Takesue2009OE}. The experimental setup and the real and imaginary parts of the reconstructed density matrix in this work are shown in Fig.\,\ref{Franson/E-TandTime-bin}(d), (e) and (f), respectively. This method has been utilized to evaluate the performance of time-bin entanglement sources \cite{Jayakumar2014NC,Chen2018APL}, and has recently been extended to realize the QST of high-dimensional time-bin entangled photon pairs \cite{Ikuta2017NJP}.

\subsubsection{Testing fundamental physics principles}
\begin{figure}[htbp]
\centering
\includegraphics[width= 0.92\textwidth]{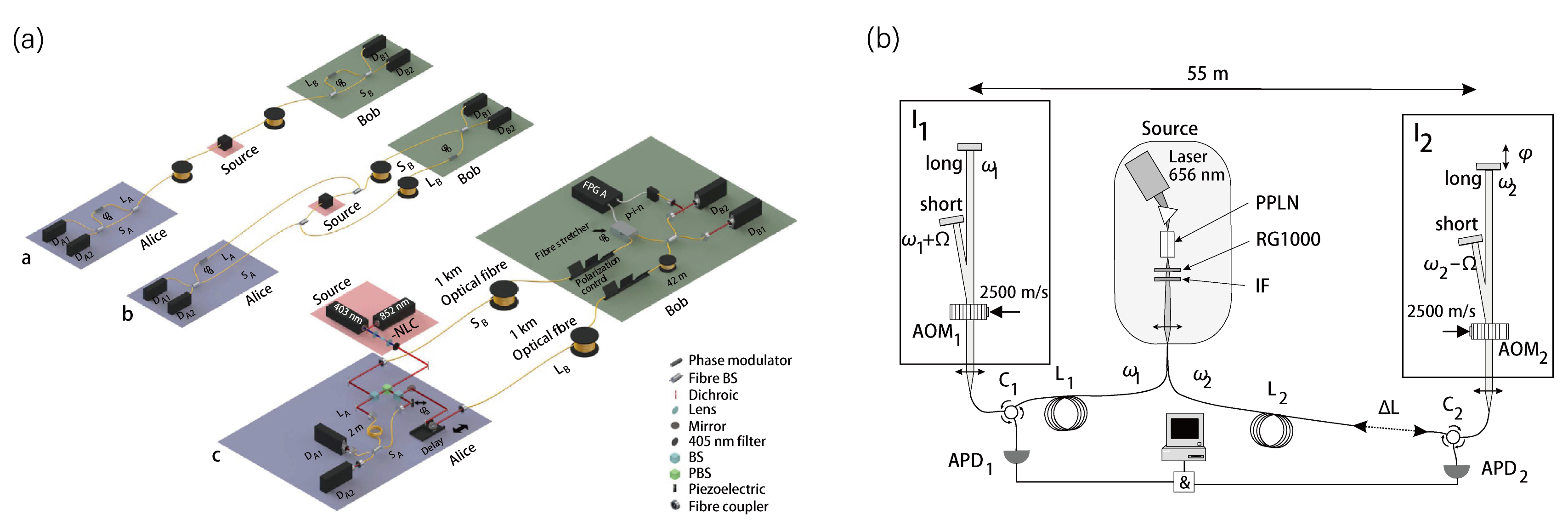}
\caption{Two schematics employing Franson interferometer in testing fundamental physics principle. (a) The evolution of the Franson interferometer from the original configuration (left one) to the hug configuration (middle one) and finally to a real experiment setup (right one) of fiber-based hug interferometer. Reprinted from Ref.\,\cite{Cuevas2013NC}. (b) Experiment setup used to test the multisimultaneity model. Reprinted from Ref.\,\cite{Stefanov2002PRL}.
}
\label{Belltest1}
\end{figure}
The test of Bell inequality is an enduring issue in the fundamental principles of quantum physics. Since the Franson interferometer was proposed to test Bell inequality \cite{Franson1989PRL}, numerous experiments have been conducted to realize this scheme, and most of these experiments have claimed the violation of Bell inequality \cite{Kwiat1993PRA,Tittel1999PRA}. However, in 1999, Aerts et al. demonstrated that the interference fringe obtained with the original Franson interferometer can also be explained by a local hidden variable model due to the necessary postselection, which would introduce a loophole \cite{Aerts1999PRL}. This loophole was later removed by Cabello et al., who proposed a modified Franson interferometer called the hug configuration. This configuration can eliminate the requirement of postselection. The relevant progress on the experiments of the hug configuration has been summarized in Sec.\,\ref{IV.B.2}, and the evolution from the original Franson interferometer to the hug configuration is illustrated in Fig.\,\ref{Belltest1}(a). In 2014, Jogenfors and Larsson theoretically revealed a general but quite complicated relationship among the original Franson interference experiment of energy-time entanglement, elements of reality, and local realism \cite{Jogenfors2014JPA}.

The Franson interferometer has also been employed in testing the multi-simultaneity model in relativistic quantum physics. This model predicts that the Franson interference fringes would disappear when the two UMZIs have relative motion, causing one interferometer to first select the output of the photons in its own inertial reference frame. In 2002, Stefanov et al. have experimentally investigated this prediction by using acousto-optic modulators in Franson interferometer to construct moving beam splitters \cite{Stefanov2002PRL}. The experimental setup is depicted in Fig.\,\ref{Belltest1}(b). However, the results of this study did not show any disappearance of correlations. In 2020, Rivera-Tapia et al. put forward an intriguing proposal to use the Franson interferometer to investigate the interaction between quantum systems and gravity in the weak-field regime \cite{Tapia2020CQG}. In this scheme, large-scale Franson interferometric arrays that can be set at low Earth orbit are employed, and the calculations based on both quantum mechanics and gravity theory suggest that the gravitational time delay in a weak gravitational field could entangle the photon pairs passing through the interferometers. The original Franson interferometer and the hug configuration are considered in this scheme, and the results indicate that the CHSH inequality could be violated in both cases.

\subsubsection{Quantum cryptography}

\begin{figure}[!h]
\centering
\includegraphics[width= 0.90\textwidth]{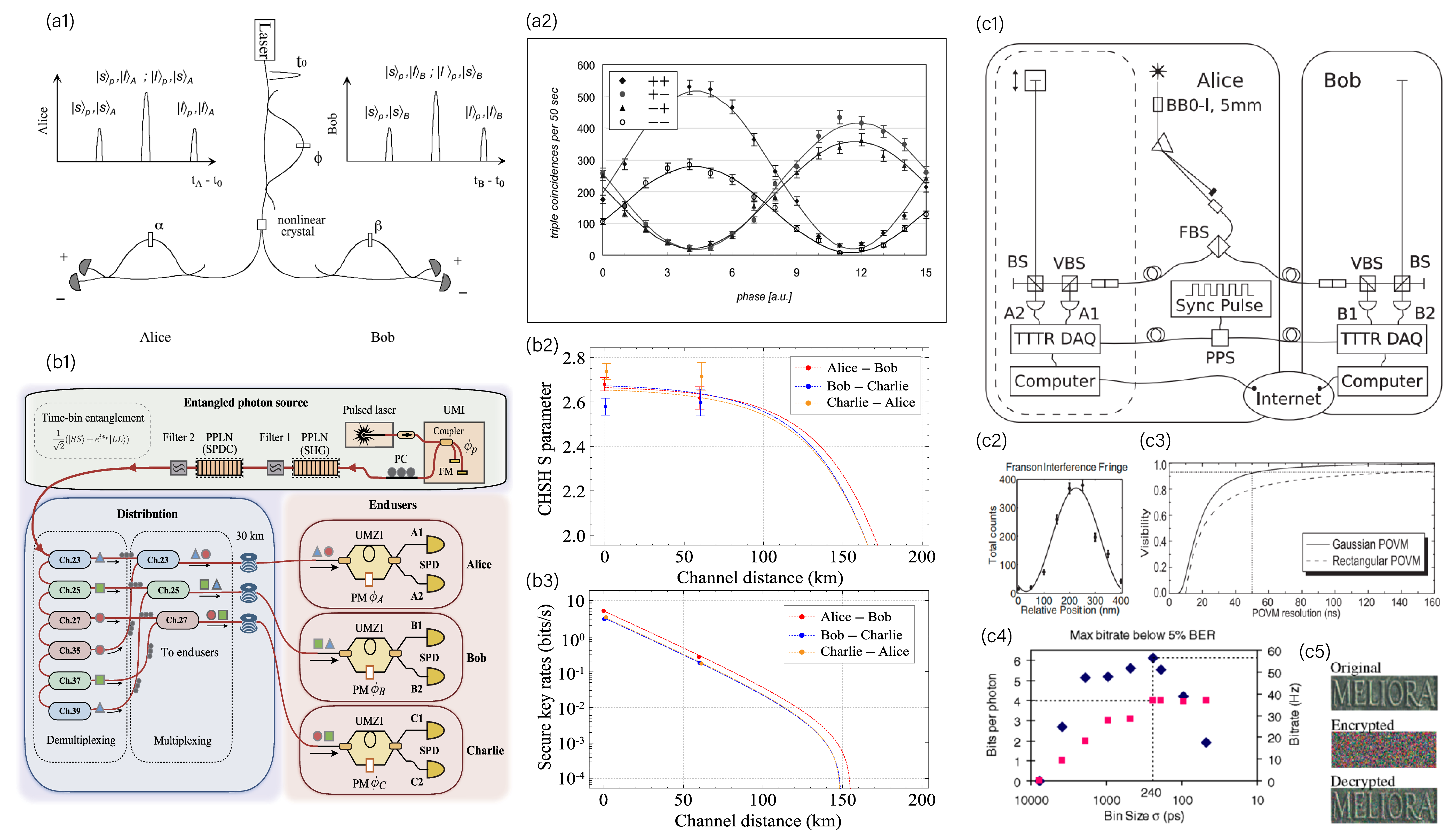}
\caption{Two experiment schematic or setup for time-bin entanglement-based QKDs and the typical results. (a1) The first schematic for time-bin entanglement-based QKD. (a2) Results of the measurement in the energy basis. Reprinted from Ref.\,\cite{Tittel2000PRL}. (b1) Experiment setup for time-bin entanglement-based QKD in quantum network with three users. For the system in (b1), (b2) and (b3) show the S parameter characterizing the violations of the CHSH inequality and the secure key rates under different channel distance, respectively. Reprinted from Ref.\,\cite{Kim2022APL}. 
The experiment setup and main results of high-dimensional QKD using energy-time entanglement. (c1) Experiment setup. (c2) Section of the measured Franson interference fringe in (c1). (c3) Calculated eavesdropper's resolution of POVM against the visibility of Franson interference. (c4) The optimal bit rate and information per photon pair versus bin size in large-alphabet protocol. (c5) Demonstration of cryptography for an image using an optimal key. Reprinted from Ref.\,\cite{AliKhan2007PRL}.
}
\label{Franson-QKD}
\end{figure}

Quantum key distribution (QKD) holds the potential for achieving unconditional security in the sharing of cryptographic keys among remote users \cite{Gisin2002RMP,Lo2014NP}. As fundamental resources, quantum entanglement can enable QKD in trusted node-free networks and over long-distances \cite{Jennewein2000PRL,Wengerowsky2018Nature}. Energy-time and time-bin entanglement-based QKDs have been experimentally demonstrated, and Franson interferometers have played a crucial role in security analysis in these implementations.

For time-bin entanglement-based QKD, Tittel et al. first proposed a scheme in 2000 and experimentally realized it. Schematics is shown in Fig.\,\ref{Franson-QKD}(a1), while the visibility of the Franson interference fringe is shown in Fig.\,\ref{Franson-QKD}(a2). In 2023, Kim et al. experimentally demonstrated multi-user QKD based on broadband time-bin entanglement and wavelength-division multiplexing technology, with the experiment setup shown in Fig.\,\ref{Franson-QKD}(b1). In this setup, each of the three users had an UMZI to perform the security test. The measured violations of the CHSH inequality and the secure key rates are depicted in Figs.\,\ref{Franson-QKD}(b2) and (b3). In 2000, Ribordy et al. established a system for energy-time entanglement-based QKD, in which the Franson interferometer is composed of a free-space UMZI for 810 nm photon and a fiber UMZI for 1550 nm photon \cite{Ribordy2000PRA}. In 2007, Ali-Khan et al. proposed and experimentally realized the first high-dimensional QKD scheme with energy-time entanglement \cite{AliKhan2007PRL}. The setup is shown in Fig.\,\ref{Franson-QKD}(c1), in which the measurements in time-basis and energy-basis are spatially separated. Figures\,\ref{Franson-QKD}(c2) and (c3) show the measured Franson interference fringe to evaluate security performance and the calculated relationship between Franson interference visibility and the eavesdropper's resolution of positive operator value measurement (POVM). In this high-dimensional QKD protocol, more than one bit could be encoded onto one coincidence under proper conditions, and the broad temporal distribution of photon in energy-time entanglement facilitated the construction of high-dimensional states. Figure\,\ref{Franson-QKD}(c4) and (c5) present the main performance of the experiment and a demonstration of image encryption/decryption in Ref.\,\cite{AliKhan2007PRL}, respectively. In 2013, Brougham et al. theoretically analyzed the use of the Franson interferometer in securing high-dimensional QKD, and found that this method is not always reliable, particularly against attacks that localize the photons to several separated times \cite{Brougham2013JPB}. In 2014, Zhang et al. proposed a method to ensure unconditional security in high-dimensional QKD with energy-time entanglement, using the visibilities from both the Franson and conjugate-Franson interferometers \cite{Zhang2014PRL}. The details of this proposal has been introduced in Sec.\,\ref{IV.B.3}. In 2015, Jorgenfors et al. demonstrated experimentally that the security test with a purely original Franson interferometer could be hacked by tailored pulses of classical light when standard avalanche photodetectors are utilized \cite{Jogenfors2015SA}. They also provided several methods to reestablish the security. In 2015, Zhong et al. achieved a photon-efficient QKD with the secure photon information efficiency reaching 8.7 bits per coincidence by introducing nonlocal dispersion cancellation into the Franson interferometer to increase the visibility of the interference fringe \cite{Zhong2015NJP}.

 \subsubsection{Entanglement-based quantum network}
 \begin{figure}[htbp]
\centering
\includegraphics[width= 0.92\textwidth]{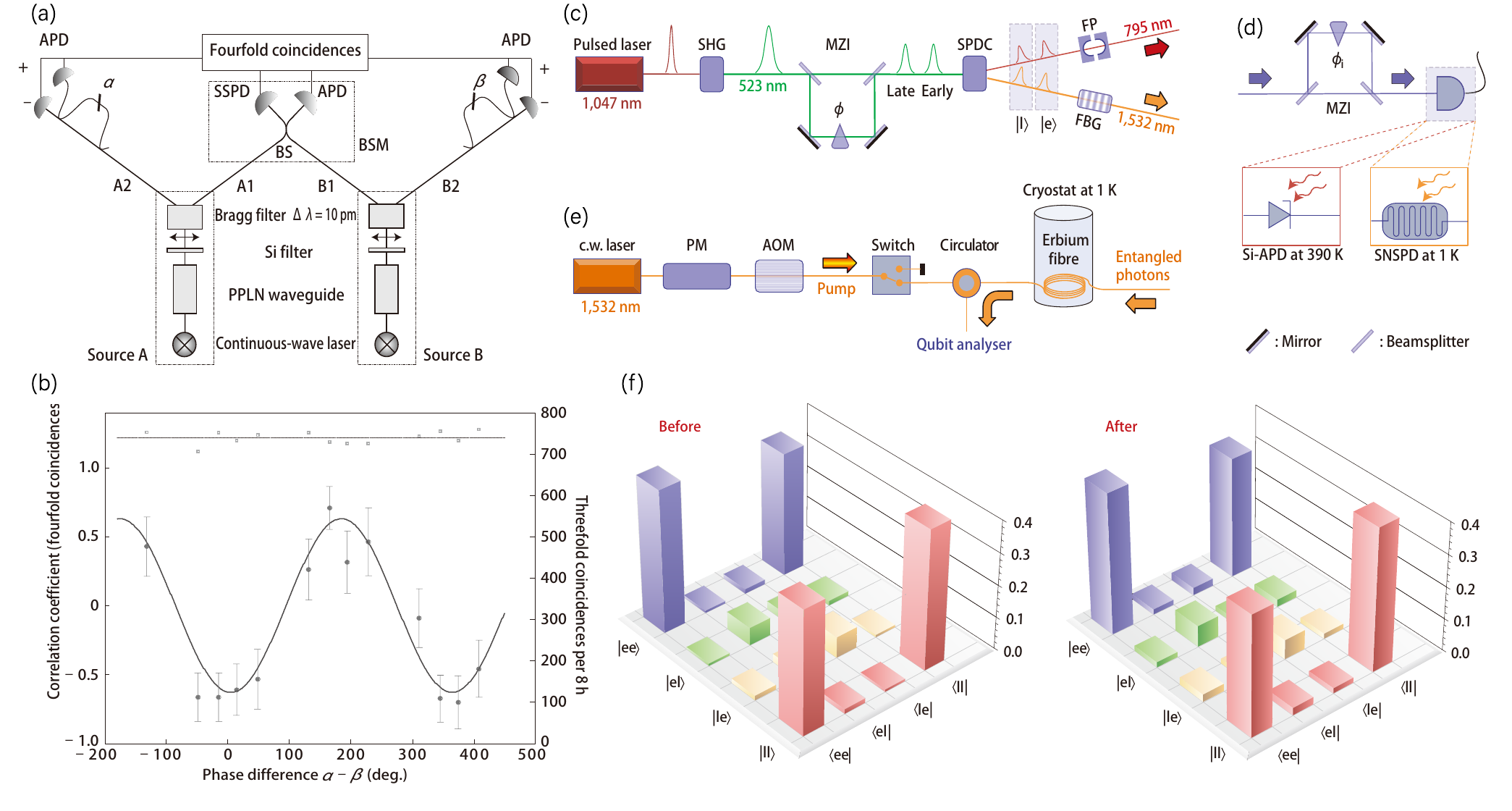}
\caption{ The experiment setups and results of two applications of Franson interferometer in entanglement-based quantum network. (a) Experiment setup for energy-time entanglement swapping. (b) Measured single counts and Franson interference fringe of the two-photon state after entanglement swapping. Reprinted from Ref.\,\cite{Halder2007NP}. The experiment setups for (c) generating time-bin entangled photons, (d) analyzing time-bin qubits, and (e) storing and retrieving one photon in time-bin entangled photon pair. (f) The real parts of the reconstructed density matrices before (left) and after (right) storage. Reprinted from Ref.\,\cite{Saglamyurek2015NP}.}
\label{Franson/entanglement-basedquantumnetwork.pdf}
\end{figure}
An ambitious goal of quantum information technology is to build a quantum network where the quantum resources, such as coherence and entanglement, can be exchanged, routed, and manipulated \cite{Kimble2008Nature,Wehner2018Science,Wei2022LPR}. However, the distribution of entangled photons among different network nodes faces a challenge due to channel loss. As the photon propagation distance between two nodes increases, the probability of successfully distributing the entanglement exponentially degrades. One approach to overcome this problem is to construct a quantum repeater, which divides the task of long-distance quantum entanglement distribution into several tasks in shorter sections and utilizes entanglement swapping between neighboring sections \cite{Sangouard2011RMP}. Numerous studies have been conducted on quantum repeaters \cite{Sangouard2011RMP,Briegel1998PRL,Krutyanskiy2023PRL}, and here we will focus on the key elements in quantum repeaters which involves Franson interferometer.

In entanglement swapping, two pairs of entangled photons send one photon to a middle station where a Bell state measurement (BSM) is performed using the two arrived photons. If the two-photon state at the BSM node is projected onto a specific Bell state, the two remaining photons become entangled. The entanglement swappings of energy time and time-bin have been carried out experimentally, and Franson interferometers have been used to evaluate the performance of the entanglement swapping \cite{Tsujimoto2020NJP}.

For energy-time entanglement swapping, the setup used by Halder et al. in Ref.\,\cite{Halder2007NP} is shown in Fig.\,\ref{Franson/entanglement-basedquantumnetwork.pdf}(a), in which two PPLN waveguides generate two pairs of entangled photons. A BS followed by two single photon detectors is used to project two uncorrelated photons onto the Bell state $\left| \varPsi ^- \right>$, and two fiber UMZIs are used to measure the Franson interference. The results in Fig.\,\ref{Franson/entanglement-basedquantumnetwork.pdf}(b) clearly show nearly constant single counts and cosine coincidence counts with a visibility of 63 $\pm$ 2$\%$, indicating successful entanglement swapping if the two photons are assumed to be in a Werner state. For time-bin entanglement swapping, Riedmatten et al. first reported an experiment entangling two non-interacting photons separated by more than 2 km of fiber \cite{Riedmatten2005PRA}. They achieved a Franson interference visibility of 80 $\pm$ 4$\%$ after swapping. In 2015, Jin et al. realized time-bin entanglement swapping with quantum-memory-compatible photons \cite{Jin2015PRA}. In 2017, Sun et al. have first implemented a field test of entanglement swapping with two independent time-bin entanglement sources located 12.5 km apart \cite{Sun2017Optica}. In this case, when the two sources are connected by fiber of 25.3 km, a Franson interference visibility of 79.9 $\pm$ 4.8$\%$ was obtained, confirming the violation of Bell inequalities after entanglement swapping \cite{Sun2017PRA}.

Quantum memory is another crucial element in a quantum repeater, and its ability to store the time-bin encoded qubits has been evaluated using Franson interferometers. In 2011, Saglamyurek et al. stored the 795 nm photon in a Ti:Tm:LiNbO$_3$ waveguide quantum memory and measured its time-bin entanglement with its delayed twin, a 1532 nm photon generated through the doubly pulsed laser-pumped SPDC process \cite{Saglamyurek2011Nature}. The two-photon density matrices before and after storage were constructed through measurements via Franson interferometer, yielding an input/output fidelity of 95.4 $\pm$ 2.9$\%$. In 2015, the same group achieved the first demonstration of telecom-band quantum memory \cite{Saglamyurek2015NP}. The setup of the experiment is illustrated in Figs.\,\ref{Franson/entanglement-basedquantumnetwork.pdf}(c), (d) and (e). In Fig.\,\ref{Franson/entanglement-basedquantumnetwork.pdf}(c), time-bin entangled photons at 795 nm and 1532 nm were generated via the SPDC process. After passing through a filter, the 1532 nm photons were coherently stored in an atomic frequency comb (AFC) prepared in erbium ions doped fiber shown in Fig.\,\ref{Franson/entanglement-basedquantumnetwork.pdf}(e). The time-bin entanglement between the 795 nm photons and the 1532 nm photons retrieved from AFC was analyzed using the Franson interferometer depicted in Fig.\,\ref{Franson/entanglement-basedquantumnetwork.pdf}(d). The reconstructed density matrices, shown in Fig.\,\ref{Franson/entanglement-basedquantumnetwork.pdf}(f), clearly demonstrated an input–output fidelity of 97.1 $\pm$ 4.9$\%$. In 2021, Rakonjac et al. demonstrated energy-time entanglement between a telecom photon and an AFC-based solid-state quantum memory capable of working in both excited-state storage and spin-wave storage \cite{Rakonjac2021PRL}. In both cases, the quality of the entanglement storage was measured using the Franson interferometer and conditional fidelities of 92$\%$ and 77$\%$ were obtained, respectively.

 \subsubsection{Quantum spectroscopy and imaging}
Similar to communication and computing, the fields of spectroscopy and imaging can also be remarkably enhanced by quantum resources, like quantum correlations and entanglements \cite{Mukamel2020JPB,Moreau2019NRP}. Several works has investigated the improvement of molecular spectroscopy by the introduction of energy-time entangled photon pairs \cite{Tabakaev2021PRA}. As a tool to measure or manipulate energy-time entanglement, it is not surprising that the Franson interferometer could be used in molecular spectroscopy. In 1989, Franson already pointed out that the interferometer may be used in coherent Raman spectroscopy \cite{Franson1989Springer}. In 2013, Raymer et al. proposed a scheme to incorporate energy-time entangled photons and a Franson interferometer into fluorescence-detected two-dimensional optical spectroscopy, in order to get some advantages not possible in classical spectroscopy \cite{Raymer2013TJPCB}. In this scheme, the entangled photon pairs passing through the Franson interferometer are used to excite the two-photon absorption in special molecules. High time and frequency resolution can be simultaneously accessed by measuring the weak fluorescence signal and the exciting fields. In the field of quantum imaging, Gao et al. proposed and experimentally realized a coincidence imaging scheme that could achieve nonlocal erasure and correction of an image of a phase object by constructing a modified Franson interferometer \cite{Gao2019APL}.

\section{Open questions and future directions}

 Quantum interferometers have proven to be powerful tools in various fields, including quantum physics, metrology, and quantum information processing.  While significant progress has been made in the development of quantum interferometers, there are still open questions and future directions that researchers are actively exploring. Here are a few of them:

1. Integrated optics:  Transitioning quantum interferometers from laboratory demonstrations to practical applications is a crucial step in realizing their full potential. Miniaturization and cost reduction are  key factors in achieving this goal. Integrated optics,  offers a promising direction for addressing these challenges \cite{Wang2020NP,Pelucchi2021NRP}.
Integrated optics involves fabricating quantum interferometers on chip-scale platforms, such as silicon or lithium niobate, leveraging established techniques from the semiconductor industry \cite{FengLanTian2023Optica,Liu2022Chip}. This approach allows for the integration of multiple components, such as waveguides, beam splitters, and detectors, onto a single chip, enabling compact and cost-effective devices. Researchers have made significant progress in this area, with demonstrations of on-chip quantum interferometers and other quantum photonic devices.

2. Quantum-enhanced measurements: Quantum interferometers can surpass the sensitivity limits of classical devices by exploiting quantum entanglement and squeezing \cite{Giovannetti2004Science}. Exploring new strategies for generating and utilizing entangled states and squeezed states can lead to further improvements in measurement precision. This includes investigating novel quantum states and developing efficient methods for their preparation and detection.

3. High-dimensional applications: Exploring high-dimensional applications of quantum interferometers is an exciting direction \cite{Erhard2020NRP}. Traditional interferometers work with two-level quantum systems, such as photons in a superposition of two paths. However, by using higher-dimensional quantum systems, such as photons in higher-dimensional spatial or frequency modes, it is possible to achieve increased information capacity and improved interference effects. High-dimensional quantum interferometry has potential applications in secure quantum communication, quantum cryptography, and quantum computing.

4. Multiparameter quantum interferometry: Extending quantum interferometry to measure multiple parameters simultaneously is an active area of research \cite{ Hou2021SA, Hou2018NC}. By using multimode quantum states and advanced interference schemes, it is possible to achieve simultaneous measurements of multiple physical quantities with enhanced precision. Developing practical methods for multiparameter quantum interferometry has applications in fields such as gravitational wave detection, magnetic field sensing, and biological sensing.

5. Integration with other quantum technologies: Quantum interferometers can be combined with other quantum technologies, such as quantum memories, quantum communication systems, or quantum computers, to enhance their capabilities and enable new applications \cite{Wei2024npj}. Exploring the integration of quantum interferometers with other quantum devices and platforms is an important direction for future research.

These are just a few examples of the open questions and future directions for quantum interferometers. Continued research in these areas will contribute to the advancement of quantum technologies and our understanding of fundamental quantum principles.

\section{Conclusion}
In conclusion, this review has examined three prominent quantum interferometers: the HOM interferometer, the N00N state interferometer, and the Franson interferometer. We first provided an overview of the theory underlying each interferometer, including both single-mode and multi-mode theories. We then categorized these interferometers into distinct categories on the basis of their unique characteristics. Finally, we explored the diverse applications of interferometers in various fields such as quantum communication, quantum computation, quantum metrology, and quantum imaging. We hope that this review paper will contribute to the ongoing development of quantum interferometer theory and foster increased interest and innovation in the exciting realm of quantum technology.

\section*{Acknowledgements}
The authors thank Prof. Zheshen Zhang and Prof. Ryosuke Shimizu for helpful discussions.  
This work was supported by the National Natural Science Foundations of China (Grant Numbers 92365106, 12074299 and 11704290) and the Natural Science Foundation of Hubei Province (2022CFA039).

%

\end{document}